%% file: main.tex
\documentclass[10pt,sigconf,letterpaper]{acmart}

\renewcommand\footnotetextcopyrightpermission[1]{} % removes footnote with conference information in first column
\renewcommand\footnotetextcopyrightpermission[1]{}
\usepackage[utf8]{inputenc}
\usepackage{color,subfigure, url, multirow, multicol}

\acmConference{}
\date{}
\author{Jelena Mirkovic}
\affiliation{\institution{USC Information Sciences Institute}
}
\email{mirkovic@isi.edu}
\author{Yebo Feng}
\affiliation{\institution{University of Oregon}
}
\email{yebof@uoregon.edu}
\author{Jun Li}
\affiliation{\institution{University of Oregon}
}
\email{lijun@cs.uoregon.edu}
\begin{abstract}

During the 2020 pandemic caused by the COVID-19 virus, many countries implemented stay-at-home measures, which led to many businesses and schools moving from in-person to online mode of operation. We analyze sampled Netflow records at a medium-sized US Regional Optical Network to quantify the changes in the network traffic due to stay-at-home measures in that region. We find that human-driven traffic in the network decreases to around 70\%, and mostly shifts to local ISPs, while VPN and online meeting traffic increases up to 5 times. We also find that networks adopt a variety of online meeting solutions and favor one but continue using a few others. We find that educational and government institutions experience large traffic changes, but aim to keep their productivity via increased online meetings. Some scientific traffic also reduces possibly leading to loss of research productivity. Businesses mostly lose their traffic and few show VPN or online meeting activity. Most network prefixes experience large loss of live addresses but a handful increase their liveness. We also find increased incidence of network attacks. 
Our findings can help plan network provisioning and management to prepare for future possible infection outbreaks and natural disasters.

\end{abstract}

\keywords{COVID-19 network traffic; traffic analysis; traffic measurement}

\begin{document}
\title{Measuring Changes in Regional Network Traffic Due to COVID-19 Stay-at-Home Measures}
\maketitle
\section{Introduction}
In early 2020 COVID-19 infections started showing up around the world. By early March, the World Health Organization declared COVID-19 a pandemic and countries around the world started implementing stay-at-home orders to slow the spread of the infections. In the United States many states implemented stay-at-home orders starting around mid-to-late March. The orders had profound effect on residents' daily lives. Among other things these orders led to closure of in-person operation of non-essential businesses, and of all educational institutions. All schools and universities, shifted to an ``online'' or ``virtual'' mode relying on the Internet to serve their students. Many businesses that could operate remotely made the same operational changes, and leveraged the Internet to deliver services to their customers (e.g., telehealth, virtual yoga classes, virtual workouts, etc.). Almost overnight, the Internet became a critical commodity, a lifeline to keep businesses afloat, students learning and employees earning their livelihood. 

We seek to quantify the network traffic changes caused by this major shift in how we work and study, in one regional network.
% We leverage sampled Netflow records collected at a mid-size regional optical network in the United States, which we will refer to as RONX in the rest of the paper.
We leverage sampled Netflow records collected at the Front Range GigaPoP (FRGP)~\cite{frgp}, a mid-size regional optical network in the United States.
This network serves regional educational, research and government institutions, local residences, as well as several small businesses and non-profits, and connects them to the rest of the Internet. Its usual load is around 120 Gbps. In this particular region, stay-at-home orders were implemented on March 26 at the state level, and between March 14 and March 30 at the level of individual schools and universities. We thus consider the period between March 14 and March 30 as transition period and study how network traffic changed before and after this transition. 

We find that human-driven traffic in the network decreases to around 70\% of its pre-transition value, and mostly shifts to local ISPs. This agrees with movement of people from education and business organizations to homes. \textit{This finding can help us identify network links that may need added capacity, such as links between local ISPs and education/government institutions.}

We further find that inbound VPN and https traffic increases for most local organizations, signaling two important channels to maintain connectivity between home and work/school during the crisis. Outbound traffic increases for most online meeting applications, signalling another important channel for connectivity. Further, since most employees and students have physically moved from local organizations (except for ISPs) this increase of online meeting traffic possibly signals lack of split VPN tunnels. Thus online media traffic in presence of VPN between home and work/school unnecessarily travels through organizations instead of going directly from home to service provider. 
\textit{This finding can help organizations invest in split VPN solutions and in educating their users on how to efficiently use these solutions. }

We also find that networks adopt a variety of online meeting solutions, and usually favor one but continue using several others. While this likely creates frustration at users, it may also provide resilience should organizations need to divest from some service providers, e.g., due to change in their business model, vulnerabilities in their software or unreliable service quality.  \textit{This finding can help organizations invest in selection and user training for two chosen online meeting solutions, to prepare for future outbreaks. }

We find that educational and government institutions experience large traffic changes, but aim to keep their productivity via increased online meetings.  Some scientific traffic also reduces, possibly leading to loss of research productivity. 
Local businesses' traffic experiences mostly reductions across the board, and we see less adoption of VPN and online meeting applications. 
\textit{This finding can help businesses invest in VPN and online meeting solutions, as well as user training, to prepare for future outbreaks. Also, we need infrastructure and processes to support continued scientific data processing and consumption, to ensure research continuity when working from home.}

Most network prefixes experience large loss of live addresses, but a handful increase their liveness, possibly due to increased use of VPN. These increases are largest at ISPs, followed by education and government institutions. \textit{This finding can help researchers that study address liveness, to draw attention to the opposite trends of liveness changes in certain, possibly dynamically allocated prefixes. }

We also find increased incidence of network attacks. 
after the transition period. \textit{This finding can help inform organizations that more effort is needed to safeguard against network attacks during stay-at-home periods in the future.} 

The rest of this paper is organized as follows.
After describing the related work in Section~\ref{sec:related},
we describe the datasets we use in Section~\ref{sec:datasets}.
We then detail the methodologies in Section~\ref{sec:methodology} and 
demonstrate our findings in Section~\ref{sec:fingings}.
We conclude the paper in Section~\ref{sec:concl}.

\section{Related Work}\label{sec:related}

In this section we discuss other works that study traffic changes due to COVID-19 and stay-at-home orders. We also discuss traffic classification approaches we considered, and the graphlet-based approach we elected to use.

\subsection{COVID-19 Traffic Changes}

Many blog posts and news articles examined impact of stay-at-home orders on network traffic~\cite{slidedeck, akamai, zoom, att, gaming, cdn, microsoft, comcast, fastly, ixp, cloudflare}. To summarize their observations: overall traffic has increased anywhere from 30\% to almost 800\% on some networks, gaming and streaming have seen large increases, and Zoom has gained in popularity -- its daily meeting participant number increased 20 times! Internet has mostly been able to cope with traffic increase without large changes in congestion~\cite{cloudflare}. Our paper looks to quantify changes at a smaller level of one regional network, and to examine how individual organizations' traffic mix has changed in response to stay-at-home orders. 

Closely related to our study is CENIC's recent report on traffic changes \cite{cenic}. Like FRGP, CENIC is a regional network serving many educational and research institutions. Our findings mostly match those in~\cite{cenic}: both networks see decrease in traffic, and both see increase in Zoom and VPN use, and in traffic being exchanged between local institutions and ISPs. But CENIC finds decrease in ingress traffic and increase in egress traffic for educational and research institutions (the ratio of ingress to egress goes from 1:3 to 1:5), while we find overall decrease in both ingress and egress traffic. We further dive deeper into various applications and traffic classes, and into specific institutions, investigating how their traffic mix has changed. Thus our research complements findings of~\cite{cenic}.

\subsection{Traffic Classification}

We use port-based and graphlet-based traffic classification to attribute traffic to applications, as described in Section \ref{sec:methodology}. Traffic classification is rich and well-explored area. Many researchers have noted that port-based classification works only for some popular applications, which have well-established ports (e.g., port 53 for DNS, port 80 for Web), but fails for emerging applications such as gaming, streaming and peer-to-peer~\cite{blinc, onthefly}. In BLINC~\cite{blinc}, Karagiannis et al. propose a multi-level traffic classification that uses a host's social behavior, 
its community behavior, and port-based features to build a model of this host's communication. Popular hosts that communicate with many others, on a few ports are servers. Graphlets are then used to represent the host's communication pattern (e.g., which ports it uses and which ports its peers use). Authors build a library of graphlets and use it to classify unknown hosts with high accuracy. We leverage their approach in our classification of media traffic (e.g., Zoom, Google Meet, etc., see Section \ref{sec:media}), with minor modifications. 

Other traffic classification approaches have been proposed, including packet-based classification~\cite{onthefly,dainotti2008classification}, clustering and machine-learning~\cite{erman, moore, mcgregor, bernaille2006early, zander, feng2019botflowmon}. These approaches require transport-level statistics such as packet inter-arrival time, or volume of traffic in each direction, which our sampled Netflow records do not have.
%\UOyebo{what about this BGPstream~\cite{bgpstream} paper? people can use BGPstream to check the historical routing of the packets, then infer the identity of packets (for example, if it is a packet towards Netflix, then it is more likely to be video streaming).}\jelena{you can see it's a packet towards Neflix by looking at the destination IP. why do you need BGPstream?} \UOyebo{1. One site can own different IPs at different times, so we need historical routing information to learn the destination in the past. BGPstream contains this info. 2. many sites employee CDNs to provide high availability and performance, IP lookup cannot accurately locate the service. For example, a Cloudflare IP could be a Netflix CDN at certain times.}

\section{Datasets}\label{sec:datasets}

In this Section we detail the datasets we use. We learn about network traffic patterns from the Netflow records collected at FRGP. We use the information about network prefixes from IPInfo.io to attribute
traffic to different organization and classify organizations by their purpose. We use BGP records from RouteViews~\cite{routeviews} to identify local prefixes in the region we observe, which are served by FRGP. 

\subsection{Network Traces}
\begin{figure}
    \centering
    \includegraphics[width=\columnwidth]{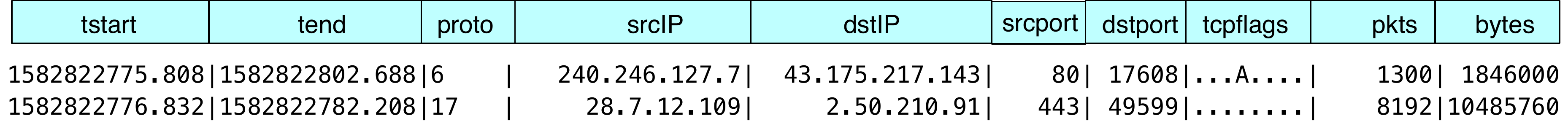}
    \caption{Netflow record fields we use}
    \label{fig:netflow}
\end{figure}

Our dataset contains sampled Netflow records from FRGP. The packets are sampled and converted into flows at four ingress points into FRGP. Two points sample 1 in 4,096 packets, and the other two sample 1 in 100 packets. Figure \ref{fig:netflow} gives an overview of the fields from a Netflow record, which we use, and two example records. A Netflow record offers a highly summarized view of the flow, including the time when the first and the last packet on the flow were observed, the IP addresses and ports of the communicating hosts, the transport protocol, the cumulative TCP flags, and the total number of packets and bytes exchanged. Netflow records are usually unidirectional, i.e., they capture traffic in one direction. One would have to pair flows in both directions to have a full view of communication. In our dataset's case, the flows are also sampled, and thus we cannot reassemble them into bidirectional communications.

The examples in Figure \ref{fig:netflow} illustrate what we can learn from a Netflow record in our dataset. In the first example, two hosts communicate using TCP protocol (proto=6). Since the first host uses a well-known port 80 and the second uses a high-numbered port we can infer that the first host is the server, and the second is the client. We see that server has sent 13 packets, and 18KB observed by the capture location, and some of these packets had acknowledgment flag set. No other flags were set. The packets were sampled with the 1:100 rate, and the processing software (\texttt{nfdump}) up-samples them to 1,300 packets and 1.8MB. In the second example, two hosts communicate using UDP protocol (proto=17). Again, the first host uses a well-known port 443 and the second uses a high-numbered port, so we infer that the first host is the server. The server host sent 2 packets and 2.5 KB. Because these were sampled with the 1:4,096 sampling rate, they are up-sampled into 8,192 packets and 10 MB.

Our dataset consists of 5-minute long Netflow record collections, from 10:20 on February 24 to 21:40 on May 21, 2020. To protect user privacy, the IP addresses in records are anonymized in a prefix-preserving manner, using CryptoPAN~\cite{cryptopan}.

\subsection{From IP Addresses to Organizations and Categories}

When we seek to attribute traffic to a specific organization, we partially deanonymize IP addresses in the following manner. The FRGP operators have allowed us to access the list of anonymized and original \slash 24 prefixes in the dataset. We use this list when needed to reverse the anonymization at the \slash 24 prefix level.
We then leverage IPinfo.io~\cite{ipinfo} to attribute each original prefix to its organization. The IPinfo.io offers information about prefixes and about autonomous systems, including the name of the organization, which owns the prefix, and the organization's category. The supported categories are ``hosting'', ``isp'', ``education'', and ``business''. 

\subsection{Local Organizations Served by FRGP}

We seek to understand changes in traffic for local organizations from the region served by FRGP. To learn identities of these organizations we collect from the RouteViews data~\cite{routeviews}  the BGP prefixes FRGP advertises. There are 9,618 advertised IPv4 \slash 24 prefixes, out of which 6,103 are visible in our dataset. More than 99.3\% of flows and 99.6\% of bytes in our dataset are sent to or by one of these prefixes. Most of them (97\% of flows and 90\% of bytes) are exchanged between a local organization and the rest of the Internet, and the rest are exchanged between two local organizations. We also analyze changes in traffic depending on the organization's category. We start from IPinfo.io categories, but then manually verify them for local organizations, since there are only 51 organizations served by FRGP. We also re-classify some ``business'' organizations as ``government'' if their domain name ends in ``.gov'' suffix. Our local organizations fall into the following categories: ``isp'', ``education'', ``business'' and ``government''. 

\subsection{Limitations}

Since our dataset contains Netflow records built from sampled traffic, we observe only a small portion  of packets flowing through FRGP and will miss short flows and infrequent sources and destinations. This limitation does not hinder our research direction, because we are interested in quantifying large traffic changes, at the organization and network level.

Another limitation is that Netflow records provide no information about individual packet dynamics, or packet payloads. We thus have to resort to using ports and IP addresses to classify traffic into applications, and to understand which host was a client and which was a server. Since we can only partially deanonymize an IP address, at the \slash 24 level, this further limits our inference accuracy. 

Yet another limitation is that Netflow records provide highly summarized TCP flag information, which introduces some uncertainty about the flow's state. For example if the \textit{tcpflags} field indicates ``...A..S.'', and the number of sampled packets is 2 or more, we cannot tell if there was a SYN packet, followed by an ACK packet or if the single packet contained both SYN and ACK flags.

Further, Netflow records do not contain information about individual packet timing, and cannot be used to evaluate quality of service, such as packet drops, delays or jitter.

Finally, we rely on IPinfo.io data to attribute \slash 24 prefixes to organizations and to classify organizations into categories. But IPinfo.io uses proprietary software to gather this data, and we have no way to independently verify its accuracy.

\subsection{Ethical Issues}
\label{sec:ethical}
Data collection is done by FRGP operators and their collaborators from a local education institution, EDX, on ongoing basis. 
We formulate a Memorandum of Agreement (MoA) with FRGP operators and the EDX team to stipulate the correct usage and accessibility of the data.

To protect the privacy of users and prevent data leakage, we set rigorous regulations for data analysis and storage. We list the regulations below:
\begin{enumerate}
    \item The IP addresses in Netflow records are anonymized in a prefix-preserving manner with CryptoPAN~\cite{cryptopan}, before storage. 
    \item All the data is stored on a restricted server by EDX team.  Besides the SSH port, all the ports on the server are closed, and no connections can be initated from the server. This minimizes the risk of accidental exfiltration of network data.  \item All data analysis is done on the restricted server.
    \item We are allowed to deanonymize the data at the \slash 24 prefix level -- this is also done only on the restricted server.
\end{enumerate}

\section{Methodology}
\label{sec:methodology}
In this section we detail our methodology for traffic classification, and for detection of traffic change direction and amount. We also explain our approach to identification of network attacks, which we analyze for changes due to stay-at-home orders.  

\subsection{Coarse Traffic Classes}

We first classify traffic into coarse classes based on protocol, port and tcpflag fields. We use protocol field to detect ICMP, TCP and UDP flows and tag the rest as ``otprot'' (other protocol).  We then classify TCP flows as ``syn'' if they have only TCP SYN flag set. These flows could be part of scans or part of short flows, where only the first packet was sampled. Unfortunately, our dataset offers no way to distinguish between these two categories. The rest of TCP flows (those that are not in ``syn'' category) and all UDP flows are potential candidates for application-based traffic classification.
More than 98\% of traffic is in this ``candidate'' class.

\subsection{Online Meeting and Gaming Traffic}
\label{sec:media}

Our next goal is to identify some popular online meeting and gaming application traffic, or ``mg-traffic'' for short, and label it as belonging to a specific application.

We start by creating a short list of popular online meeting and gaming applications. We focus on BlueJeans, Zoom, Skype, Google Meet, GoTo Meeting and WebEx for online meetings, and Steam for gaming. We chose to focus on these applications,  because they have seen wide usage in FRGP.

Classifying mg-traffic into its proper application is challenging, because each application may use a wide range of ports and servers. Each application provider publicly lists the ports that their application uses on their Web page~\cite{webwebex, webzoom, webskype, webbj, webgoto, webgmeet, webvalve}, as well as the list of IP prefixes hosting the application servers. We call these lists ``ground-truth ports'' and ``ground-truth prefixes,'' respectively. Providers publish ground-truth port and prefix information to help new customers adjust their firewall rules to allow application traffic. The ground-truth information is a good start for application traffic classification, but it is not sufficient. 

\begin{figure}
    \centering
    \includegraphics[width=\columnwidth]{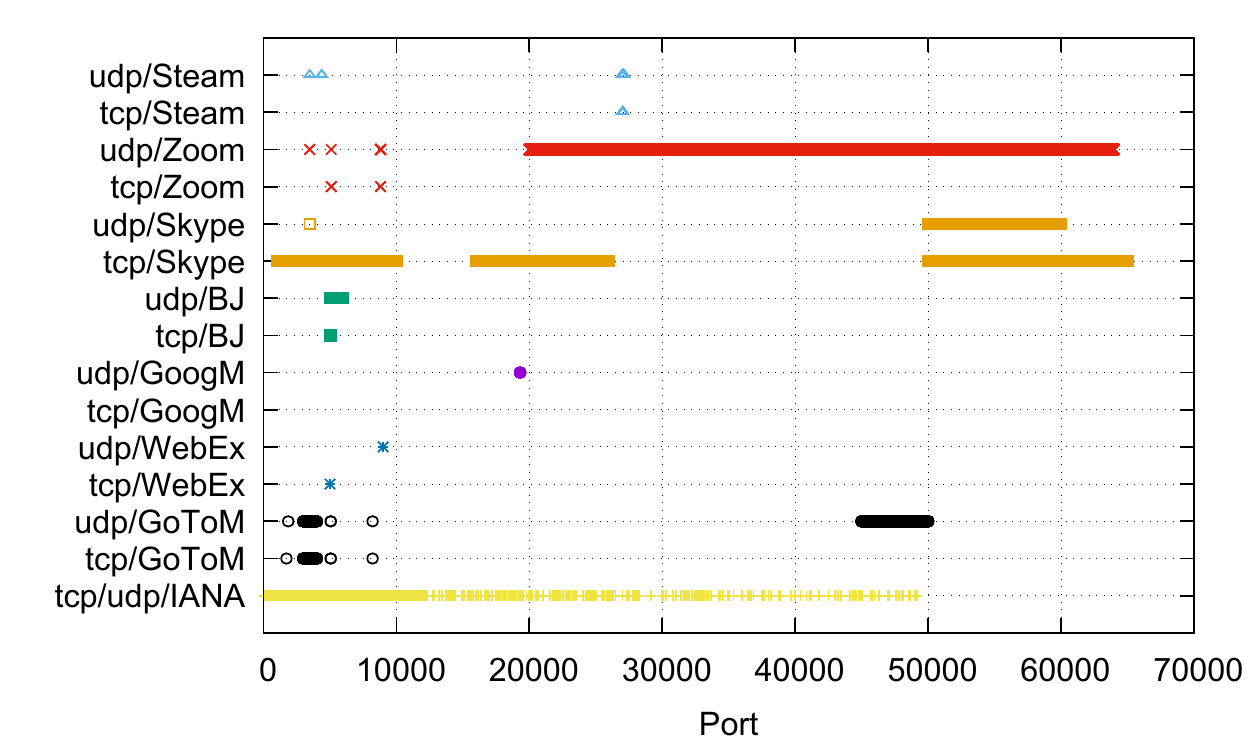}
    \caption{Online meeting and Steam ports, along with service ports allocated by IANA. There is a lot of overlap, with many applications using same port numbers.}
    \label{fig:mediaports}
\end{figure}

The ground-truth prefix list may not be exhaustive. A provider may strike a business deal with an ISP to host its servers, or it may deploy its servers in a cloud to scale up with the demand. The provider may update only some of its customers with this information, without updating their public Web page. Second, it is tempting to use ground-truth port information for port-based mg-traffic classification. But the ground-truth ports have a great overlap between applications. 
We illustrate this in the Figure~\ref{fig:mediaports}, where we show the IANA-allocated service ports (yellow line) and the ports that each application provider lists on their Web page as required for their product to work. Each application's ports are shown separately based on the transport protocol, but are colored the same color. The lower lines denote TCP ports, and the upper lines denote UDP ports. There is a significant overlap between applications, and almost no ports are exclusively used by a single application. 

\begin{figure*}
    \centering
    \includegraphics[width=5in]{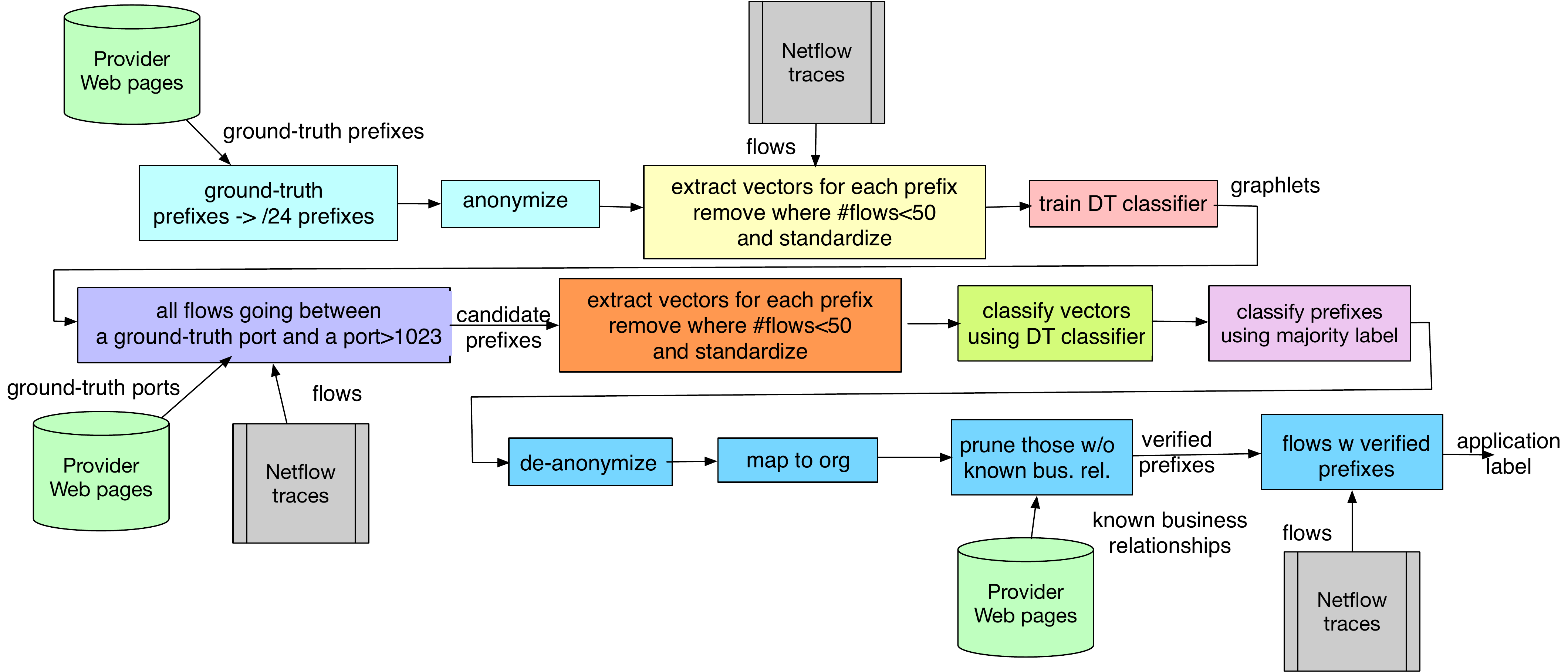}
    \caption{Our approach to mg-traffic classification.}
    \label{fig:mcclass}
\end{figure*}

We adopt the graphlet-based approach outlined in~\cite{blinc}, to classify flows into applications for mg-traffic. Graphlet-based approach classifies flows by first identifying application servers, clustering the servers based on the similarity of the count and the numbers of their open ports, and then labeling all traffic to these servers with a given application's label. We identify application servers by using a multi-step process, as illustrated in Figure~\ref{fig:mcclass}.  First, we create the list of all the /24  ground-truth prefixes. If any prefix is listed on the provider's Web site as smaller than  /24, we extend it into its covering /24 prefix. We then map these /24 prefixes into anonymized prefixes, using FRGP-provided mapping information. Second, we extract port numbers associated with each prefix, for each hour of the dataset, and the number of flows going to that prefix in an hour. We denote this representation of the prefix's activity in a given hour ``a vector''. We then remove vectors that had fewer than 50 flows, to ensure we have high-quality data for training and testing. This removes 152 out of 723 ground-truth prefixes, but these prefixes account only for 0.3\% of traffic volume exchanged with all ground-truth prefixes. 
Third, we order the port numbers in this vector
 in increasing order, and pad it in front with zeros to standardize vectors to the same size. We label each vector with its application. Instead of manually devising graphlets we use a decision-tree classifier on this set of vectors, to learn how to classify vectors into applications. We randomly divide our vector set into 50\% training and 50\% testing. The classifier achieves 99.4\% accuracy on the vector test set.  To further quantify the accuracy of our classification, we classify our entire dataset (training and testing). When the prefix appears in multiple vectors we take the majority vote for its final label. This way, we correctly classify
568 out of the remaining 571 prefixes, and correctly classify 99.9\% of all traffic exchanged with ground-truth servers. Our graphlets are thus able to accurately classify mg-traffic into applications.
Fourth, we use the ground-truth list of ports to identify all \textit{candidate} prefixes in our dataset. These are prefixes that receive traffic to, or send traffic from a port on ground-truth list for a given application, and the other port in that communication is a user or dynamic port ($>$ 1,023). We identify 2,386,514 candidate prefixes contributing to 40 times more traffic than ground-truth prefixes. Fifth, for each candidate prefix we extract vectors using the same approach as for ground-truth prefixes. We discard vectors with fewer than 50 flow samples per hour. This leaves us with 137,335 candidate prefixes contributing to 37 times more traffic than ground-truth prefixes. Sixth, we apply our decision tree classifier to this set of strong candidate prefixes and label them with application labels using majority vote over the labels of their associated vectors. However, there is high likelihood of mislabeling in this step, since our classifier does not have ``no-application'' class. Thus candidate prefixes that do not have servers serving any of our chosen applications will be mislabeled. We prune these candidates in our seventh and final step. 
We de-anonymize each strong candidate prefix, and map it to the organization that owns it. We then keep only those prefixes whose ownership is consistent with their application label. This may mean that the owner of the prefix is also the owner of the application (e.g., Google for Google Meet or Cisco for WebEx) or that there is a known business relationship between the prefix's owner and the application provider. For example Cisco is owner of the WebEx application, but it is also promoter of Zoom, since it offers Zoom Connector service for Cisco~\cite{ciscoconector}. As another example, BlueJeans lists AT\&T and Centurylink on its partner list~\cite{bjbus}.
Table \ref{tab:business} lists all the business relationships we considered and the sources of this information. 
At the end of this step we end up with 21,447 verified prefixes that contribute around 50\% of traffic contributed by ground-truth prefixes. Jointly the verified and ground-truth prefixes, form the list we denote \textit{known-mg-prefixes}. % media destinations. Our final count of prefixes is shown in Table \ref{tab:finalmedia}.
Table \ref{tab:finalmedia} shows the counts of ground-truth and verified prefixes.

\begin{table}
    \centering
    \begin{tabular}{c|c|c}
        \textbf{application} & \textbf{bus. partner} & \textbf{source}  \\ \hline
        BlueJeans & AT\&T & \cite{bjbus} \\
        BlueJeans & CenturyLink & \cite{bjbus} \\
        BlueJeans & Level3 & \cite{bjl3} \\
BlueJeans & Microsoft & \cite{bjbus} \\
Google Meet & Google & owner org \\
Zoom & Amazon & \cite{zoomam} \\
Zoom & Cisco & \cite{ciscoconector} \\
WebEx & Cisco & owner org\\
Webex & AT\&T & \cite{webexatt} \\
Webex & CenturyLink & \cite{webexcl} \\
Webex & Amazon & \cite{webexam} \\
GoTo & Logmein & owner org \\
Skype & Microsoft & owner org \\
    \end{tabular}
    \caption{Online meeting applications and known business partners}
    \label{tab:business}
\end{table}

\begin{table}[]
    \centering
    \begin{tabular}{c|c|c}
        \textbf{Application} & \multicolumn{2}{c}{\textbf{/24 prefixes}} \\ \cline{2-3}
        & \textbf{Ground-truth} & \textbf{Verified}  \\ \hline
        BlueJeans & 76 & 56 \\
        Google Meet & 3 & 90 \\
        GoTo Meeting & 527 & 0 \\
        Skype & 6,360 & 3,119 \\
        Webex & 1,121 & 3,255 \\
        Zoom & 574 & 14,936 \\
        Steam & 147 & 0 
    \end{tabular}
    \caption{Final counts of known-mg-prefixes}
    \label{tab:finalmedia}
\end{table}

\subsection{Application Traffic}

We next focus only on the ports whose traffic comprises a significant portion of the volume in our dataset. 
We  identify those applications (based on ports or our known-mg-prefix list)  whose flows contribute at least 1\% of the total volume for at least 1/7th of time before or after the transition period. We chose 1/7th to be able to represent applications that may be large enough only on certain days of the week. These applications are: SSH (port 22), Web (port 80), Unidata (port 388), https (port 443), rsync (port 873), Zoom (various ports, but one of the endpoints is from known-mg-prefix list for zoom), VPN (ports 4500 and 4501) and Perfsonar (port 5201). Other than Zoom, the rest of the applications use only the limited set of well-known ports and we thus believe port-based identification is accurate for them.  
We further extend our list of applications of interest to include related applications with the similar purpose: E-mail (ports 25, 110, 995, 143, 993, 2525, 465) telnet (port 23, related to SSH), FTP (ports 20 and 21, related to SSH and rsync), and our chosen online meeting and gaming applications (Skype, BlueJeans, WebEx, GoTo Meeting, Google Meet and Steam). We also include DNS and NTP traffic. DNS accompanies many other protocols, such as Web, https, SSH, telnet, FTP, etc., and is mostly human-driven. NTP is automatically generated and thus should exhibit different dynamics. 
We also include application traffic to non-standard ports for our chosen applications: Web (81, 82, 8080, 8090), VPN (4502) and https (4433). Jointly, traffic associated with our chosen applications represents more than 
90\% of the traffic in our dataset.

Finally, we tag traffic that is exchanged between two ports, whose numbers are both higher than 10,000 as ``highhigh'' class. While IANA list of services defines many services in that number range, they are not very popular. On the other hand other streaming or gaming applications, or peer-to-peer services can use these ports. We analyze highhigh traffic in Section \ref{sec:highhigh}.

\subsection{Anomaly Detection}

As the Netflow data we analyze are sampled with a very low rate (1/4096), we can only hope to detect large anomalies such as Distributed Denial of Service (DDoS) attacks and bandwidth usage anomalies.

We employ two existing approaches to detect the anomalies in the network. Our first approach uses ASTUTE-based anomaly detection program~\cite{silveira2010astute}.
We use a sliding time window to monitor the traffic of several protocols and applications---the overall traffic, NTP traffic, DNS traffic, ICMP traffic, and SYN/SYN-ACK flows. 
The ASTUTE approach assumes that in legitimate cases the volume change of flows over short timescales tend to cancel out each other, and identifies strongly correlated traffic flow changes as anomalies.
Our second approach uses FastNetMon~\cite{odintsovfastnetmon}, a threshold-based commercial DDoS detection software that offers fast detection for many types of DDoS attacks, such as SYN flood, SYN-ACK flood, ICMP flood, and NTP amplification attack.

We consider a detected anomaly as true positive if it satisfies one of the following conditions: 
(1) both detection approaches detected the anomaly, or
(2) only one approach detected the anomaly,  but the peak volume of the anomaly is at least twice the expected traffic volume. 
We focus on five types of anomalies: DNS amplification attack, NTP amplification attacks, ICMP floods, SYN/SYN-ACK floods, and other unspecific increases in bandwidth usage.

To better quantify the traffic change during an anomaly, we define \textbf{peak intensity index} $\zeta$, calculated as $\zeta = V_{peak} / V_{exp}$, where $V_{peak}$ denotes the peak volume of the anomaly and $V_{exp}$ denotes the expected traffic volume. For the anomaly with a short duration (within 30 minutes), we treat the traffic volume right before the anomaly as $V_{exp}$. For the anomaly with a longer duration, we calculate $V_{exp}$ by statistically averaging legitimate traffic volumes at the same time in the surrounding seven days.

\subsection{Statistical Tests}
\label{sec:statis_test}

Throughout our study we ask for each traffic type and each communication of interest ``were there large changes due to stay-at-home orders''. One way to answer these questions would be to compare statistical measures of traffic (e.g., means or medians) directly. However, some traffic classes experience very strong diurnal and weekly patterns, which introduces significant variance. 
We thus use hypothesis testing to detect significant changes in traffic and to quantify them.

We use the Wilcoxon-Mann-Whitney (WMW) test, which is a ``nonparametric test of the null hypothesis that it is equally likely that a randomly selected value from one population will be less than or greater than a randomly selected value from a second population''~\cite{wikipedia}. In other words, this test helps us establish if two measurements -- one taken before and one taken after the transition period -- are drawn from the distributions that have same or different means. In case when the test concludes that the means are different, it can also inform us if the ``before'' values have a mean that is smaller or larger that the ``after'' values.

The WMW test does not make any assumptions about the nature of the distributions it measures, but it requires that values are independent of each other, which is true in our case.
The WMW test measures a null hypotheses that samples from two populations have no difference in means. The test produces a $p$-value, which should be below $0.5$ to reject the null hypotheses.

We run the WMW test using R software, with parameters ``less'' and ``greater'' to detect the direction of change, if any. If both tests produce a $p$-value higher than $0.5$ we conclude that the distributions are the same. Otherwise, the test that produced the $p<0.5$ tells us the nature of the traffic's change (increased or decreased). 
To quantify the change, we adopt different approaches, depending if we work with the 5-minute or daily values. For 5-minute values we 
calculate ratio of the \textit{medians} after and before the transition period. We use the medians because 5-minute traffic is bursty and occasional high values can skew the mean, while median is much more stable. For daily values comparison we calculate ration of the \textit{means} after and before the transition period.

\begin{figure}
    \centering
    \includegraphics[width=\columnwidth]{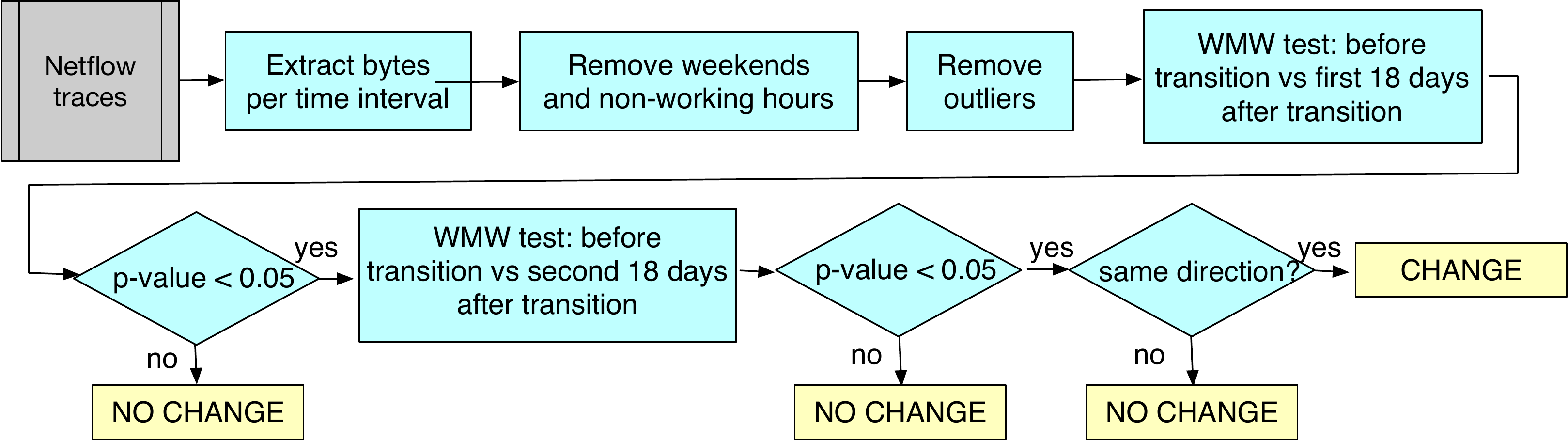}
    \caption{Workflow: how we establish if traffic of a certain type has changed}
    \label{fig:workflow}
\end{figure}

\subsection{Limitations}

There are several limitations to our traffic classification approach. First, we extend ground-truth prefixes smaller than /24 to its covering /24. This may lead to us collecting more than just that application's traffic in step 2 and creating an imprecise classifier. Unfortunately, since we can only deanonymize our dataset at /24 prefix level we cannot avoid this limitation.

The second limitation lies in the way we identify candidate prefixes. While our classifier's accuracy is high between different application classes, we have no way to identify prefixes that host other applications, which we do not model. Thus, it is possible that we misclassify e.g., a prefix hosting a streaming service like Hulu as one of our chosen applications. We hope that we have sufficiently mitigated this risk by post-filtering our candidate prefixes using known business relationships between providers. However, we have no ground truth to quantify how many verified prefixes may still be misclassified. 

The third limitation comes from our port-based classification of traffic other than mg-traffic class. It is well known that some applications may masquerade themselves as others by using non-standard ports. This is especially true for applications overloading web and https ports, since these ports are usually open in firewalls.

The fourth limitation comes from the detection of anomalies. As data is sampled at high rate we may miss some lower-rate anomalies. We focus on volumetric anomalies such as DDoS attacks and bandwidth usage anomalies, which can be identified with sampled data. We also lack ground truth about anomalies and cannot estimate the accuracy of our approach.  We utilize two approaches to anomaly detection to ameliorate this limitation. 

While all these limitations are serious, we believe they equally affect all data in our dataset, i.e., before and after the transition. Since we study traffic \textit{changes}, we believe our findings still hold in spite of the above limitations.

\section{Findings}
\label{sec:fingings}

We start by discussing large traffic changes and then delve deeper into applications and organizations whose traffic patterns have changed.

\subsection{Large Traffic Categories}

\begin{figure*}
\subfigure[All traffic]{
   \includegraphics[width=.3\textwidth]{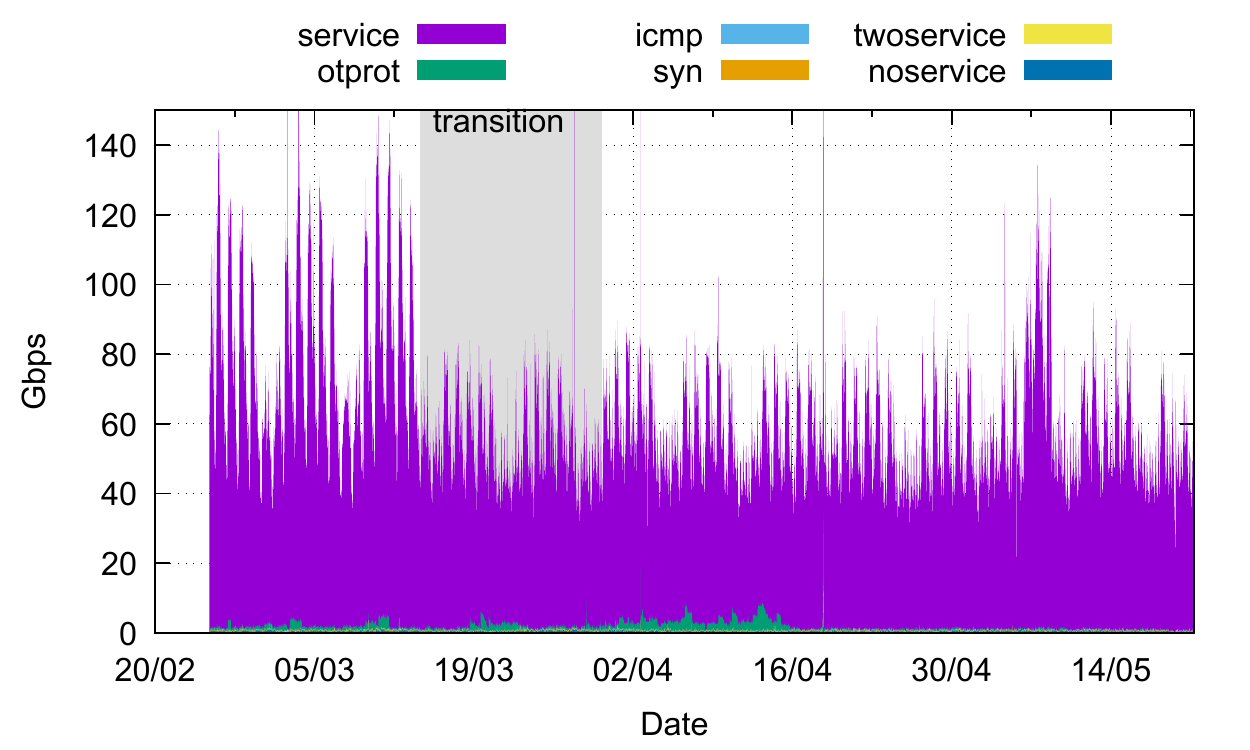}
   \label{fig:all}
 } 
 \subfigure[Zoomed -- smaller traffic categories]{
   \includegraphics[width=.3\textwidth]{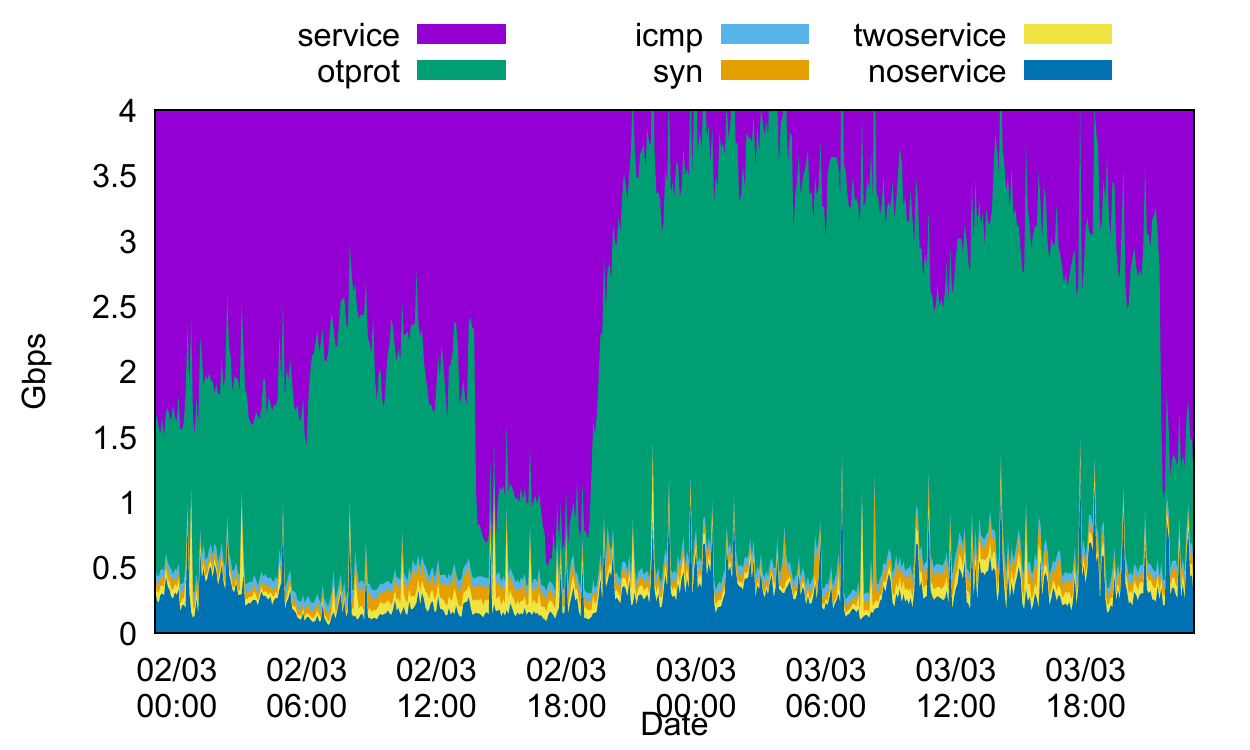}
   \label{fig:zoom}
 } 
 \subfigure[Diurnal patterns]{
   \includegraphics[width=.3\textwidth]{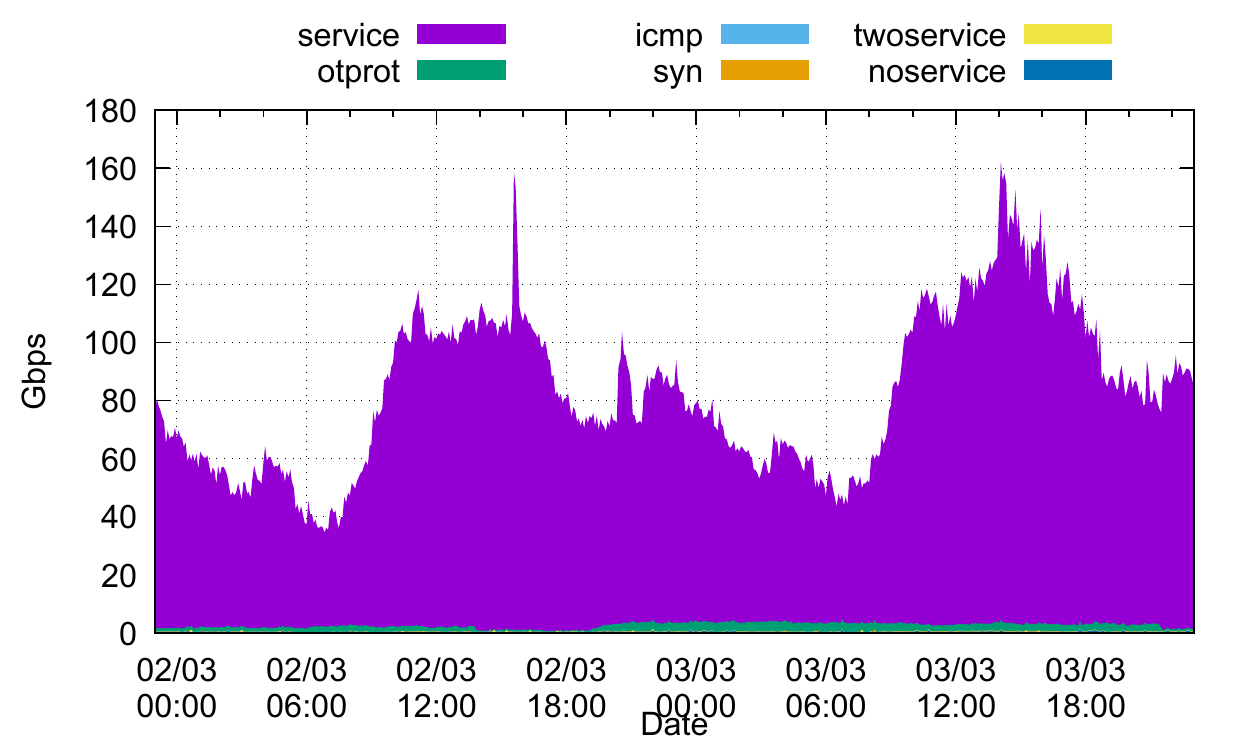}
   \label{fig:diurn}
 } 
\caption{Traffic volume per category. Service and ``otproto'' traffic decreases after the transition period.}\label{fig:total}
\end{figure*}

Figure \ref{fig:all} shows the total volume per second over our monitoring period. More than 97\% of traffic is service traffic. That traffic shows strong diurnal patterns, and reduces rapidly as soon as the transition starts. Other types of traffic are at much-lower levels, with the largest category being ``otproto'' traffic, which is mostly IPv6. That traffic also exhibits diurnal patterns, and also reduces drastically a few weeks after the transition.

We perform the WMW test on each traffic category. It detects that service traffic reduces to 70\% of the before-value, syn traffic reduces to 94\%, noservice traffic reduces to 66\%, twoservice reduces to 66\% and icmp traffic reduces to 97\%. 
The ``otproto'' traffic reduces only in the second test period, and thus we also label it as ``no change''. 
%To illustrate the sensitivity of WMW test we show the time-based plots and the cdf of service and twoservice categories of traffic in Figure \ref{fig:cdfs}. The service traffic clearly reduces and this is visible both in the time-based plots and in the cdf. The twoservice traffic's change is subtle in time-based plots and easy to miss, however the cdf plots reveal that the ``after'' distributions are clearly lower than the ``before'' distribution -- this is the quality measured by the WMW test. 

%\begin{figure*}
%\subfigure[Service traffic time plot]{
%   \includegraphics[width=.22\textwidth]{figs/service.time.pdf}
%   \label{fig:str}
% } 
% \subfigure[Service traffic cdf]{
%   \includegraphics[width=.22\textwidth]{figs/service.cdf.pdf}
%   \label{fig:scdf}
% } 
% \subfigure[Twoservice traffic time plot]{
%   \includegraphics[width=.22\textwidth]{figs/twoservice.time.pdf}
%   \label{fig:tstr}
% } 
% \subfigure[Twoservice traffic cdf]{
%   \includegraphics[width=.22\textwidth]{figs/twoservice.cdf.pdf}
%   \label{fig:tscdf}
% } 
%\caption{Time-based and cdf-based plots of service and twoservice traffic's change before and after the transition period.}\label{fig:cdfs}
%\end{figure*}

\subsection{Application Traffic}

We next focus on understanding how large application traffic has changed. 

\begin{figure}
    \centering
    \includegraphics[width=\columnwidth]{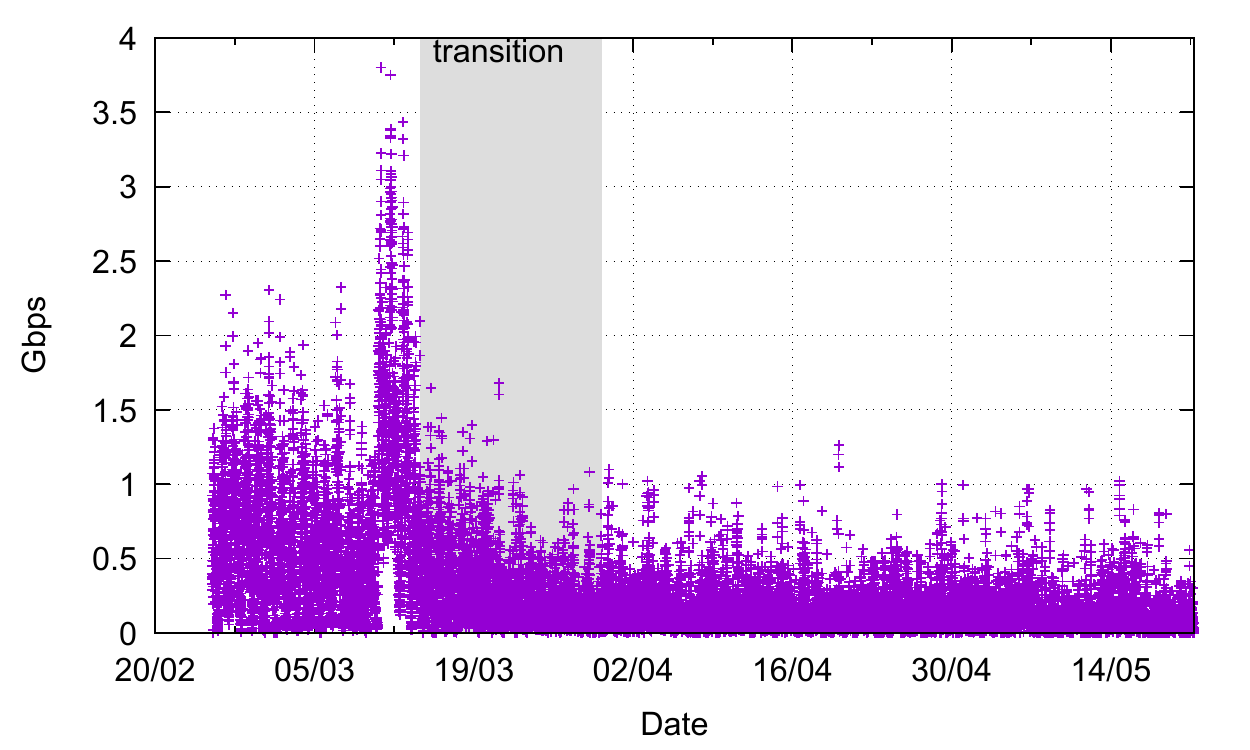}
    \caption{Valve traffic change (all hours)}
    \label{fig:valve}
\end{figure}

Table \ref{tab:applist} shows the list of these applications and their contribution to total volume before the transition period. We then apply MWM test on these applications and show the direction and percentage of change in the last four columns of Table  \ref{tab:applist}: columns 3 and 4 show change for work-hour traffic, and columns 5 and 6 show change for rest-hour traffic. Human-driven application traffic, such as https, web, gaming (Steam), etc. reduces, due to stay-at-home orders. We expect this traffic simply shifted with users. As users moved from local education and business networks, to their homes, the traffic moved with them and is now being sourced by residential ISPs. Since these ISPs usually connect to service providers directly through upstream ISPs, this traffic bypasses our vantage point after the transition. We illustrate the magnitude and dynamics of the change in Figure \ref{fig:valve} by showing the change in the Steam traffic. The traffic visibly decreases at the start of the transition period. 

Data-driven traffic mostly decreases too, possibly shifting with users. This is true for all categories, except for Unidata, where traffic slightly increases. Automated traffic  changes very little. Other traffic exhibits a mixed pattern of change. Some categories reduce, similar to human-driven traffic, while others change very little, similar to automated traffic.

\begin{figure*}
\subfigure[Google Meet traffic -- rest hours]{
   \includegraphics[width=.22\textwidth]{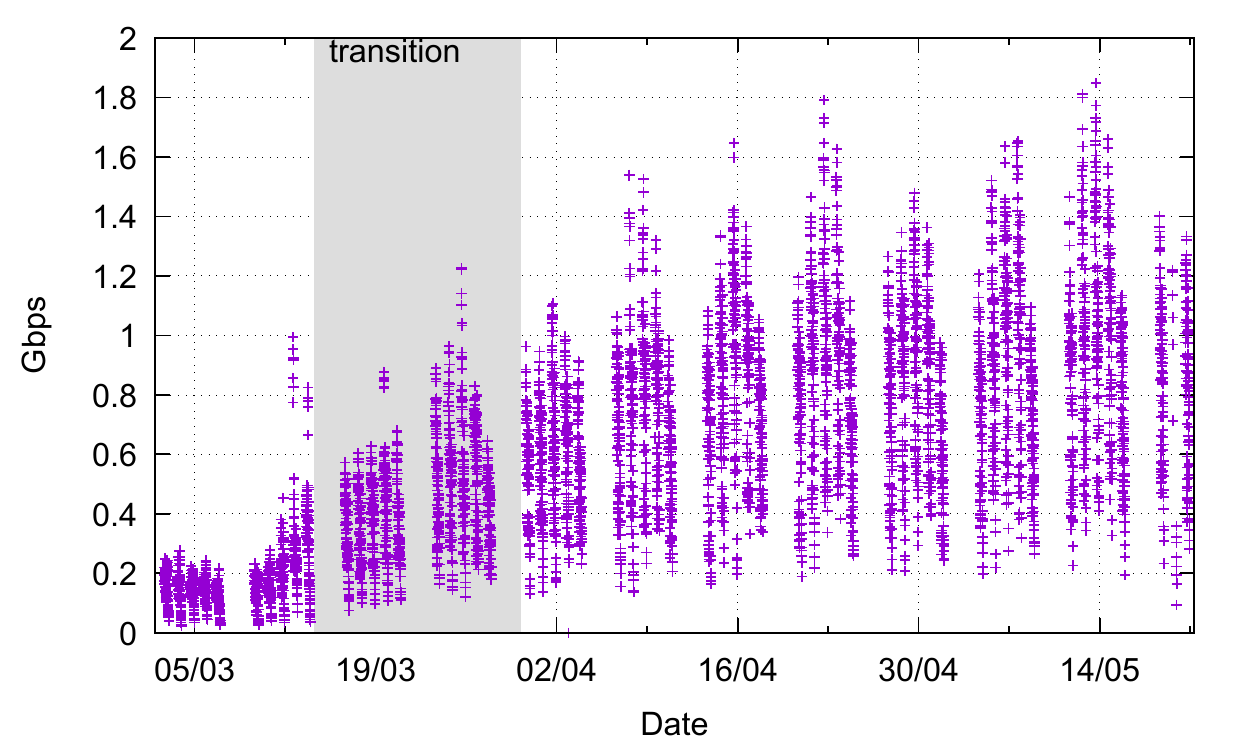}
   \label{fig:gwh}
 } 
 \subfigure[Google Meet traffic -- rest hours]{
   \includegraphics[width=.22\textwidth]{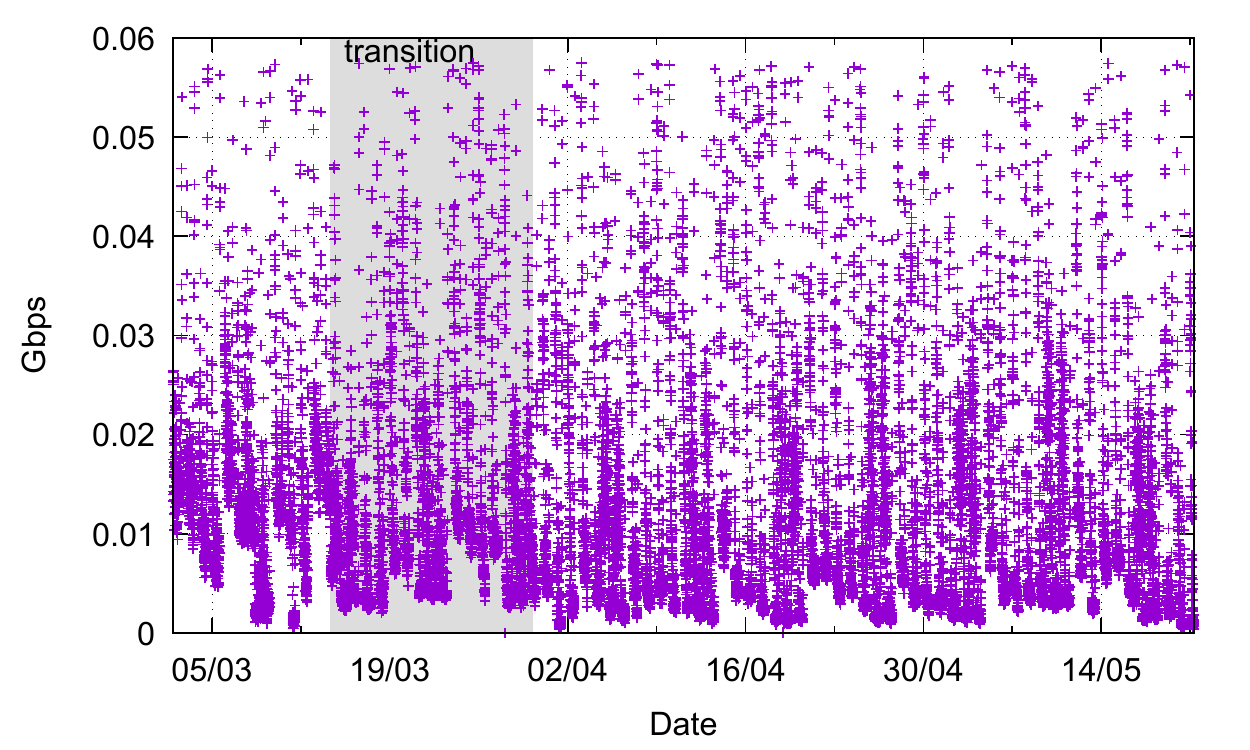}
   \label{fig:gnwh}
 } 
 \subfigure[VPN traffic -- work hours]{
   \includegraphics[width=.22\textwidth]{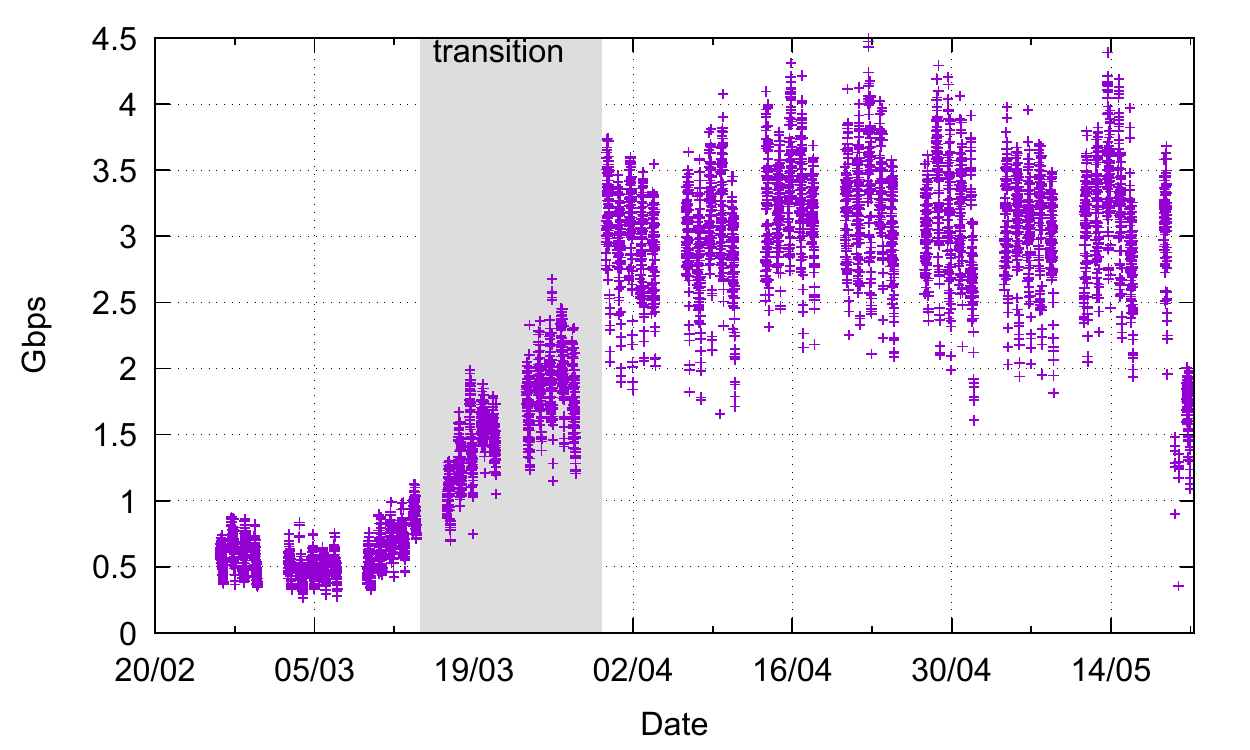}
   \label{fig:vwh}
 } 
 \subfigure[VPN traffic -- rest hours]{
   \includegraphics[width=.22\textwidth]{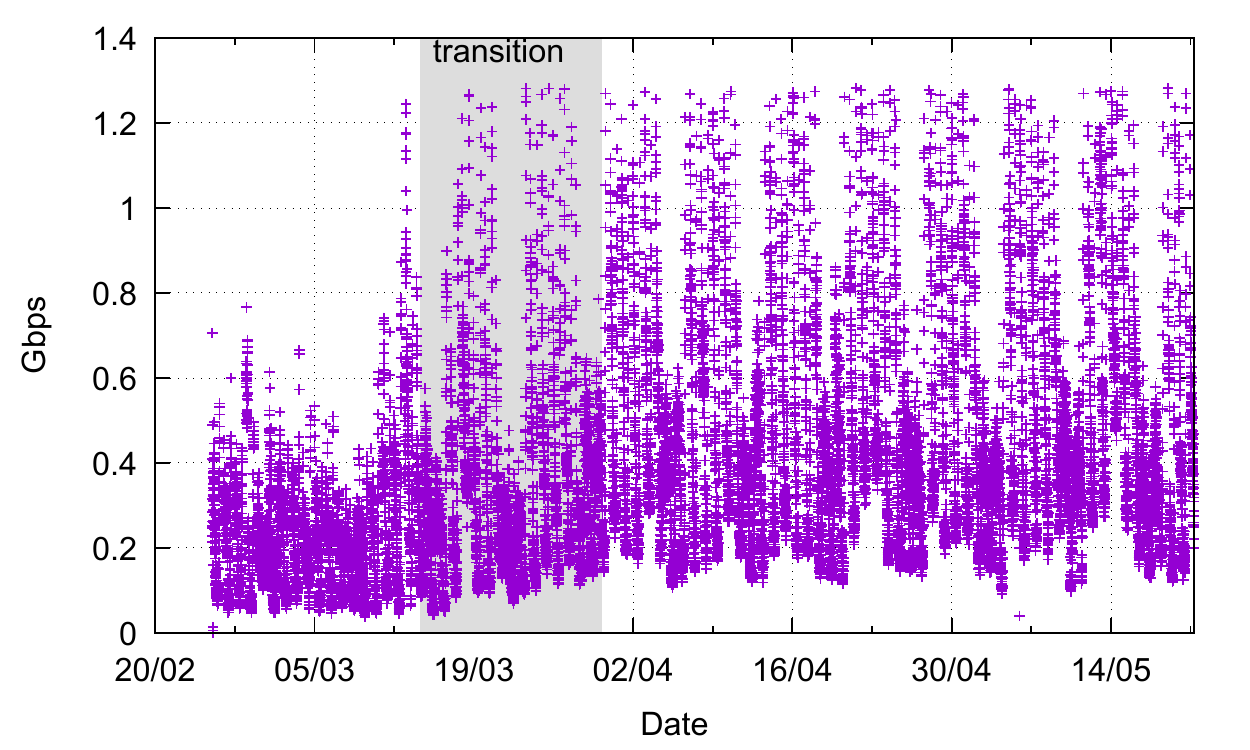}
   \label{fig:vnwh}
 } 
\caption{Illustration of traffic changes during work hours and rest hours. Google Meet traffic increases during work hours and decreases during rest hours, while VPN traffic increases during both work hours and rest hours.}\label{fig:changeil}
\end{figure*}

Online meeting and VPN traffic exhibits large and different changes, depending on the application and time of observation. All applications but BlueJeans and Skype increase their traffic manifold during work hours, ranging from 187\% (WebEx) to 554\% (VPN). During rest-hours the trend sometimes reverses. BlueJeans traffic increases slightly (101\%), while Google Meet and WebEx traffic decreases (60\% and 70\%) compared to usage during rest-hours before the transition. Figure \ref{fig:changeil} illustrates the dynamics of work-hour and rest-hour traffic change for Google Meet and VPN traffic.

\small{
\begin{table}
    \centering
    \begin{tabular}{c|c|c|c|c|c}
    \textbf{k/a} & \textbf{\% volume} & \multicolumn{2}{c}{\textbf{work-hours}} &
    \multicolumn{2}{c}{\textbf{rest-hours}} \\ \cline{2-6}
    & & change & aft/bef & change & aft/bef \\ \hline
    \multicolumn{4}{c}{\textbf{human-driven}} \\ \hline
     https & 55.9\% & $\downarrow$ & 46\% & $\downarrow$ & 64\% \\
     web & 11.6\% & $\downarrow$ & 35\% & $\downarrow$ & 47\%\\
      dns & $<0.1$\% &$\downarrow$  & 50\% & $\downarrow$  & 50\% \\
     email & 0.2\% & $\downarrow$ & 58\% & $\downarrow$  & 76\%  \\
     telnet & $<0.1$\% & $\downarrow$ & 41\% & &  \\
     ssh & 2.4\% &  & & &\\
    steam & 0.9\% & $\downarrow$ & 14\% &  $\downarrow$ & 19\% \\ \hline
    \multicolumn{4}{c}{\textbf{Data-driven}} \\ \hline
     unidata & 8.9\% & $\uparrow$ & 104\% & $\uparrow$ & 104\%  \\
     rsync & 0.5\% & $\downarrow$ & 0.2\% & $\downarrow$ & 0.5\% \\
     ftp & 0.1\% & $\downarrow$ & 73\% & $\downarrow$ & 85\% \\ \hline
     \multicolumn{4}{c}{\textbf{Automated}} \\ \hline
     Perfsonar & 2.3\% & $\uparrow$ & 113\% & $\uparrow$ & 109\% \\
     icmp & 0.1\% & $\downarrow$ & 98\% & \\
          ntp & 0.3\% & $\downarrow$ & 96\% & $\downarrow$ & 92\% \\ \hline
       \multicolumn{4}{c}{\textbf{Online meeting and VPN}} \\ \hline
     BlueJeans & $<0.1$\% & $\downarrow$ & 81\% & $\uparrow$ & 101\% \\
  GoTo Meeting & $<0.1$\% & $\uparrow$ & 238\% & $\uparrow$ & 189\% \\
      Google Meet & $<0.1$\% & $\uparrow$ & 512\% & $\downarrow$ & 60\%\\
     Zoom & 0.4\% & $\uparrow$  & 254\% & $\uparrow$ & 144\%\\
     Webex &  $<0.1$\% &  $\uparrow$ & 187\% & $\downarrow$ & 70\%\\
      Skype & 0.3\% & $\downarrow$ & 65\% & $\downarrow$ & 48\%\\
       VPN & 0.4\% & $\uparrow$ & 554\% & $\uparrow$ & 207\%\\ \hline
       \multicolumn{4}{c}{\textbf{Other}} \\ \hline
       highhigh & 10.4\% & &&  $\downarrow$ & 97\%\\
 syn & 0.1\% &$\downarrow$ & 69\% && \\
          otprot & 1.6\% && & & \\
   noservice & 0.5\% & && $\downarrow$& 82\% \\
     twoservice & 0.1\% & $\downarrow$ & 75\% & $\downarrow$ & 63\% \\   
     \end{tabular}
    \caption{Traffic categories we analyze}
    \label{tab:applist}
\end{table}
}
 \normalsize

\subsubsection{Case study: Zoom}

\begin{figure}
    \centering
    \includegraphics[width=.5\textwidth]{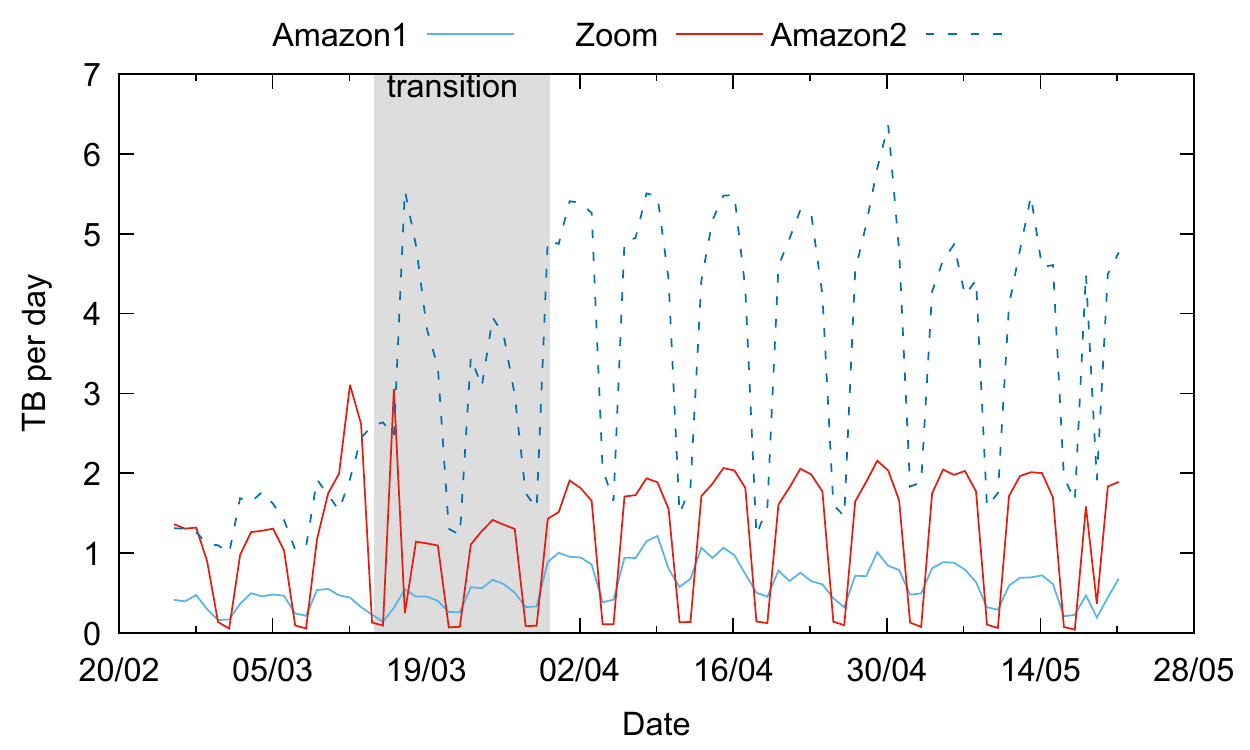}
    \caption{Zoom traffic shifting to Amazon as load increases.}
    \label{fig:zoomdsts}
\end{figure}

In this section we dissect Zoom's traffic per destination AS. Figure \ref{fig:zoomdsts} shows the daily volume of traffic reaching three largest Zoom traffic destination autonomous systems (ASes) in our dataset. As traffic increases in the week before stay-at-home order Zoom shifts most of the increase to the Amazon cloud (Amazon2 in the Figure). Traffic to its other Amazon cloud enclave (Amazon1) remains unchanged. This finding is possible due to our identification of verified mg-traffic prefixes. Without these, we would only be able to observe the red line in the Figure, and would conclude that Zoom traffic did not change due to stay-at-home orders.

\subsubsection{Case study: Online Meeting Choices}

\begin{figure*}
\subfigure[Government inst G08]{
   \includegraphics[width=.32\textwidth]{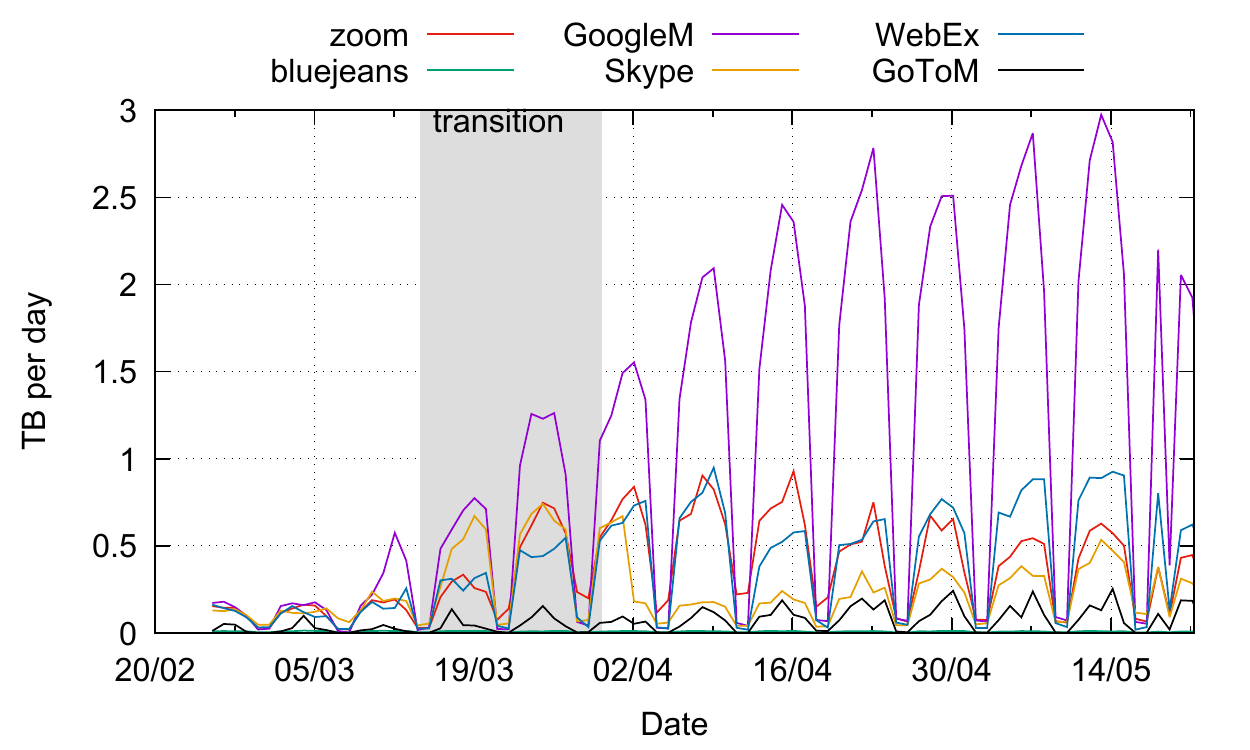}
   \label{fig:stateco}
 } 
 \subfigure[Educational inst. E19]{
   \includegraphics[width=.32\textwidth]{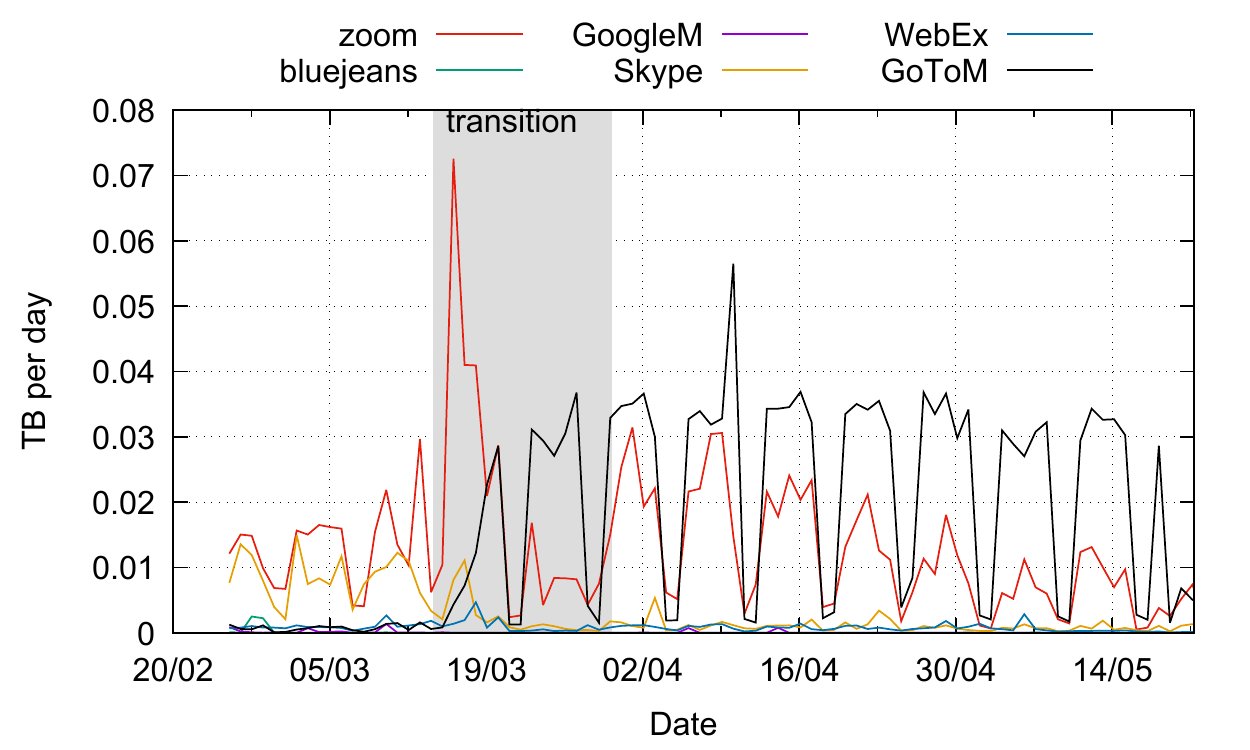}
   \label{fig:pueblo}
 } 
 \subfigure[Educational inst. E09]{
   \includegraphics[width=.32\textwidth]{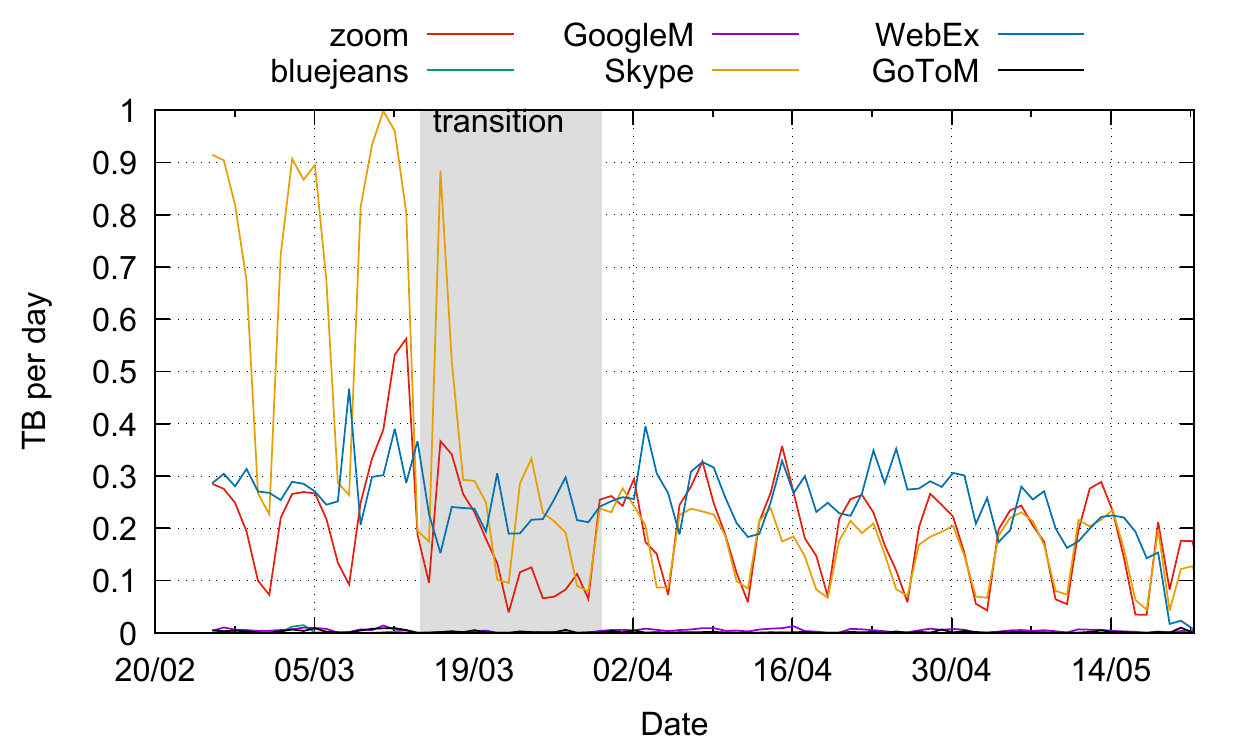}
   \label{fig:ucar}
  } 
\caption{Dynamics of choosing the right online meeting application.}\label{fig:media}
\end{figure*}

In this section we look into how networks decide between different applications for online meetings,  and how this impacts their network load. Figure \ref{fig:stateco} shows a government network that seems to have tried Google Meet and Skype in the first week of transition, and then chose Google Meet. Other online meeting applications remain present at much smaller levels, perhaps due to outside collaborators that use these applications.
Figure \ref{fig:pueblo} shows a small education institution that seems to have tried Zoom before the shutdown but has decided to adopt GoTo Meeting instead. Zoom still remains in use, but to a much lesser extent, and its use declines over the weeks. Finally, Figure \ref{fig:ucar} shows a large educational institution that seems to use Zoom, WebEx and Skype prior to transition, with Skype dominating over other applications. A week before the transition, it increases its use of Zoom, possibly testing it out. Then after the transition Zoom ``wins,'' but Skype and WebEx remain in significant, albeit reduced, use.

\begin{figure*}
\subfigure[Online meeting -- large volume]{
   \includegraphics[width=.45\textwidth]{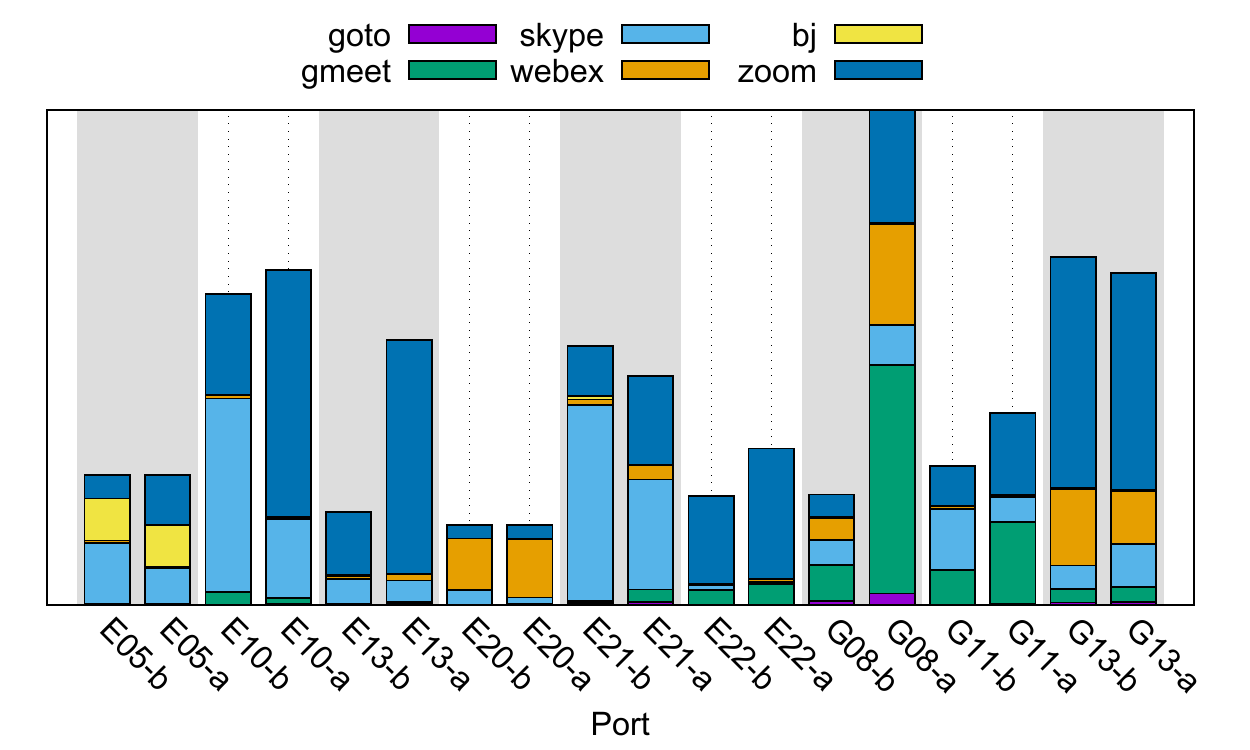}
   \label{fig:bmhist}
 } 
 \subfigure[Online meeting -- small volume]{
   \includegraphics[width=.45\textwidth]{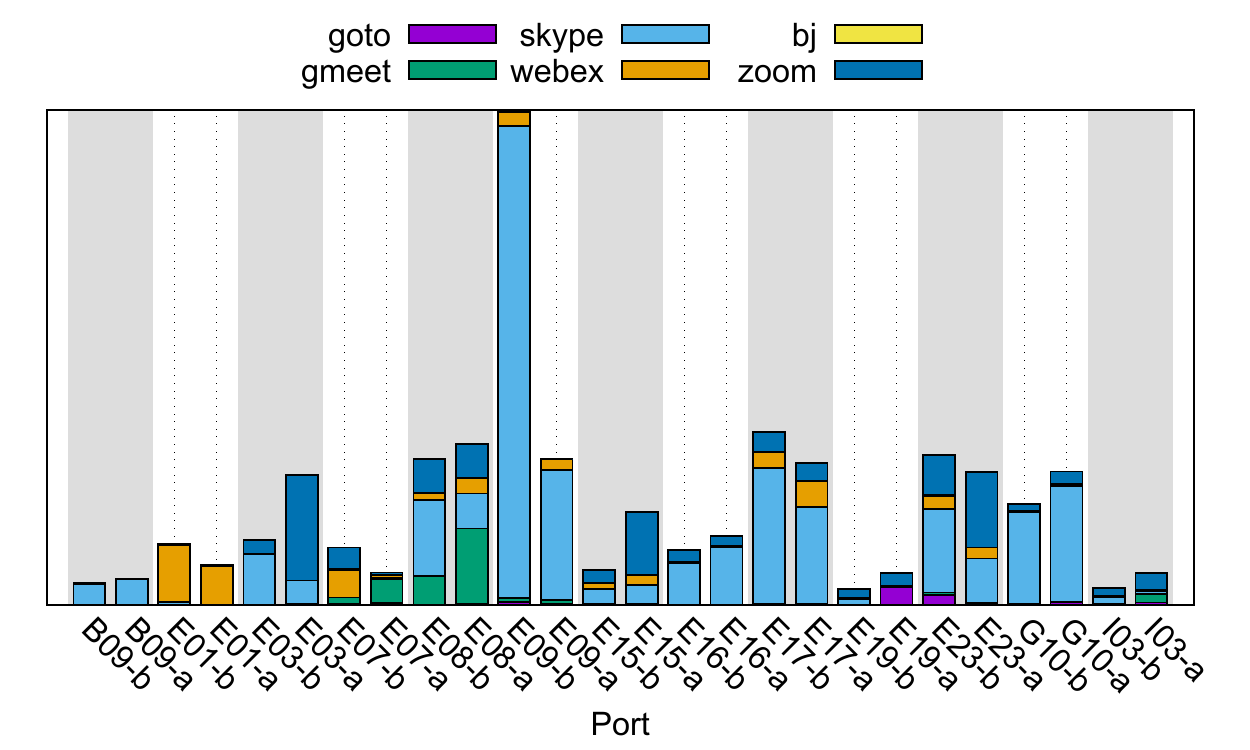}
   \label{fig:smhist}
 } 
\caption{Online meeting traffic changes. The "-b" on x-axis denotes data in two weeks before the transition and "-a" denotes data in two weeks after the transition.}\label{fig:mediachanges}
\end{figure*}

We further plot online meeting traffic mix in two weeks preceeding and two weeks following the transition period, using side-by-side histograms. We show separately large-media organizations, which exchange more than 5 TB (Figure \ref{fig:bmhist}), and small-media organizations, which exchange less than 5 but more than 0.5 TB (Figure~\ref{fig:smhist}) of media traffic after the transition. We see a large mix of behaviors. Among large-media organizations, many adopt Zoom and increase its usage, but keep using other applications as well. Most use three or more applications. Organization G08 is striking in its use of a variety of applications in large proportions. Among small-media organizations, we see prevalence of Skype and WebEx and to smaller extent Zoom.

\subsection{Large traffic changes}

We now look deeper into how traffic changed for different organizations. We calculate daily traffic exchanged on flows originated to and from each organization, and run MWW test to compare results before and after the transition period. 

Tables \ref{tab:trafto} and \ref{tab:traffrom} in the Appendix show the ratio of traffic change for each organization and per application. We summarize our findings here.

For inbound connections to local organizations: (1) https and VPN flow volume has increased consistently, (2) web and SSH flow volume has decreased for some organizations and greatly increased for others, (3) highhigh traffic has mostly decreased. For outbound connections from local organizations: (1) Zoom, WebEx and Google Meet have increased consistently, while other online meeting traffic has increased for some organizations and decreased for others, (2) human-driven traffic (web, https, ssh, steam) have decreased for everyone, (3) highhigh traffic has also decreased for everyone except for a handful of organizations. We will look deeper into this traffic category in Section~\ref{sec:highhigh}.

\subsubsection{Traffic Shifts Between Organization Types}

Table \ref{tab:peershifts} in the appendix shows how the traffic between peers changed. We summarize our findings here.  Almost all organizations saw increase in total volume of traffic on the flows originated towards them (i.e., the server is at a local organization), and decrease in total volume of traffic on the flows they originate (i.e., client is at a local organization). Most of the traffic increase was between ISPs and local organizations, which shows that traffic shifted with users. When users moved from local organizations to homes served by ISPs, so did the traffic. Most of the traffic decrease was on flows originated by local organizations towards businesses, illustrating the large impact on business from stay-at-home orders. However, it is possible that some of that traffic bypassed our vantage point and simply moved with users, to be exchanged between their home ISPs and businesses.

There are some giants among local organizations that source or sink much more traffic than the others. E09 loses 50 TB of traffic per day in inbound flows. 
Most of this loss is due to loss traffic from educational institutions. This organization focuses on scientific data aggregation, processing and delivery. This traffic loss may indicate loss of research opportunity due to stay-at-home orders.

\subsubsection{Case study: VPN}

\begin{figure}
    \centering
    \includegraphics[width=\columnwidth]{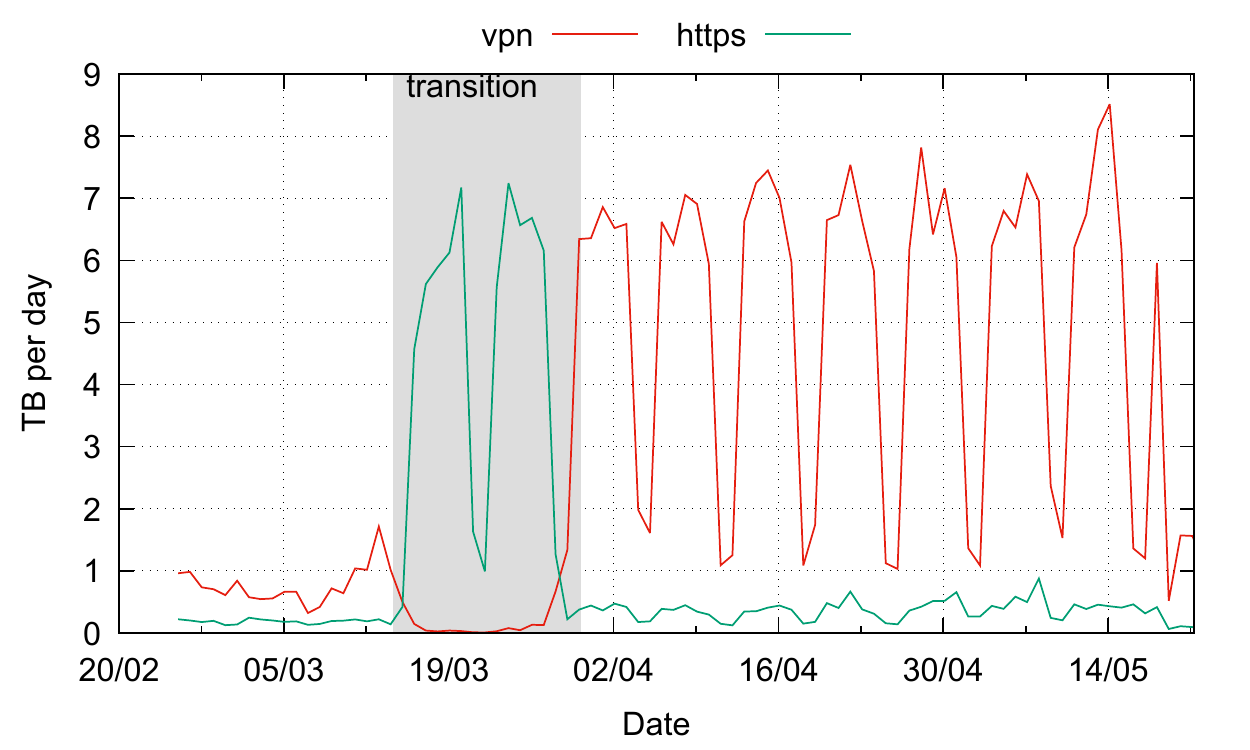}
    \caption{VPN and https traffic on flows to E13 -- these two trends complement each other, possibly indicating a temporary shift from traditional VPN to SSL-based VPN during transition.}
    \label{fig:vpn}
\end{figure}

From Table~\ref{tab:peershifts} we see that most organizations see increase in inbound VPN and https flows. Yet in 15 out of 35 organizations, either VPN or https increase but not both. We hypothesize that this may be due to some organizations using SSL-based VPN service, while others use a traditional VPN. 
Figure \ref{fig:vpn} shows traffic on VPN and https flows initiated to the organization E13. During the transition period they seem to experience a problem with their VPN service and switch to SSL-based VPN. The VPN usage goes down and https-usage goes up. After the transition they switch places, https usage declines sharply and VPN usage increases. 

\subsection{Live Addresses}

\begin{figure}
    \centering
    \includegraphics[width=.5\textwidth]{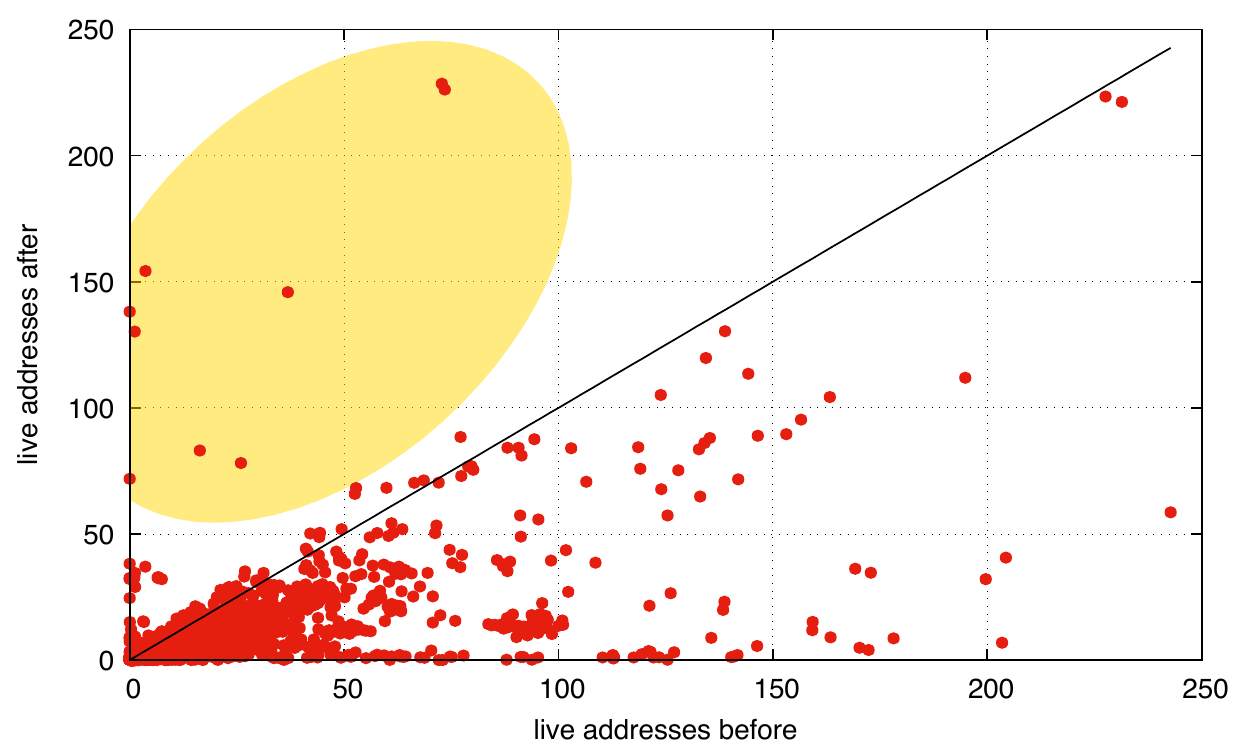}
    \caption{Prefix activity before and after the transition period: we show only those prefixes whose liveness has changed. The yellow ellipse highlights those prefixes whose liveness increased significantly.}
    \label{fig:liveness}
\end{figure}

We measure the number of live addresses in each \slash 24 prefix that is announced by FRGP. In total there are 9,618 prefixes and 6,103 appear in our dataset. We record the maximum number of live addresses per prefix per day and then run WMW test to detect changes. There are 2,435 prefixes or 40\% that exhibit a change in liveness: 1,962 have decreased activity and 473 have increased activity.
Figure \ref{fig:liveness} shows the average daily number of active IP addresses before and after the transition period, only for those prefixes whose liveness has changed. We see that most prefixes that increased their liveness experienced a small change, while most prefixes that decreased their liveness (points below x=y line) did so dramatically. However, there are several prefixes whose liveness increased a lot after the transition period -- these prefixes are highlighted by yellow ellipse in the Figure. We hypothesize that this may be due to increased use of VPN, without support for split tunneling, which then forces user traffic to third parties to go through local organizations and necessitates higher use of dynamic addresses. Out of 18 organizations whose inbound VPN traffic has increased 15 also experienced increased liveness of certain prefixes. However we lack further information needed to prove or disprove our hypothesis.

Table \ref{tab:livepertype} shows the count of prefixes in each of the liveness change categories -- same, increased or decreased --  per organization type (education, government, business or isp). ISPs saw the highest increase in liveness, followed by universities and government organizations. Conversely, largest percentages of decreased activity were in education, while government and business prefixes mostly maintained their liveness. 

\begin{table}[]
    \centering
    \begin{tabular}{c|c|c|c}
        \textbf{type} &\textbf{inc}  &\textbf{same} & \textbf{dec}  \\ \hline
edu & 286 (9\%)& 1,630 (48\%) & 1,446 (43\%) \\
gov & 153 (7\%) &  1,674 (73\%)& 446 (20\%) \\
bus & 14 (3\%) & 347 (81\%) & 66 (16\%) \\
isp & 20 (49\%) & 17 (41\%) & 4 (10\%) \\
total & 473 (8\%) & 3,668 (60\%) & 1,962 (32\%)
    \end{tabular}
    \caption{Count (and percentage) of prefixes that changed liveness per organization type}
    \label{tab:livepertype}
\end{table}

\begin{figure}
    \centering
    \includegraphics[width=\columnwidth]{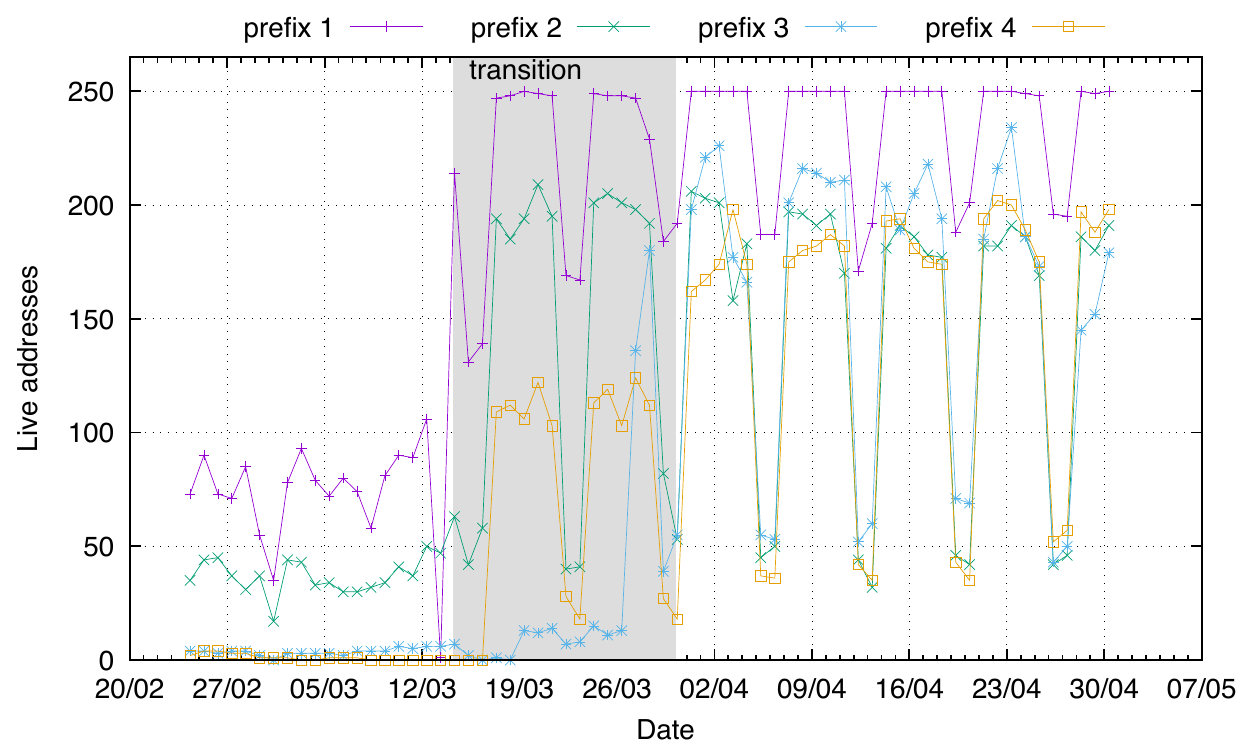}
    \caption{Four sample prefixes that increased their liveness during the transition period}
    \label{fig:sampleprefixes}
\end{figure}

Figure \ref{fig:sampleprefixes} illustrates the dynamics of change for four prefixes that have greatly increased their liveness. Prefix 1 experiences sudden change at the start of the transition period. Prefix 2 changes its liveness a few days later. Prefix 3 increases its liveness from zero to around 100 for the next two weeks and then increases it further to around 200 IPs. Prefix 4 sees slight increase during the transition and then dramatic increase in the last few days of the transition period. All prefixes exhibit weekly patterns, having fewer live addresses on weekends than on weekdays.

\subsection{Understanding High-high Traffic}
\label{sec:highhigh}

We now analyze high-high traffic to understand its nature changes. Prior to transition, the bulk of that traffic (72\%) is exchanged between education or government local institutions and business and education remote institutions, and 66\% involves an education institution, either local or remote. After transition, similarly, 69\% of traffic is exchanged between education or government local institutions and business and education remote institutions, with 62\% of traffic involving an education institution. 
Since several local institutions and one local government institution collect and analyze large amounts of scientific data, and offer it publicly for consumption, we hypothesize that most of the high-high traffic we see is actually scientific data streaming.
Both before and after the transition the highest volume of daily high-high traffic is exchanged with a single local educational institution focused on scientific data production about weather, climate, geological phenomena, etc, followed by another local educational institution, a national physics laboratory, several remote universities and NIST. 

Due to stay-at-home orders local government organizations reduce their daily high-high traffic by 25 TB (across all organizations) and local education organization reduce it by almost 200 TB. If we are correct in our hypothesis that most of the high-high traffic was scientific data traffic, this traffic reduction potentially means significant loss of research due to stay-at-home orders.

\input{daily_analysis}
\input{anomaly}

\section{Conclusion}
\label{sec:concl}
It is difficult to predict disasters and pandemics, but studying how networks and organizations adapted to this pandemics can help us prepare for the future ones. We find large traffic changes in online meeting software and VPN usage, and large shifts of traffic from education and government institutions to local ISPs. We also find a variety of online meeting solutions that may frustrate users, while providing some resiliency to organizations. We further find that research productivity may suffer during stay-at-home period because scientific data traffic reduces between education and government institutions. Finally, we find that network anomalies increase during stay-at-home. These findings can help us invest in appropriate software and user training, and in security of our network infrastructure, to be better prepared for the future emergencies. 

\bibliographystyle{plain}
\bibliography{refs.bib}

\appendix
\input{ip_appendix}
\input{jappendix}
\input{daily_appendix}

\end{document}

%% file: daily_analysis.tex
\section{Traffic changes within a day}
\label{sec:daily}

In the previous Sections, we focused on traffic changes over a long term. 
In this section, we look at a particular workday and weekend to study the traffic changes at a finer granularity.
For the workday traffic comparisons, we select a Thursday (March 5) before the transition period, and a Thursday (April 9) after the transition period. 
For the weekend traffic comparisons, we select a Saturday (March 7) before the transition period and a Saturday (April 11) after the transition period.
We then extract the traffic generated by different protocols and applications from these days to study their distributions and changes.

\begin{figure*}

 \subfigure[https traffic volume]{
   \includegraphics[width=.23\textwidth]{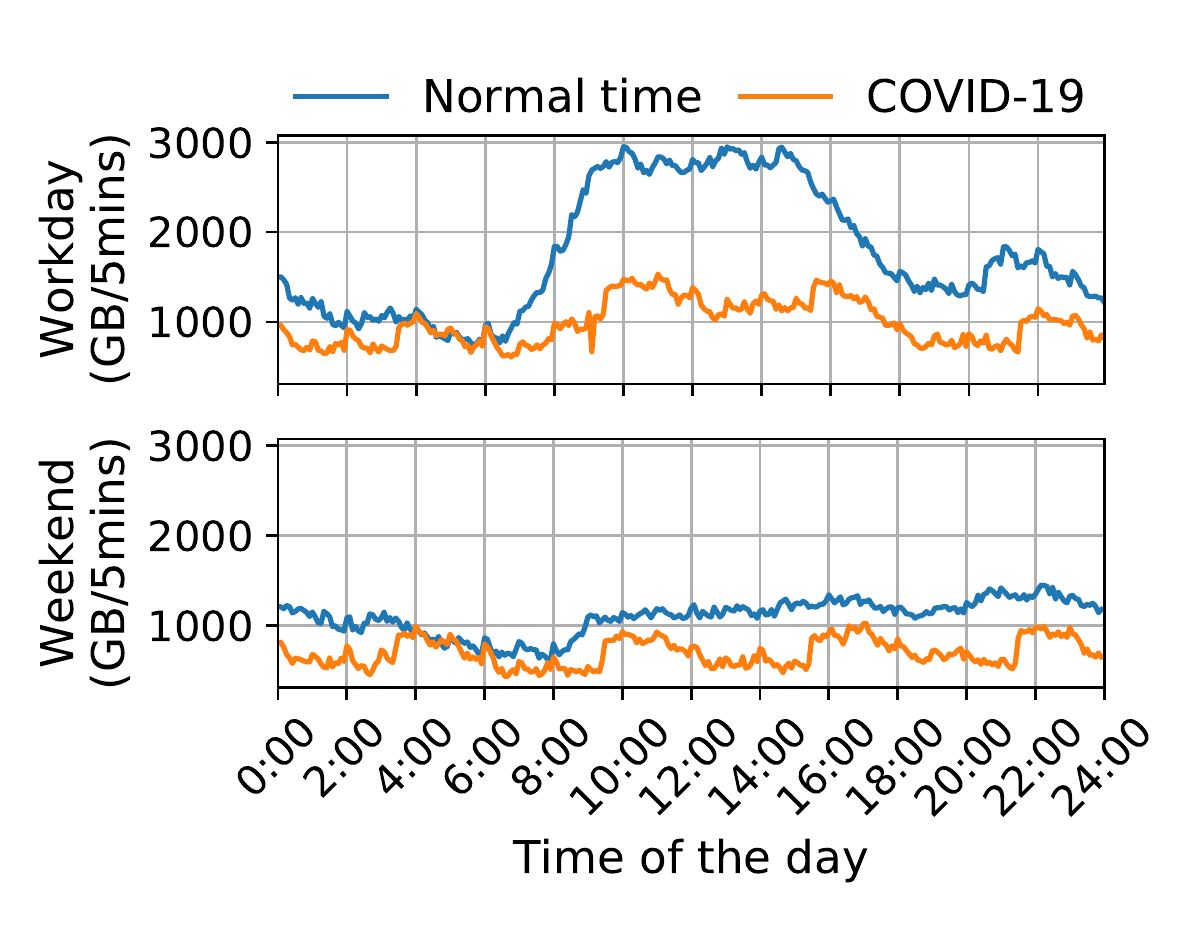}
   \label{fig:daily_https}
 } 
 \subfigure[Zoom traffic volume]{
   \includegraphics[width=.23\textwidth]{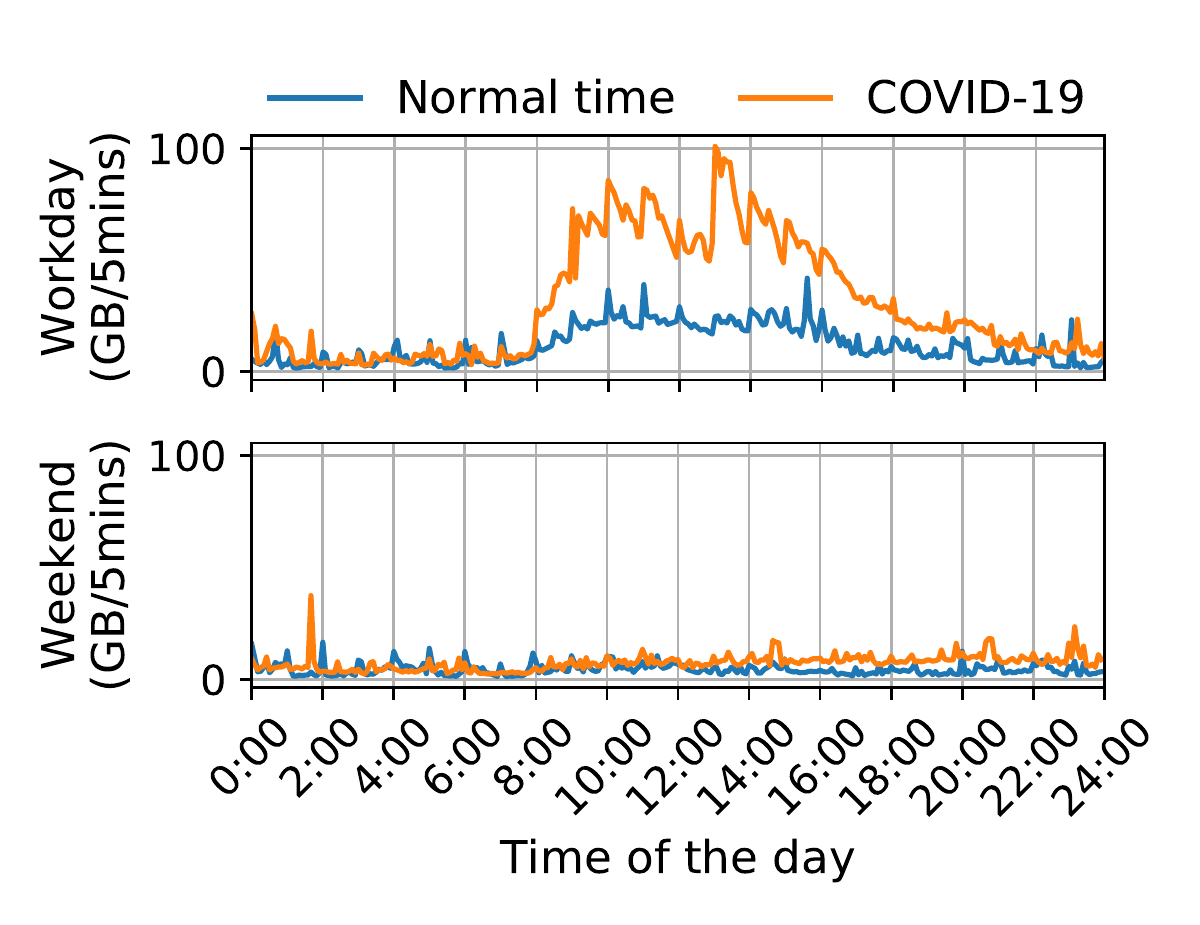}
   \label{fig:daily_zoom}
 }
 \subfigure[VPN traffic volume]{
   \includegraphics[width=.23\textwidth]{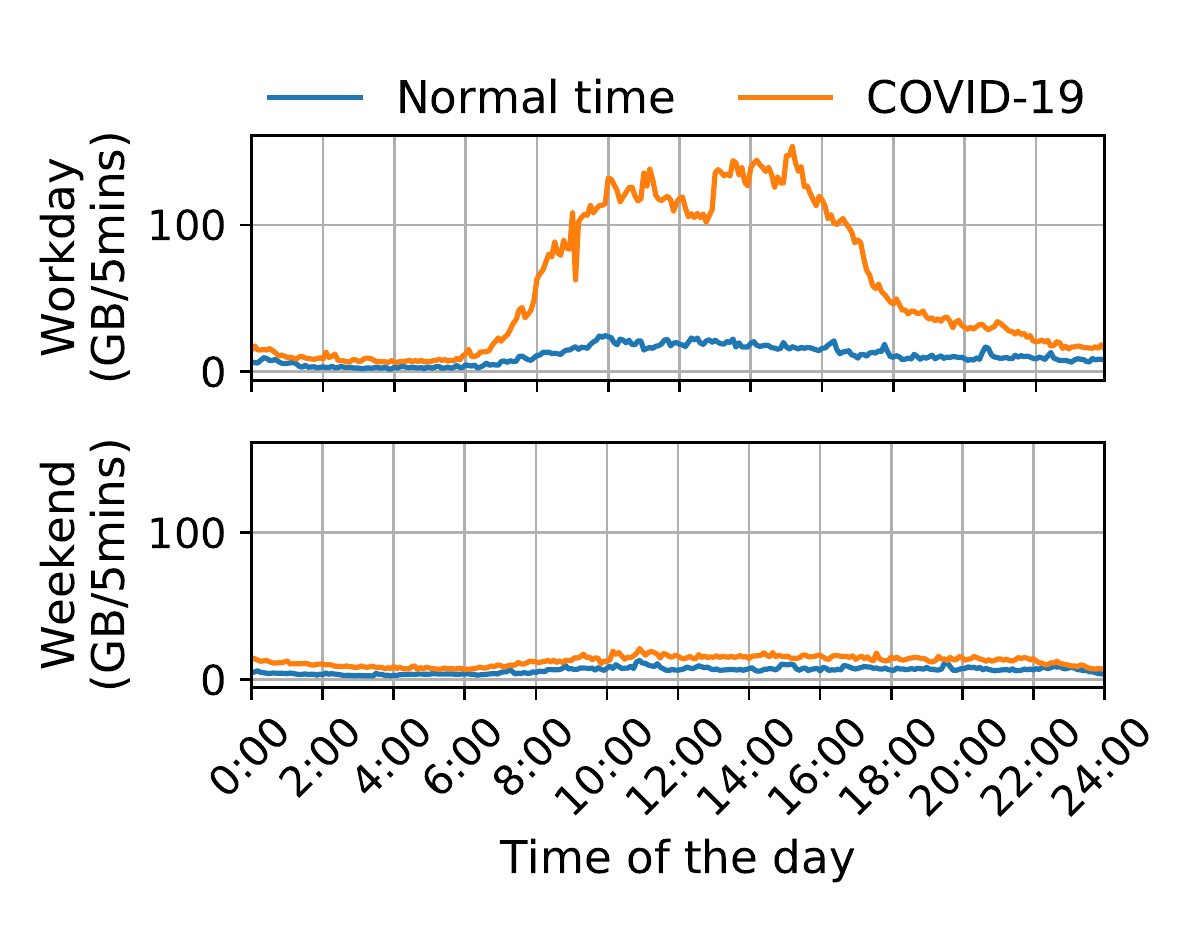}
   \label{fig:daily_vpn}
 }
 \subfigure[Steam traffic volume]{
   \includegraphics[width=.23\textwidth]{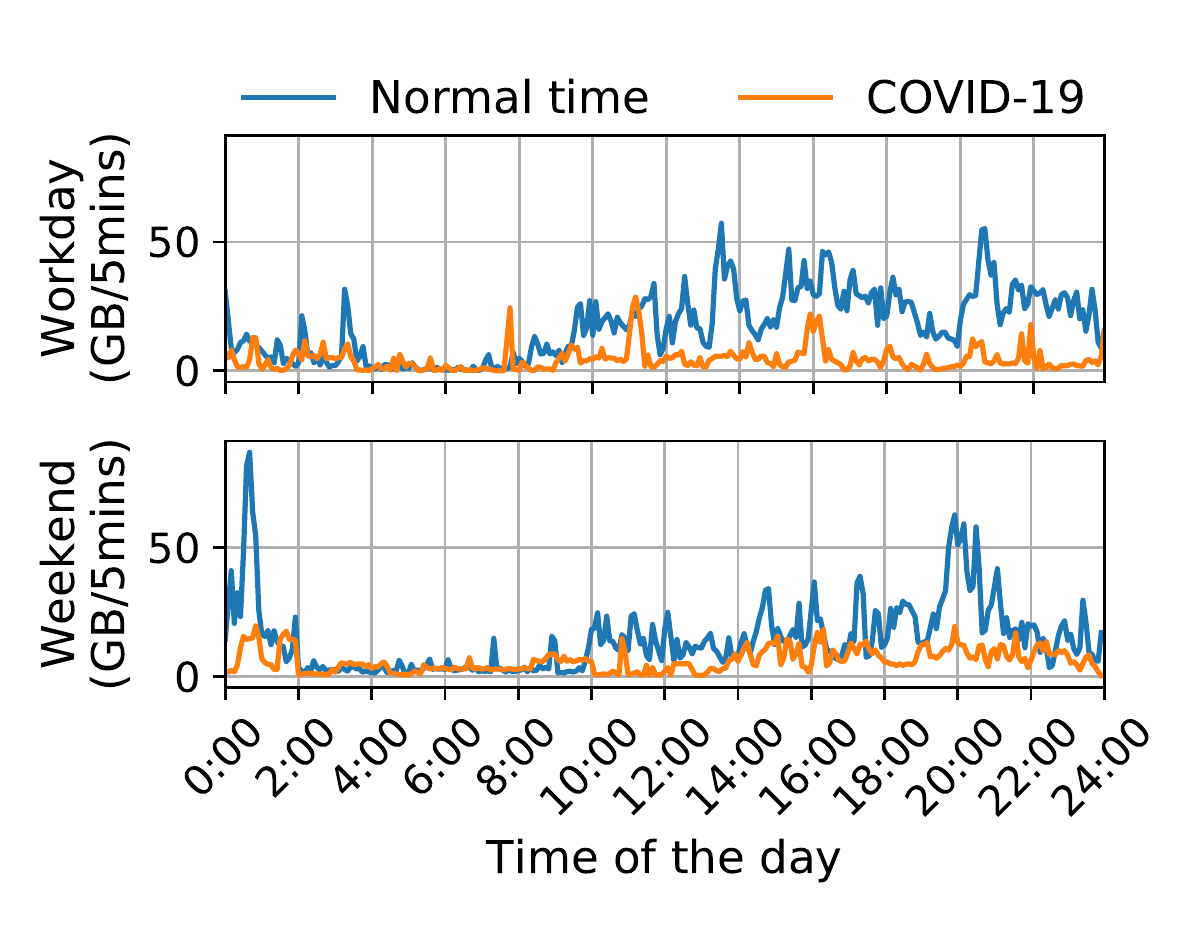}
   \label{fig:daily_valve}
 }
\caption{Traffic changes within workdays and weekends}
\label{fig:daily}
\end{figure*}

Figures~\ref{fig:daily} illustrate the traffic changes for different protocols and applications during a particular day.
The upper subplots show the traffic distributions during a workday and the bottom subplots show the traffic distributions during a weekend.
We also give more examples in Appendix~\ref{sec:daily_appendix}.

We observe the following trends: 
(1) Verified with the WMW test in Section~\ref{sec:statis_test}, the traffic changes during these days are consistent with the traffic changes in the long term (Table~\ref{tab:applist}); 
(2) Many applications' traffic distributions correspond to people's daily work schedule. For example, after the transition period, the traffic of online meeting applications, such as VPN, Zoom, Google Meet, swell from 8 am to 5 pm on the workday, but remain at a low level during the rest hours;
(3) Besides the working hours, the overall traffic volume reaches a small peak from 8 pm to 9 pm. This trend is also obvious for https and Steam traffic. We believe that this phenomenon matches people's leisure habits, of playing video games or watching movies before sleep. However, the evening peak becomes smaller after the transition period, possibly due to students moving out of the dorms;
(4) The traffic distributions during the weekend are more even than the traffic distributions during the workday, which indicates that people are more flexible about their activities during weekends.
% \jun{Points 2-4 are not before vs. after comparison.}
% \UOyebo{Revised a little bit on 2 and 3.}

%% file: anomaly.tex
\subsection{Measurement of Anomalies}
\label{sec:anomaly}

We further study the anomalies that happened from February 24 to May 21 to investigate whether the pandemic had any impact on network security or not.

\begin{figure}
\includegraphics[width=.45\textwidth]{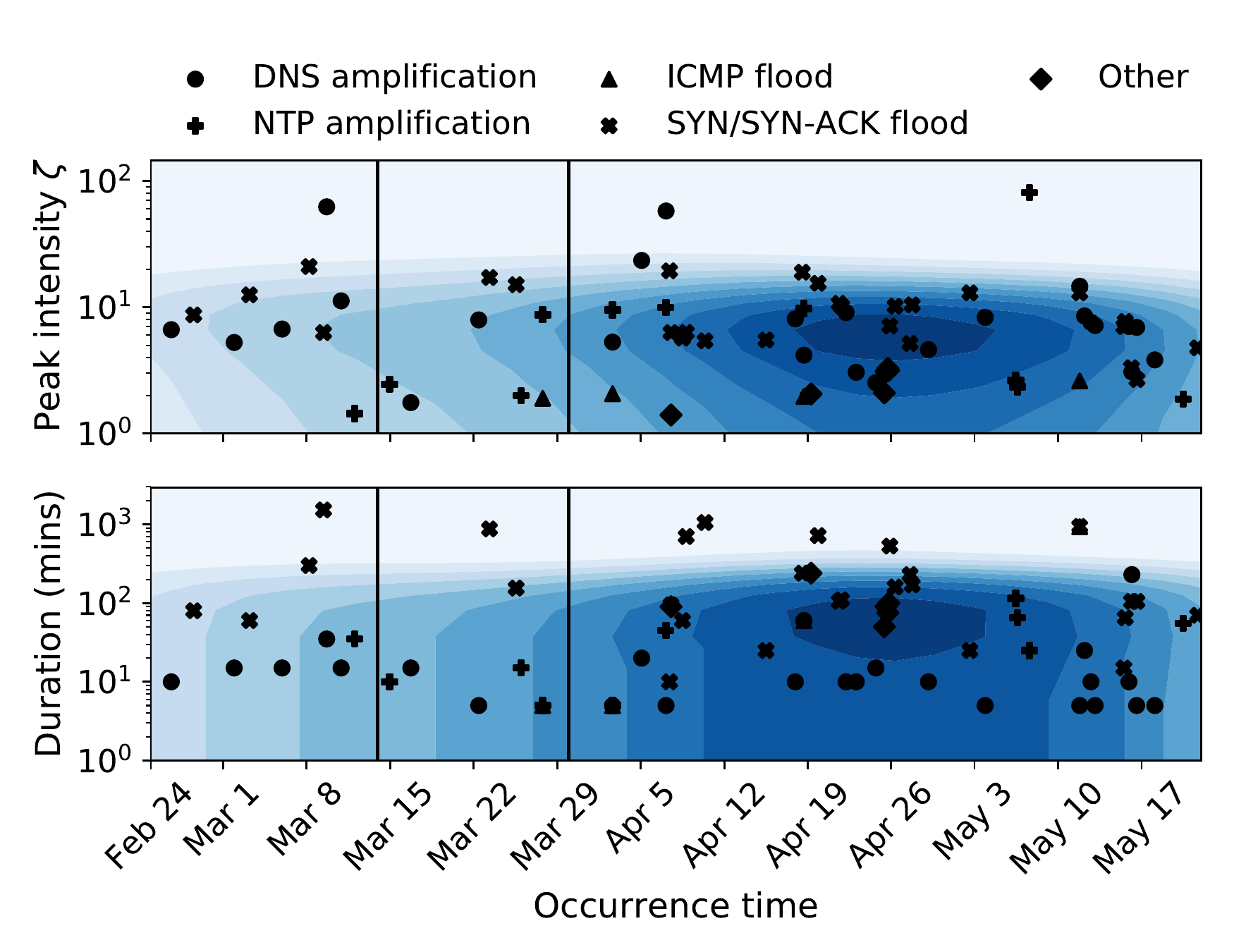}
\caption{Illustration of the detected anomalies, including their duration, peak intensity indexes $\zeta$, and occurrence time.}
\label{fig:anomaly_two}
\end{figure}

Figure~\ref{fig:anomaly_two} shows the anomaly detection results. 
The upper subplot illustrates the peak intensities index $\zeta$ of the anomalies occurred at different times.
The lower subplot illustrates the duration of the detected anomalies at different times.
Table~\ref{tab:anomaly} quantities the changes of anomaly frequency, anomaly duration, and peak intensity $\zeta$. 
Since there are no ICMP floods and other anomalies from February 24 to March 14, the $after/before$ values for these attacks are not applicable.

We observe the following trends from the results:
(1) The number of anomalies increases significantly after the transition period.
All four types of anomalies happen more frequently than before;
(2) The change of anomaly duration varies according to the type of anomaly.
Some anomalies such as DNS attack and NTP attack tend to last longer during the pandemic, while SYN floods have shorter durations during the pandemic;
(3) The average peak intensity index of NTP attacks dramatically increases after the transition, while the peak intensity indexes of DNS attacks and SYN flood attacks decrease.
Clearly, with the stay-at-home order in effect, network attacks became more active than before.
% Therefore, ISPs should concentrate more on anomaly detection and attack defense to provide reliable to the users.

{\small
\begin{table}
    \centering
    \begin{tabular}{c|c|c|c|c|c|c}
        \textbf{Anomaly}  & \multicolumn{2}{c}{\textbf{Frequency}} & \multicolumn{2}{c}{\textbf{Duration}} & \multicolumn{2}{c}{\textbf{Peak int. $\zeta$}}\\ \cline{2-7}
%        & change & aft/bef & change & aft/bef & change & aft/bef \\ \hline
        & change & $\frac{after}{before}$ & change & $\frac{after}{before}$ & change & $\frac{after}{before}$ \\ \hline
DNS amp & $\uparrow$ & 131\% & $\uparrow$ & 137\% & $\downarrow$ & 56\% \\
NTP amp & $\uparrow$ & 255\% & $\uparrow$ & 151\% & $\uparrow$ & 1163\%\\
ICMP flood & $\uparrow$ & N/A & $\uparrow$ & N/A & $\uparrow$ & N/A\\
S/SA flood & $\uparrow$ & 192\% & $\downarrow$ & 54\% & $\downarrow$ & 74\% \\
Other  & $\uparrow$  & N/A & $\uparrow$ & N/A & $\uparrow$ & N/A
    \end{tabular}
    \caption{Changes of anomalies}
    \label{tab:anomaly}
\end{table}
}
\normalsize

%% file: ip_appendix.tex
\section{Traffic changes of individual IP addresses}
\label{sec:more_ind_ip}

In this appendix, we track some individual IP addresses to analyse traffic changes from a user's perspective. 
Here, we only select IPs that are consistent and active, i.e., whose main purpose (e.g., Web server) does not change, and whose activity remains considerable during our observation period. 

We further classify the selected IPs into \emph{servers} and \emph{clients} by observing the distribution of their traffic over port numbers.
If the majority of traffic from an IP address are from a well-known server port towards a variety of high-numbered ports, we consider the IP address as a server IP. Otherwise it is a client IP. 

We filter out a set of qualified IP addresses, and then we select a handful of IPs with diverse purpose. 
Figures~\ref{fig:individual_ip} in Appendix illustrate the traffic volume changes of six server IPs and six client IPs  with Table~\ref{tab:ip_to_obse} summarizing their traffic changes.
The shaded areas show weekly traffic volume, and the lines show hourly traffic volume.

We observe the following phenomena:
(1) The traffic changes of individual servers are very diverse, not necessarily following the application's overall direction of traffic change;
(2) Residential clients tend to generate more traffic after the transition, probably because the stay-at-home order focuses people's activities at home;
(3) The government and education clients tend to be less active after the transition; and
(4) The traffic transition of our select education clients happened before the traffic transition of our select government clients, due to the combined effect of both the stay-at-home measure and the school spring break.

\begin{table*}
    \centering
    \begin{tabular}{c|c|c|c|c|c|c}
        \textbf{IP} &\textbf{Nature}  & \multicolumn{2}{c}{\textbf{work-hours}} & \multicolumn{2}{c}{\textbf{rest-hours}} \\ \cline{3-6}
        & & change & aft/bef & change & aft/bef \\ \hline
        A & HTTPS server & $\downarrow$  & 59\% & $\downarrow$ & 62\% \\
        B & Unidata server & $\uparrow$  & 116\% & $\downarrow$ & 98\% \\
        C & Streaming server & $\uparrow$ & 1671\% & $\uparrow$ & 656\% \\
        D & HTTPS server & $\uparrow$   & 886\% & $\uparrow$ & 397\% \\
        E & HTTPS server & $\downarrow$  & 46\% & $\downarrow$ & 53\% \\
        F & SSH server & $\uparrow$  & 129\% & $\uparrow$ & 119\% \\
        G & Residential client & $\uparrow$  & 4318\% & $\uparrow$ & 45296\% \\
        H & Education client & $\downarrow$ & 11\% & $\downarrow$ & 14\% \\
        I & Government client & $\downarrow$  & 89\% & $\downarrow$ & 21\% \\
        J & Residential client & $\uparrow$  & 188\% & $\downarrow$ & 71\% \\
        K & Education client & $\downarrow$  & 22\% & $\downarrow$ & 21\% \\
        L & Government client & $\downarrow$  & 90\% & $\downarrow$ & 64\% 
    \end{tabular}
    \caption{IP addresses to observe and their traffic changes}
    \label{tab:ip_to_obse}
\end{table*}

\begin{figure*}
  \subfigure[IP A --- https server]{
   \includegraphics[width=.3\textwidth]{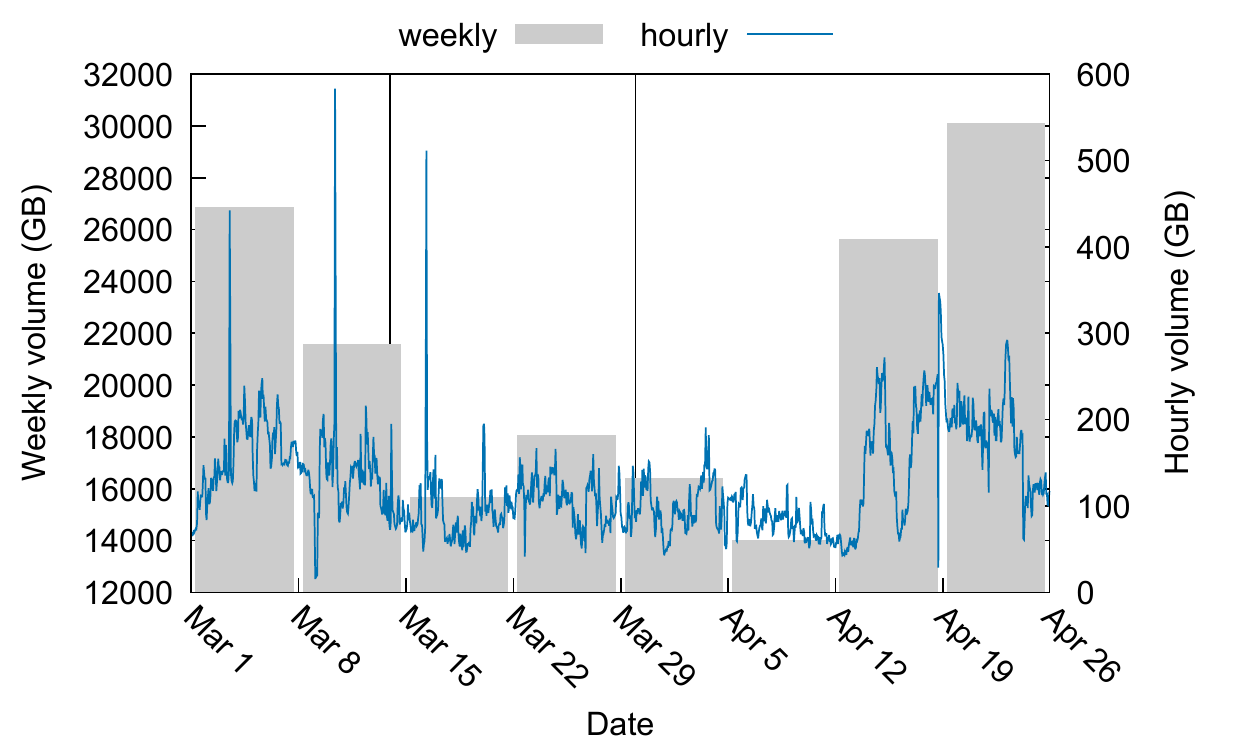}
   \label{fig:14.181.162.228}
 } 
 \subfigure[IP B --- Unidata server]{
   \includegraphics[width=.3\textwidth]{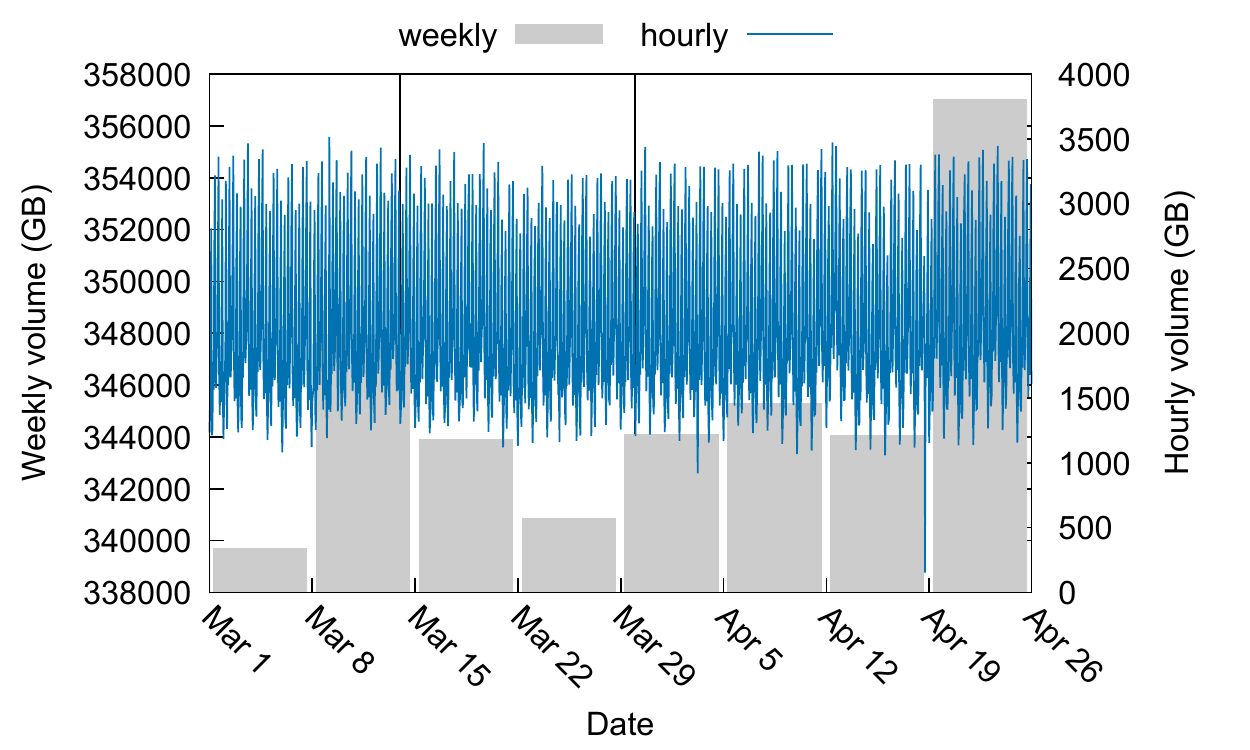}
   \label{fig:14.14.121.114}
 }
 \subfigure[IP C --- streaming server]{
   \includegraphics[width=.3\textwidth]{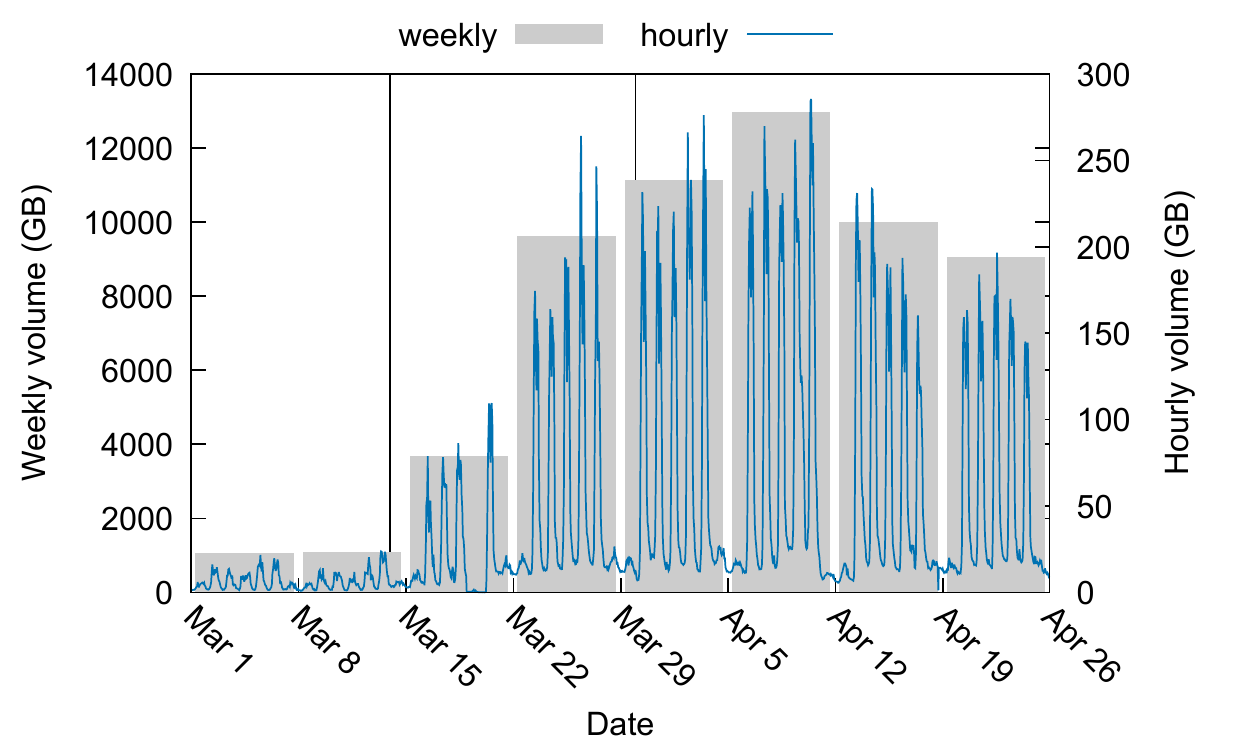}
   \label{fig:42.128.166.100}
 }
 \subfigure[IP D --- HTTPS server]{
   \includegraphics[width=.3\textwidth]{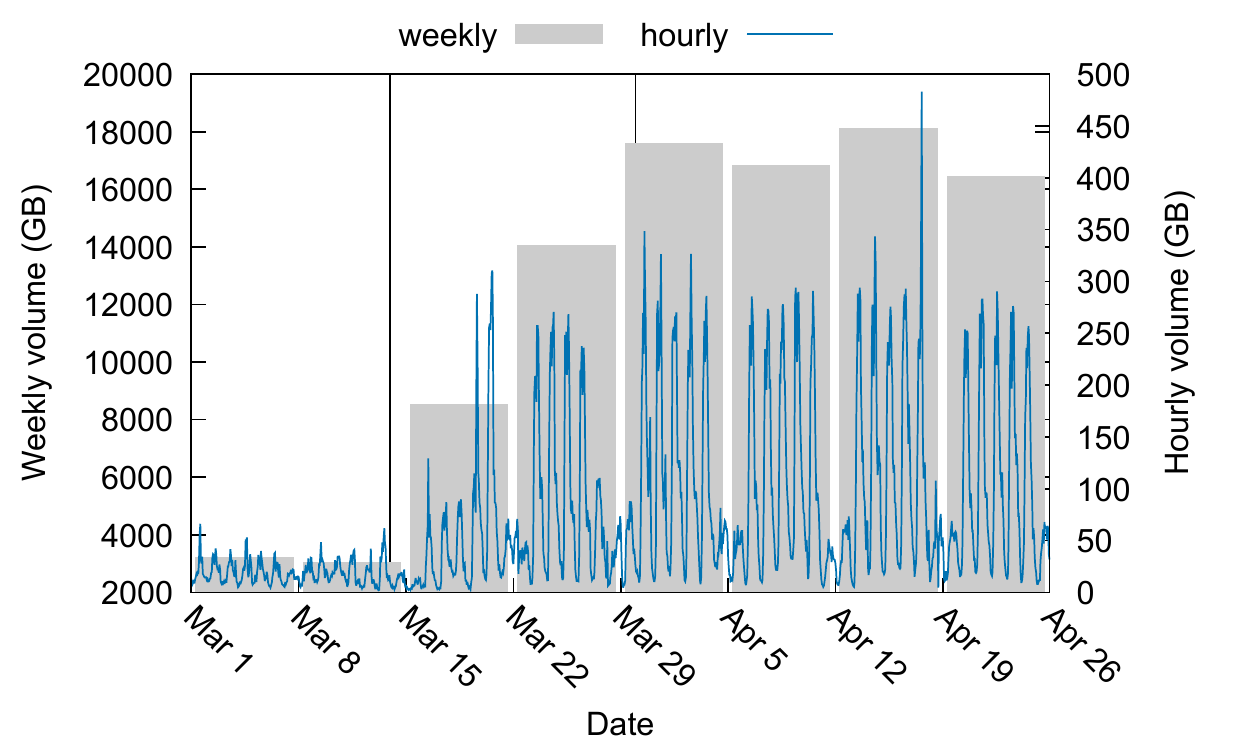}
   \label{fig:69.59.57.38}
 }
 \subfigure[IP E --- HTTPS server]{
   \includegraphics[width=.3\textwidth]{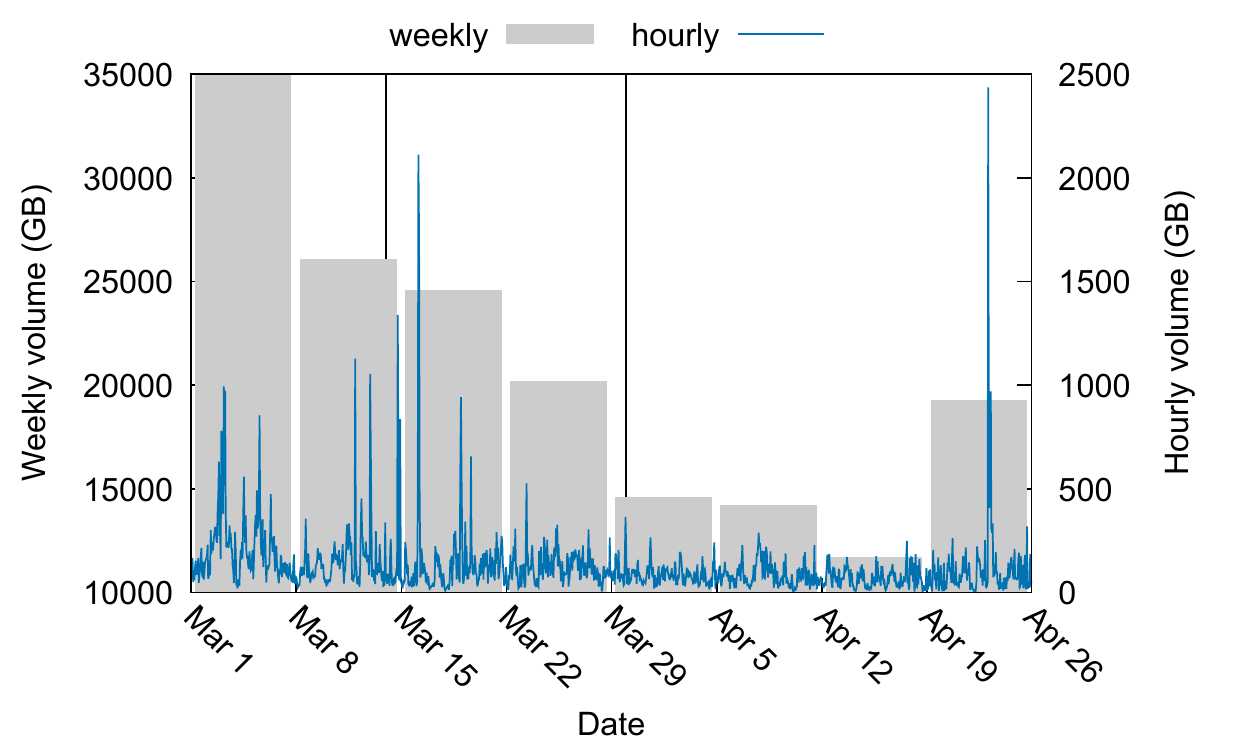}
   \label{fig:38.40.230.228}
 }
 \subfigure[IP F --- SSH server]{
   \includegraphics[width=.3\textwidth]{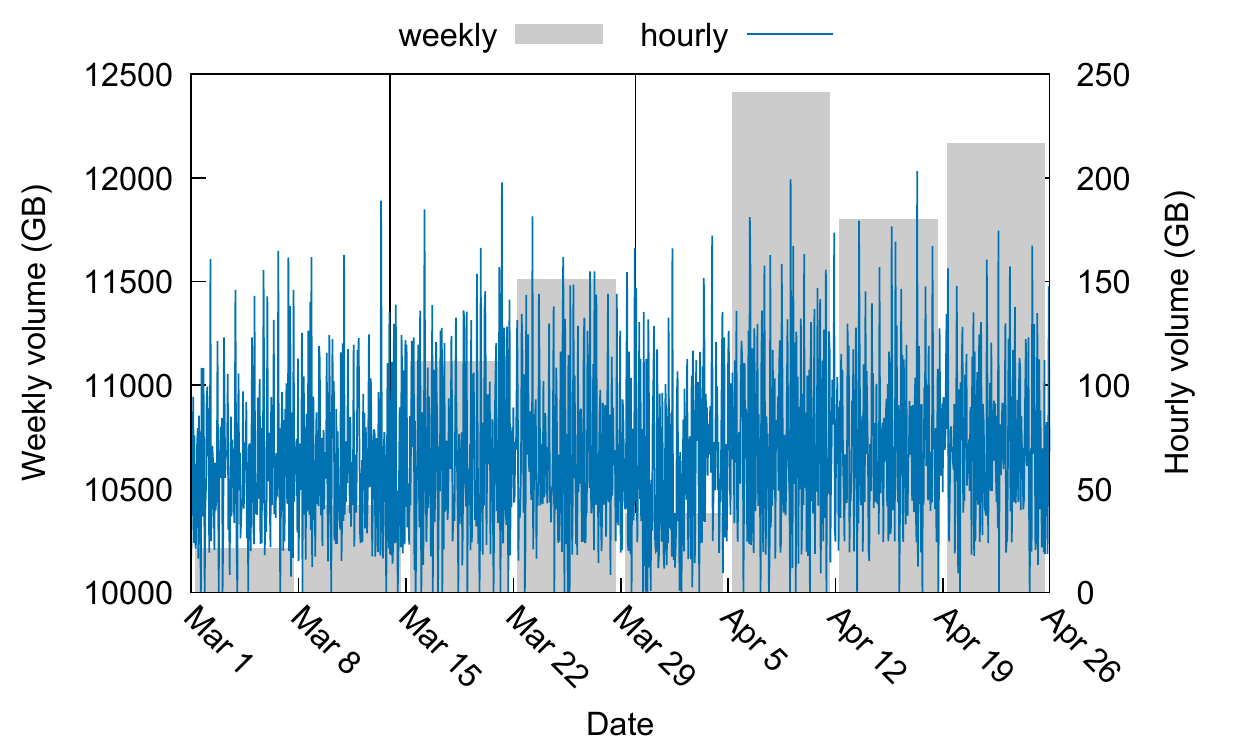}
   \label{fig:65.58.115.246}
 }
 \subfigure[IP G --- residential client]{
   \includegraphics[width=.3\textwidth]{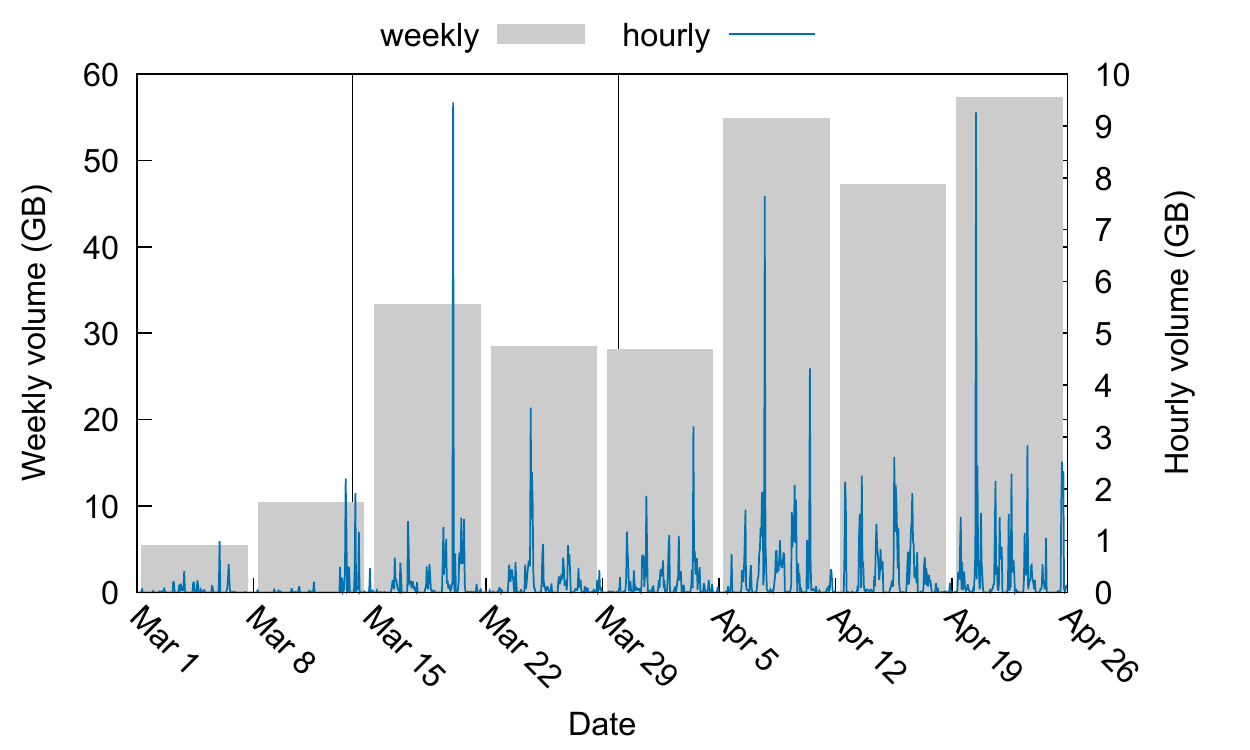}
   \label{fig:69.11.3.238}
 } 
 \subfigure[IP H --- education client]{
   \includegraphics[width=.3\textwidth]{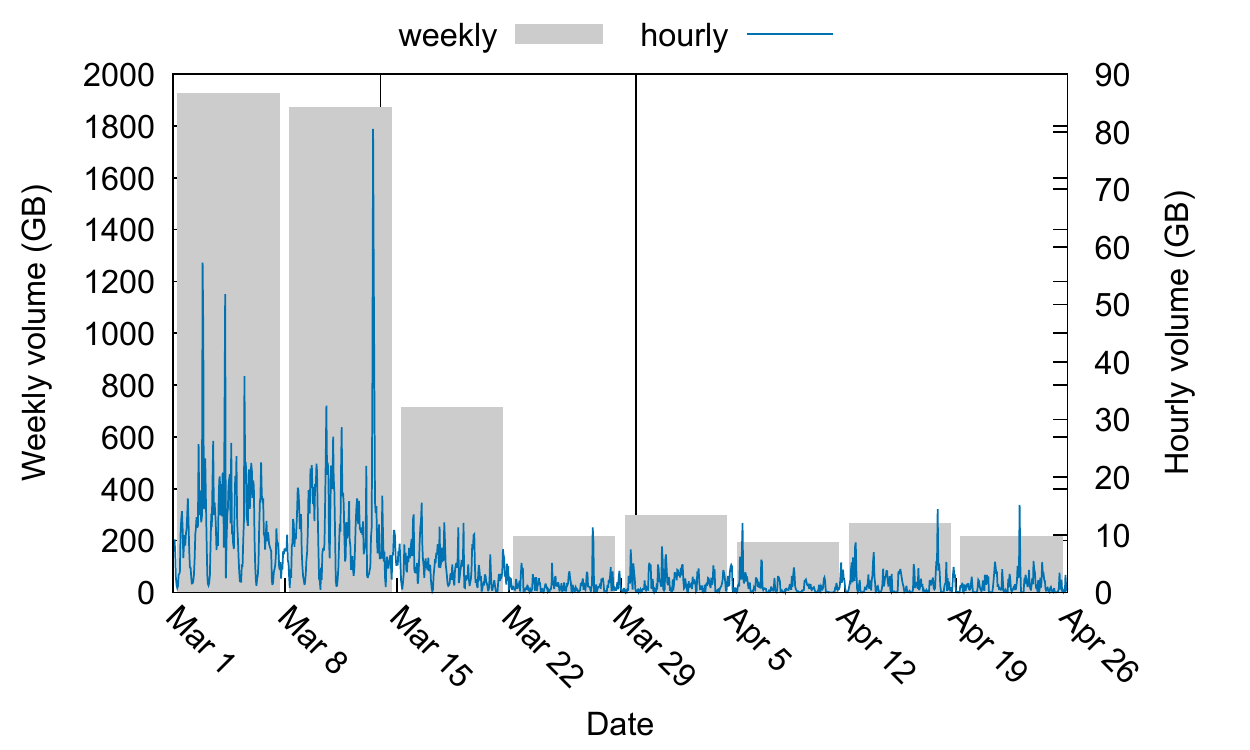}
   \label{fig:14.181.157.122}
 }
 \subfigure[IP I --- government client]{
   \includegraphics[width=.3\textwidth]{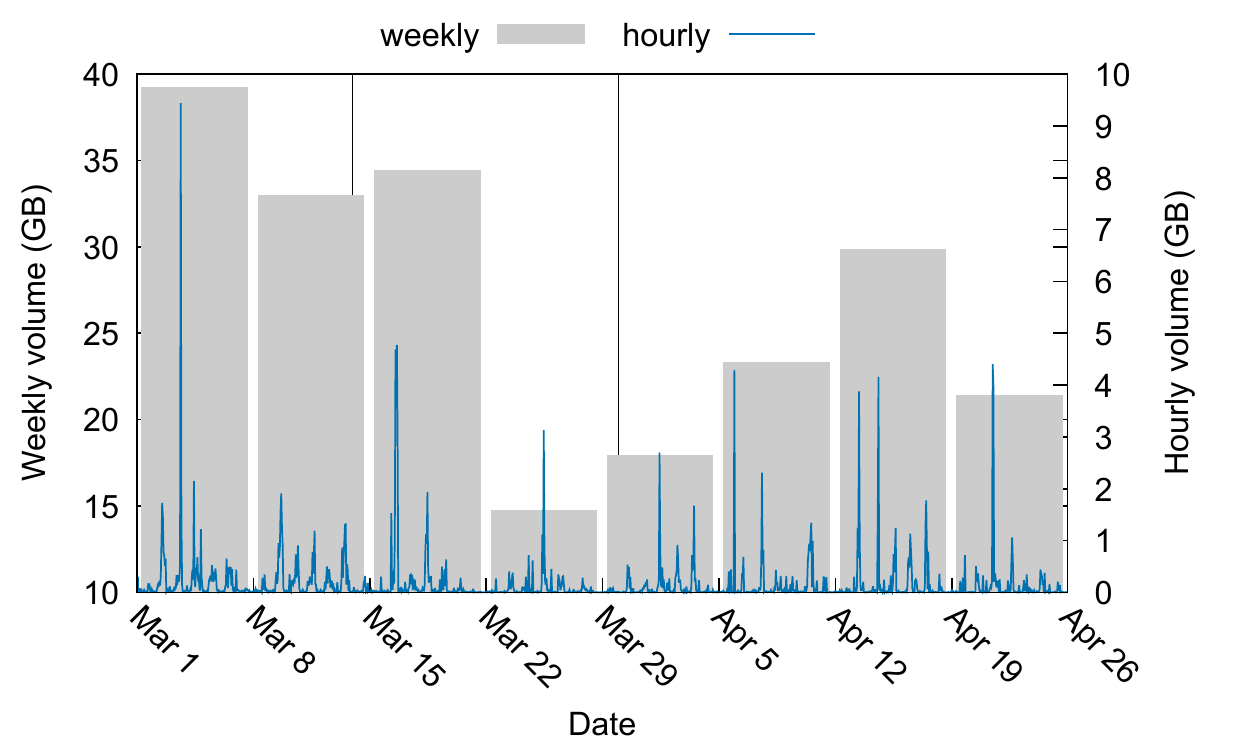}
   \label{fig:14.181.156.169}
 }
 \subfigure[IP J --- residential client]{
   \includegraphics[width=.3\textwidth]{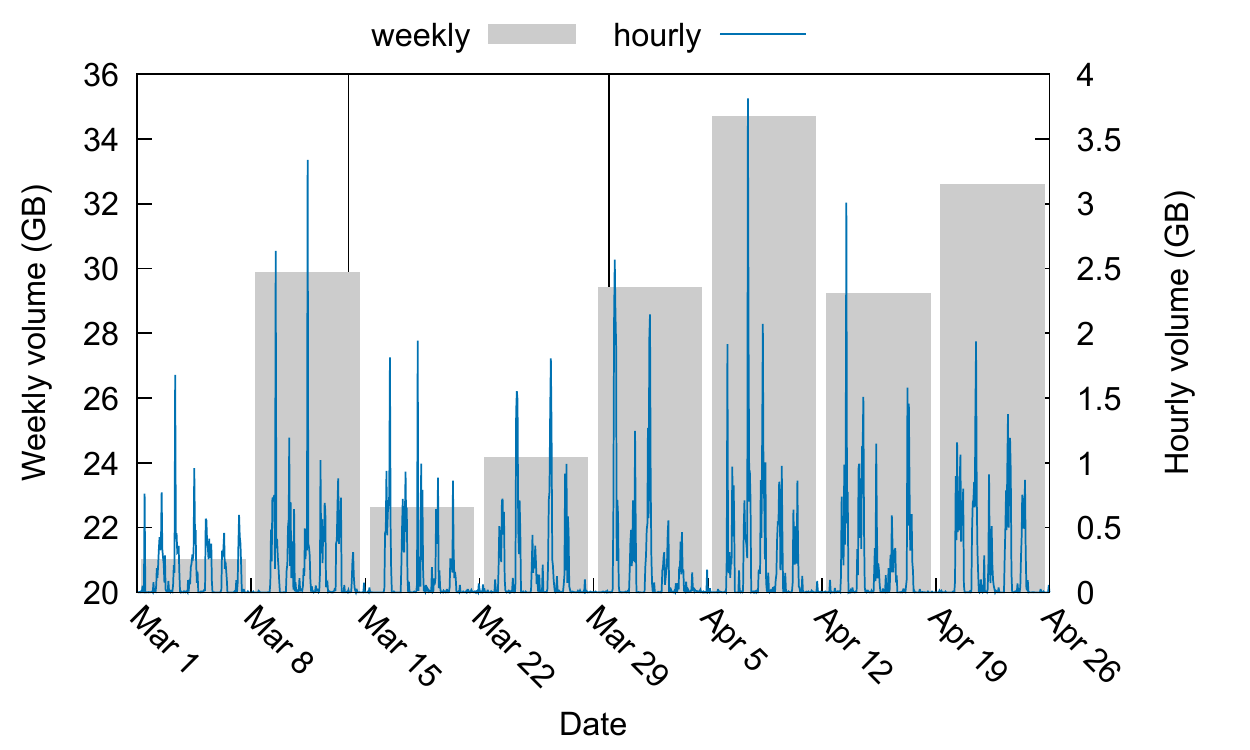}
   \label{fig:7.29.13.139}
 }
 \subfigure[IP K --- education client]{
   \includegraphics[width=.3\textwidth]{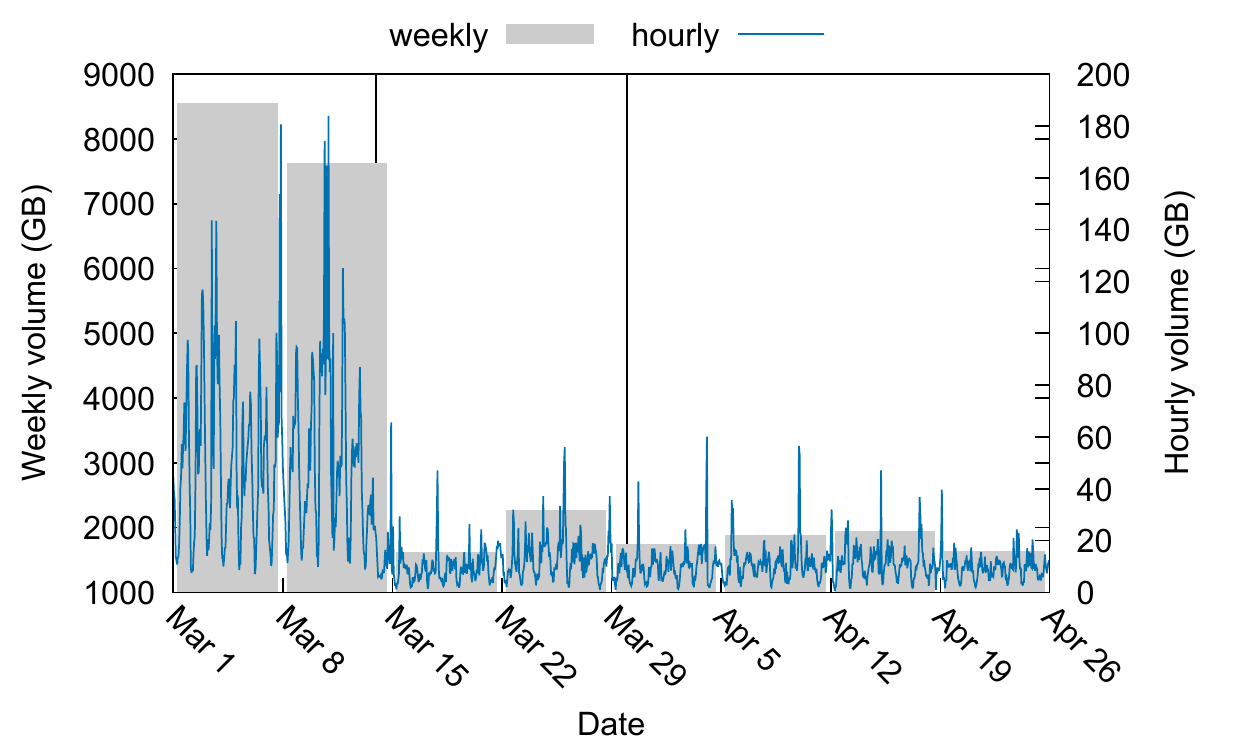}
   \label{fig:15.11.49.126}
 }
 \subfigure[IP L --- government client]{
   \includegraphics[width=.3\textwidth]{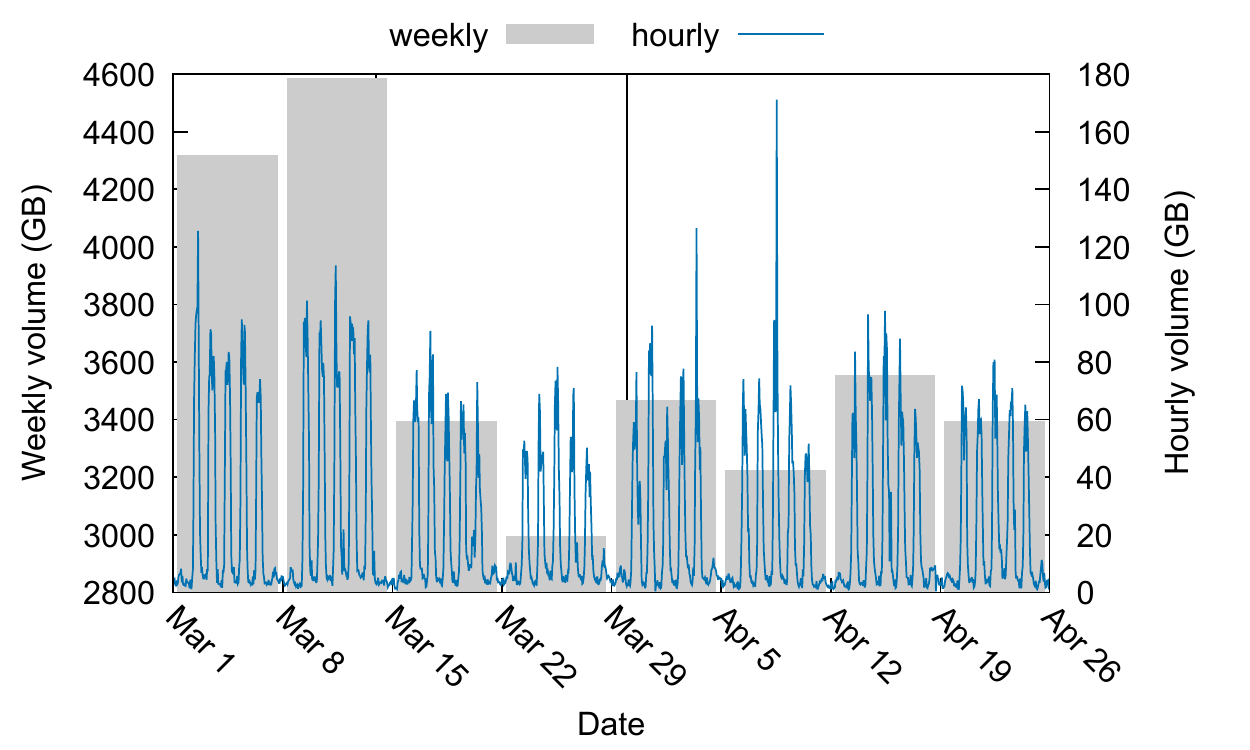}
   \label{fig:42.128.254.237}
 }
 \caption{More examples --- illustration of individual IPs' traffic changes}
\caption{Illustration of individual IPs' traffic changes. Shaded areas show weekly volume, and the lines show hourly volume of traffic.}
\label{fig:individual_ip}
\end{figure*}

% Here, we give more examples of the individual IP addresses' traffic volume changes.
% We pick the candidate IP addresses with the rules in Section~\ref{sec:individual}.
% We then randomly choose some IP addresses from the candidate IP set to demonstrate.
% Figures~\ref{fig:more_individual_ip} illustrate the traffic volume changes of three random server IPs and three random client IPs from representative ASes, with Table~\ref{tab:more_ip_to_obse} summarizing their traffic changes.

% \begin{table*}
%     \centering
%     \begin{tabular}{c|c|c|c|c|c|c}
%         \textbf{IP} &\textbf{Nature}  & \multicolumn{2}{c}{\textbf{work-hours}} & \multicolumn{2}{c}{\textbf{rest-hours}} \\ \cline{3-6}
        
%     \end{tabular}
%     \caption{More examples --- IP addresses to observe and their traffic changes}
%     \label{tab:more_ip_to_obse}
% \end{table*}

% \begin{figure*}

%  \label{fig:more_individual_ip}
% \end{figure*}

%% file: jappendix.tex
\section{Traffic Shifts per Organization}

Tables \ref{tab:trafto} and \ref{tab:traffrom} in the Appendix show the ratio of traffic change for each organization and per application. 
For clarity, we only show cells where MWW test established the change and where average daily traffic's change was larger than 1 TB. For ease of comparison we color the traffic increase as red, and traffic decrease as blue. We also remove rows and columns that show no change.

% {\scriptsize
\begin{table*}
    \centering
    \begin{tabular}{c|c|c|c|c|c|c|c|c|c|c|c|c|c|c|c}
\textbf{type} & \textbf{AS}  & \textbf{ftp}  & \textbf{ssh}  & \textbf{web}  & \textbf{https}  & \textbf{vpn}  & \textbf{highhigh}  \\ \hline 
business & B04  &  &   & \textcolor{red}{4.546} &  &   & \textcolor{blue}{0.062} \\ 
business & B05  &   & \textcolor{blue}{0.013} &  &  &   & \textcolor{blue}{0.183} \\ 
business & B06  &  &  &   & \textcolor{red}{2.061} &   & \textcolor{red}{38.712} \\ 
business & B07   & \textcolor{blue}{0.511} &  &  &   & \textcolor{red}{1.345}  & \textcolor{blue}{0.244} \\ 
business & B09  &  &  &  &   & \textcolor{red}{8.407} &  \\ 
education & E01  &  &  &   & \textcolor{red}{4.157} &   & \textcolor{blue}{0.184} \\ 
education & E03  &  &  &   & \textcolor{blue}{0.611} &  &  \\ 
education & E05  &  &   & \textcolor{blue}{0.434}  & \textcolor{blue}{0.564}  & \textcolor{red}{5.733}  & \textcolor{blue}{0.843} \\ 
education & E06  &  &  &  &  &   & \textcolor{blue}{0.08} \\ 
education & E07  &  &  &   & \textcolor{red}{2.549} &  &  \\ 
education & E08  &  &  &   & \textcolor{red}{4.118}  & \textcolor{red}{1755058.231}  & \textcolor{blue}{0.102} \\ 
education & E09  &  &   & \textcolor{blue}{0.262}  & \textcolor{blue}{0.316}  & \textcolor{red}{4.613}  & \textcolor{blue}{0.298} \\ 
education & E10  &   & \textcolor{red}{1.653} &   & \textcolor{red}{1.831}  & \textcolor{blue}{0.304} &  \\ 
education & E12  &  &  &   & \textcolor{red}{1.873}  & \textcolor{red}{58.124}  & \textcolor{red}{2.786} \\ 
education & E13  &   & \textcolor{blue}{0.276} &   & \textcolor{red}{2.174}  & \textcolor{red}{7.043} &  \\ 
education & E15  &  &  &   & \textcolor{red}{1.704} &   & \textcolor{blue}{0.483} \\ 
education & E16  &  &  &   & \textcolor{red}{6.212} &   & \textcolor{blue}{0.347} \\ 
education & E17  &  &  &   & \textcolor{red}{1.798}  & \textcolor{red}{7.475}  & \textcolor{blue}{0.148} \\ 
education & E18  &  &  &  &  &  &  \\ 
education & E19  &  &  &   & \textcolor{red}{1.904} &   & \textcolor{blue}{0.301} \\ 
education & E20  &  &  &   & \textcolor{red}{5.787}  & \textcolor{red}{2.236}  & \textcolor{blue}{0.25} \\ 
education & E21  &  &  &   & \textcolor{blue}{0.858}  & \textcolor{red}{2.404} &  \\ 
education & E22   & \textcolor{red}{10.402} &   & \textcolor{red}{1.616}  & \textcolor{red}{1.25}  & \textcolor{red}{3.78} &  \\ 
education & E23  &  &  &  &   & \textcolor{red}{7.499}  & \textcolor{blue}{0.405} \\ 
gov & G02  &  &  &   & \textcolor{red}{1.2} &  &  \\ 
gov & G05  &  &  &   & \textcolor{blue}{0.092} &  &  \\ 
gov & G08  &  &  &   & \textcolor{red}{2.517}  & \textcolor{red}{11.019} &  \\ 
gov & G09  &  &  &   & \textcolor{red}{1.36}  & \textcolor{red}{3.813} &  \\ 
gov & G10  &  &  &   & \textcolor{red}{3.158} &  &  \\ 
gov & G11  &  &  &   & \textcolor{red}{1.5}  & \textcolor{red}{4.681}  & \textcolor{blue}{0.428} \\ 
gov & G12  &   & \textcolor{blue}{0.21} &   & \textcolor{red}{1.421}  & \textcolor{red}{12437285.305} &  \\ 
gov & G13   & \textcolor{blue}{0.595}  & \textcolor{red}{3.037}  & \textcolor{red}{1.241}  & \textcolor{red}{1.057} &   & \textcolor{blue}{0.109} \\ 
isp & I02  &   & \textcolor{blue}{0.299}  & \textcolor{blue}{0.396}  & \textcolor{blue}{0.382}  & \textcolor{red}{6.04}  & \textcolor{red}{3.107} \\ 
isp & I03  &  &   & \textcolor{red}{6.87}  & \textcolor{red}{2.367}  & \textcolor{red}{18.144}  & \textcolor{blue}{0.49} \\ 
isp & I05  &  &  &   & \textcolor{blue}{0.313} &  &  \\ 
    \end{tabular}
    \caption{Traffic on flows initiated to local organizations.}
    \label{tab:trafto}
\end{table*}
% }

{\scriptsize
\begin{table*}
    \centering
    \begin{tabular}{c|c|c|c|c|c|c|c|c|c|c|c|c|c|c|c}
\textbf{type} & \textbf{AS}  & \textbf{ftp}  & \textbf{ssh}  & \textbf{web}  & \textbf{https}  & \textbf{vpn}  & \textbf{bj}  & \textbf{goto}  & \textbf{gmeet}  & \textbf{skype}  & \textbf{webex}  & \textbf{zoom}  & \textbf{steam}  & \textbf{highhigh}  \\ \hline 
business & B02  &   &   &    & \textcolor{red}{2.517} &   &   &   &   &   &   &   &   &   \\ 
business & B03  &   &    & \textcolor{blue}{0.429}  & \textcolor{blue}{0.723} &   &   &   &   &   &   &   &   &   \\ 
business & B04  &   &    & \textcolor{blue}{0.348}  & \textcolor{blue}{0.283} &   &   &   &   &   &   &    & \textcolor{blue}{0.093} &   \\ 
business & B05  &   &   &    & \textcolor{blue}{0.485} &   &   &   &   &   &   &   &   &   \\ 
business & B06  &   &   &   &    & \textcolor{blue}{0.391} &   &   &   &   &   &   &    & \textcolor{red}{43219681280} \\ 
business & B07  &    & \textcolor{blue}{0.421}  & \textcolor{blue}{0.902} &    & \textcolor{red}{44.424} &   &   &   &   &    & \textcolor{red}{5.952} &    & \textcolor{blue}{0.278} \\ 
business & B09  &   &   &   &    & \textcolor{red}{11.832} &   &   &   &   &   &   &   &   \\ 
education & E01  &   &    & \textcolor{blue}{0.232}  & \textcolor{blue}{0.357} &   &   &   &    & \textcolor{blue}{0.179}  & \textcolor{blue}{0.439} &    & \textcolor{blue}{0.11}  & \textcolor{blue}{0.384} \\ 
education & E03  &   &    & \textcolor{blue}{0.524}  & \textcolor{blue}{0.688} &   &   &   &    & \textcolor{blue}{0.518} &    & \textcolor{red}{8.36} &   &   \\ 
education & E04  &   &   &    & \textcolor{blue}{0.019} &   &   &   &   &   &   &   &   &   \\ 
education & E05  &    & \textcolor{blue}{0.28}  & \textcolor{blue}{0.256}  & \textcolor{blue}{0.312}  & \textcolor{blue}{0.27} &   &   &    & \textcolor{blue}{0.558} &    & \textcolor{red}{2.057}  & \textcolor{blue}{0.353}  & \textcolor{blue}{0.069} \\ 
education & E06  &   &    & \textcolor{blue}{0.12}  & \textcolor{blue}{0.162} &   &   &   &   &   &   &   &   &   \\ 
education & E08  &    & \textcolor{blue}{0.078}  & \textcolor{blue}{0.111}  & \textcolor{blue}{0.032}  & \textcolor{blue}{0.051} &   &    & \textcolor{red}{9.629}  & \textcolor{blue}{0.504}  & \textcolor{red}{1.86} &   &   & \textcolor{blue}{0.359} \\ 
education & E09   & \textcolor{blue}{0.015}  & \textcolor{blue}{0.399}  & \textcolor{blue}{0.325}  & \textcolor{blue}{0.377}  & \textcolor{red}{3.118} &    & \textcolor{blue}{0.296} &    & \textcolor{blue}{0.243}  & \textcolor{blue}{0.731} &    & \textcolor{blue}{0.18}  & \textcolor{blue}{0.316} \\ 
education & E10  &    & \textcolor{blue}{0.373}  & \textcolor{blue}{0.345}  & \textcolor{blue}{0.227}  & \textcolor{blue}{0.278}  & \textcolor{blue}{0.493} &   &    & \textcolor{blue}{0.371} &    & \textcolor{red}{2.473}  & \textcolor{blue}{0.111} &   \\ 
education & E11  &   &    & \textcolor{blue}{0.32}  & \textcolor{blue}{0.36} &   &   &   &   &   &   &   &   &   \\ 
education & E12  &   &    & \textcolor{blue}{0.386}  & \textcolor{blue}{0.227}  & \textcolor{red}{654.819} &   &   &    & \textcolor{blue}{0.198} &    & \textcolor{blue}{0.258} &  &   \\ 
education & E13  &   &    & \textcolor{blue}{0.361}  & \textcolor{blue}{0.421}  & \textcolor{blue}{0.076} &    & \textcolor{red}{2.132} &   &    & \textcolor{red}{1.842}  & \textcolor{red}{4.161}  & \textcolor{blue}{0.196} &   \\ 
education & E15  &   &    & \textcolor{blue}{0.264}  & \textcolor{blue}{0.209}  & \textcolor{blue}{0.052} &   &   &    & \textcolor{red}{1.451}  & \textcolor{red}{2.382}  & \textcolor{red}{4.621}  & \textcolor{blue}{0.446}  & \textcolor{blue}{0.534} \\ 
education & E16  &   &   & \textcolor{blue}{0.183}  & \textcolor{blue}{0.18}  & \textcolor{blue}{0.502} &   &   &    & \textcolor{red}{1.47} &   &    & \textcolor{blue}{0.227}  & \textcolor{blue}{0.339} \\ 
education & E17  &   &    & \textcolor{blue}{0.051}  & \textcolor{blue}{0.087}  & \textcolor{red}{11.494} &   &   &    & \textcolor{blue}{0.7}  & \textcolor{red}{1.723} &    & \textcolor{blue}{0.026}  & \textcolor{blue}{0.228} \\ 
education & E18  &   &    & \textcolor{blue}{0.293}  & \textcolor{blue}{0.103} &   &   &   &   &   &   &    & \textcolor{blue}{0.01}  & \textcolor{red}{60.664} \\ 
education & E19  &   &    & \textcolor{blue}{0.126}  & \textcolor{blue}{0.165}  & \textcolor{blue}{0.113} &    & \textcolor{red}{32.08} &    & \textcolor{blue}{0.106} &   &    & \textcolor{blue}{0.033}  & \textcolor{blue}{0.313} \\ 
education & E20  &   &    & \textcolor{blue}{0.594}  & \textcolor{blue}{0.249}  & \textcolor{blue}{0.079} &   &   &    & \textcolor{blue}{0.578} &   &   &  &   \\ 
education & E21  &   &    & \textcolor{blue}{0.312}  & \textcolor{blue}{0.394} &    & \textcolor{blue}{0.082}  & \textcolor{red}{1.617}  & \textcolor{red}{7.261}  & \textcolor{blue}{0.559}  & \textcolor{red}{3.485}  & \textcolor{red}{1.544}  & \textcolor{blue}{0.285}  & \textcolor{blue}{0.544} \\ 
education & E22  &   &   &   &    & \textcolor{blue}{0.014} &   &    & \textcolor{red}{1.486} &    & \textcolor{red}{2.033}  & \textcolor{red}{2.291} &    & \textcolor{blue}{0.447} \\ 
education & E23  &   &    & \textcolor{blue}{0.131}  & \textcolor{blue}{0.169}  & \textcolor{blue}{0.44} &    & \textcolor{blue}{0.047} &    & \textcolor{blue}{0.536} &    & \textcolor{red}{1.671}  & \textcolor{blue}{0.106}  & \textcolor{blue}{0.3} \\ 
gov & E07  &   &   &   &   &   &   &    & \textcolor{red}{5.132} &   &   &   &   &   \\ 
gov & G02   & \textcolor{red}{746.609} &   &   &   &   &   &   &   &   &   &   &   &   \\ 
gov & G05  &   &   &    & \textcolor{blue}{0.378} &   &   &   &   &   &   &   &   &   \\ 
gov & G06  &   &    & \textcolor{blue}{0.125}  & \textcolor{blue}{0.121} &   &   &   &   &   &   &   &   &   \\ 
gov & G08  &    & \textcolor{blue}{0.294}  & \textcolor{blue}{0.597}  & \textcolor{blue}{0.736} &   &    & \textcolor{red}{4.047}  & \textcolor{red}{9.017} &    & \textcolor{red}{4.718}  & \textcolor{red}{4.338} &    & \textcolor{blue}{0.796} \\ 
gov & G09  &   &    & \textcolor{blue}{0.683} &   &   &   &   &   &   &   &   &   &   \\ 
gov & G10  &   &    & \textcolor{blue}{0.606}  & \textcolor{blue}{0.564}  & \textcolor{blue}{0.245} &    & \textcolor{red}{4.311} &    & \textcolor{red}{1.358} &    & \textcolor{red}{1.87} &   &   \\ 
gov & G11  &   &    & \textcolor{blue}{0.35}  & \textcolor{blue}{0.419}  & \textcolor{blue}{0.44}  & \textcolor{blue}{0.15} &    & \textcolor{red}{2.7}  & \textcolor{blue}{0.479} &    & \textcolor{red}{2.269}  & \textcolor{blue}{0.135}  & \textcolor{red}{2.618} \\ 
gov & G12  &   &    & \textcolor{blue}{0.411}  & \textcolor{blue}{0.51} &   &   &   &   &    & \textcolor{red}{20.661} &   &    & \textcolor{blue}{0.53} \\ 
gov & G13  &   &   &    & \textcolor{blue}{0.615}  & \textcolor{blue}{0.018} &   &   &    & \textcolor{red}{1.967} &   &   &   &   \\ 
isp & I01  &   &   &    & \textcolor{red}{1.508} &   &   &   &   &   &   &   &   &   \\ 
isp & I02   & \textcolor{blue}{0.203}  & \textcolor{red}{6.924}  & \textcolor{blue}{0.479}  & \textcolor{red}{2.414}  & \textcolor{red}{6.965} &   &   &   &   &   &   &    & \textcolor{blue}{0.753} \\ 
isp & I03  &   &    & \textcolor{blue}{0.473}  & \textcolor{blue}{0.143}  & \textcolor{blue}{0.039} &   &    & \textcolor{red}{26.044} &   &    & \textcolor{red}{1.816} &  &   \\ 
isp & I06   & \textcolor{blue}{0.005} &   &    & \textcolor{blue}{0.853}  & \textcolor{red}{4.128} &   &   &   &   &   &   &   &   \\ 
    \end{tabular}
    \caption{Traffic on flows initiated by the local organizations.}
    \label{tab:traffrom}
\end{table*}
}

Table \ref{tab:peershifts} shows how the traffic between peers changed. We only show those changes that amounted to more than 1 TB change per day. 

{\scriptsize
\begin{table*}
    \centering
    \begin{tabular}{c|c|c|c|c|c|c|c|c|c|c|c}
\textbf{type} & \textbf{AS}  & \textbf{total change} & \multicolumn{4}{c|}{\textbf{From these types to AS -- change rto (daily TB)}} & \textbf{total change} & \multicolumn{4}{|c}{\textbf{From AS to these types -- change rto (daily TB)}}  \\ \cline{4-7} \cline{9-12}
& & \textbf{daily TB}  & \textbf{business}  & \textbf{education}  & \textbf{isp}  & \textbf{hosting}  & \textbf{daily TB}  & \textbf{business}  & \textbf{education}  & \textbf{isp}  & \textbf{hosting}  \\ \hline 
business & B05  & \textcolor{blue}{-2}  &  & \textcolor{blue}{$\sim$ 0} (-2)  &  &  &  &  &  &  &  \\ 
education & E05  &  &  &  &  &  & \textcolor{blue}{-16}   & \textcolor{blue}{0.3} (-8)  & \textcolor{blue}{0.6} (-3)  & \textcolor{blue}{0.3} (-3)  & \textcolor{blue}{0.2} (-2) \\ 
education & E06  &  &  &  &  &  & \textcolor{blue}{-1}  &  &  &  &  \\ 
education & E08  &  &  &  &  &  & \textcolor{blue}{-17}  & \textcolor{blue}{$\sim$ 0} (-11)  & \textcolor{blue}{$\sim$ 0} (-2)   & \textcolor{blue}{0.1} (-1) & \textcolor{blue}{$\sim$ 0} (-2)  \\
education & E09  & \textcolor{blue}{-50}  &   & \textcolor{blue}{0.2} (-47)  & \textcolor{blue}{0.8} (-1) &  & \textcolor{blue}{-28}   & \textcolor{blue}{0.2} (-16)  & \textcolor{red}{1.6} (1)  & \textcolor{blue}{0.2} (-3)  & \textcolor{blue}{0.4} (-9) \\ 
education & E10  & \textcolor{red}{11}   & \textcolor{red}{1.5} (5)  & \textcolor{blue}{0.8} (-1)  & \textcolor{red}{3.4} (6) &  & \textcolor{blue}{-53}   & \textcolor{blue}{0.3} (-28)  & \textcolor{blue}{0.4} (-9)  & \textcolor{blue}{0.1} (-9)  & \textcolor{blue}{0.1} (-5) \\ 
education & E11  &  &  &  &  &  &  &  &  &  &  \\ 
education & E12  &  &  &  &  &  & \textcolor{blue}{-1}  &  &  &  &  \\ 
education & E13  & \textcolor{red}{4}  &  &   & \textcolor{red}{5.6} (4) &  &   & \textcolor{red}{1.3} (2)  & \textcolor{blue}{0.4} (-1)  & \textcolor{blue}{0.3} (-1)  & \textcolor{blue}{0.3} (-1) \\ 
education & E14  &  &  &  &  &  &  &  &  &  &  \\ 
education & E15  &  &  &  &  &  & \textcolor{blue}{-7}   & \textcolor{blue}{0.2} (-3)  & \textcolor{blue}{0.1} (-1)  & \textcolor{blue}{0.1} (-1)  & \textcolor{blue}{0.2} (-1) \\ 
education & E16  &  &  &  &  &  & \textcolor{blue}{-11}   & \textcolor{blue}{0.1} (-4)  & \textcolor{blue}{0.2} (-3)  & \textcolor{blue}{0.2} (-1)  & \textcolor{blue}{0.1} (-1) \\ 
education & E17  &  &  &  &  &  & \textcolor{blue}{-15}   & \textcolor{blue}{0.1} (-6) & \textcolor{blue}{$\sim$ 0} (-3)   & \textcolor{blue}{0.1} (-3) & \textcolor{blue}{$\sim$ 0} (-2)  \\ 
education & E18  &  &  &  &  &  &  &  &  &  &  \\ 
education & E19  &  &  &  &  &  & \textcolor{blue}{-4}   & \textcolor{blue}{0.1} (-1)  & \textcolor{blue}{0.1} (-1)  & \textcolor{blue}{0.2} (-1) &  \\ 
education & E20  &  &  &  &  &  & \textcolor{blue}{-3}   & \textcolor{blue}{0.4} (-2) &  &  &  \\ 
education & E21  & \textcolor{red}{1}   & \textcolor{blue}{0.5} (-1)  & \textcolor{red}{1.3} (1)  & \textcolor{red}{1.3} (1) &  & \textcolor{blue}{-42}   & \textcolor{blue}{0.4} (-21)  & \textcolor{blue}{0.5} (-8)  & \textcolor{blue}{0.3} (-7)  & \textcolor{blue}{0.3} (-5) \\ 
education & E22  & \textcolor{red}{4}   & \textcolor{blue}{0.7} (-3) &   & \textcolor{red}{1.5} (8) &  & \textcolor{blue}{-10}   & \textcolor{blue}{0.4} (-12)  & \textcolor{blue}{1} (1) &  &  \\ 
education & E23  &  &  &  &  &  & \textcolor{blue}{-14}   & \textcolor{blue}{0.1} (-7)  & \textcolor{blue}{0.3} (-2)  & \textcolor{blue}{0.1} (-2)  & \textcolor{blue}{0.1} (-2) \\ 
gov & G08  & \textcolor{red}{6}  &  &   & \textcolor{red}{3.2} (6) &  & \textcolor{blue}{-2}  &  &  &  &  \\ 
gov & G09  &  &  &  &  &  &  &  &  &  &  \\ 
gov & G10  &  &  &  &  &  &  &  &  &  &  \\ 
gov & G11  &  &  &  &  &  & \textcolor{blue}{-18}   & \textcolor{blue}{0.5} (-10)  & \textcolor{blue}{0.4} (-2)  & \textcolor{blue}{0.3} (-3)  & \textcolor{blue}{0.2} (-2) \\ 
gov & G12  &  &  &  &  &  &  &  &  &  &  \\ 
gov & G13  & \textcolor{blue}{-7}  &   & \textcolor{blue}{0.6} (-9)  & \textcolor{red}{1.1} (2) &  &  &  &  &  &  \\ 
isp & I02  &  &  &  &  &  & \textcolor{red}{3}   & \textcolor{red}{2.3} (1)  & \textcolor{red}{3.1} (2) &  &  \\ 
isp & I03  &  &  &  &  &  & \textcolor{blue}{-2}   & \textcolor{blue}{0.1} (-1) &  &  &  \\ 
    \end{tabular}
    \caption{Traffic shifts between peer types}
    \label{tab:peershifts}
\end{table*}
}

%% file: daily_appendix.tex
\section{More examples of traffic changes within a day}
\label{sec:daily_appendix}

\begin{figure*}
  \subfigure[All traffic volume]{
   \includegraphics[width=.23\textwidth]{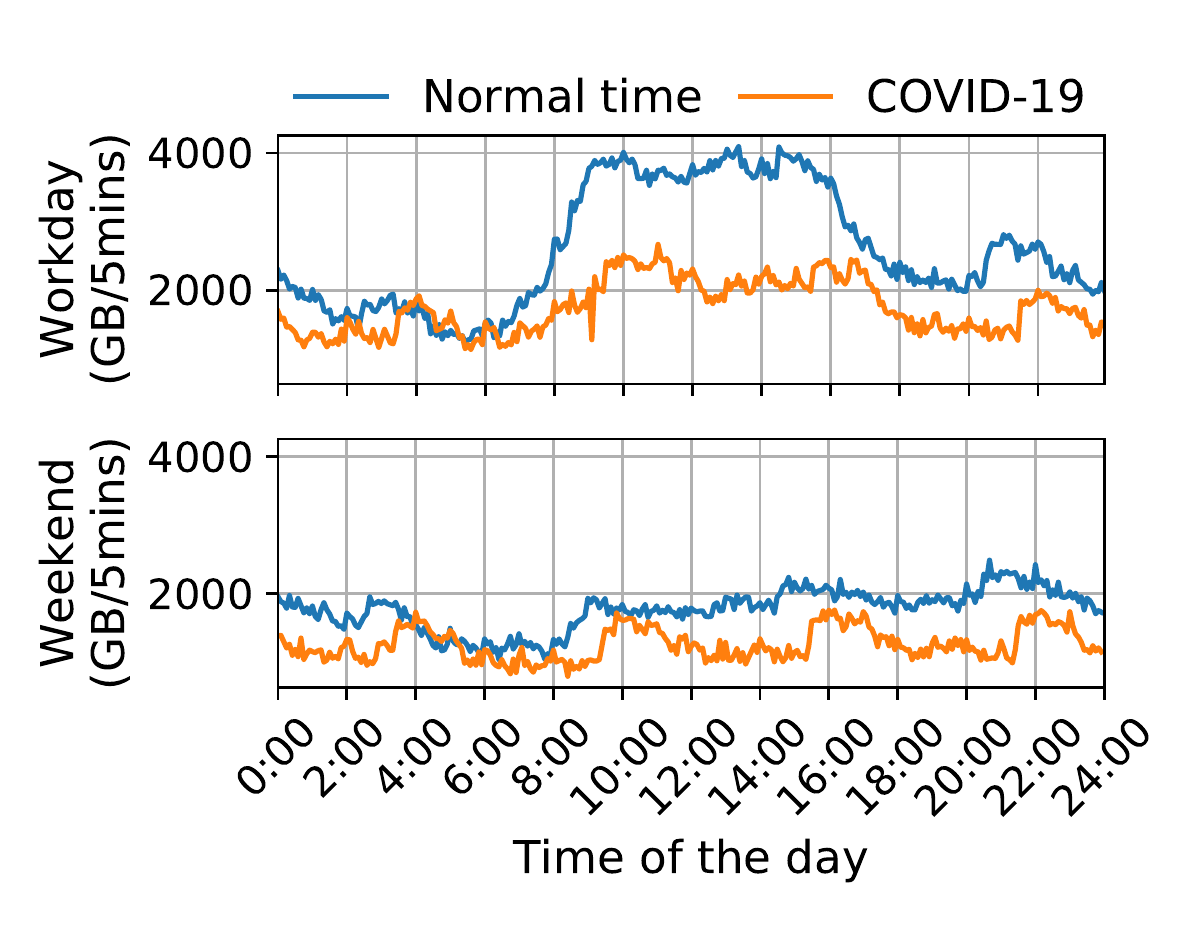}
   \label{fig:daily_all}
 } 
  \subfigure[DNS traffic volume]{
   \includegraphics[width=.23\textwidth]{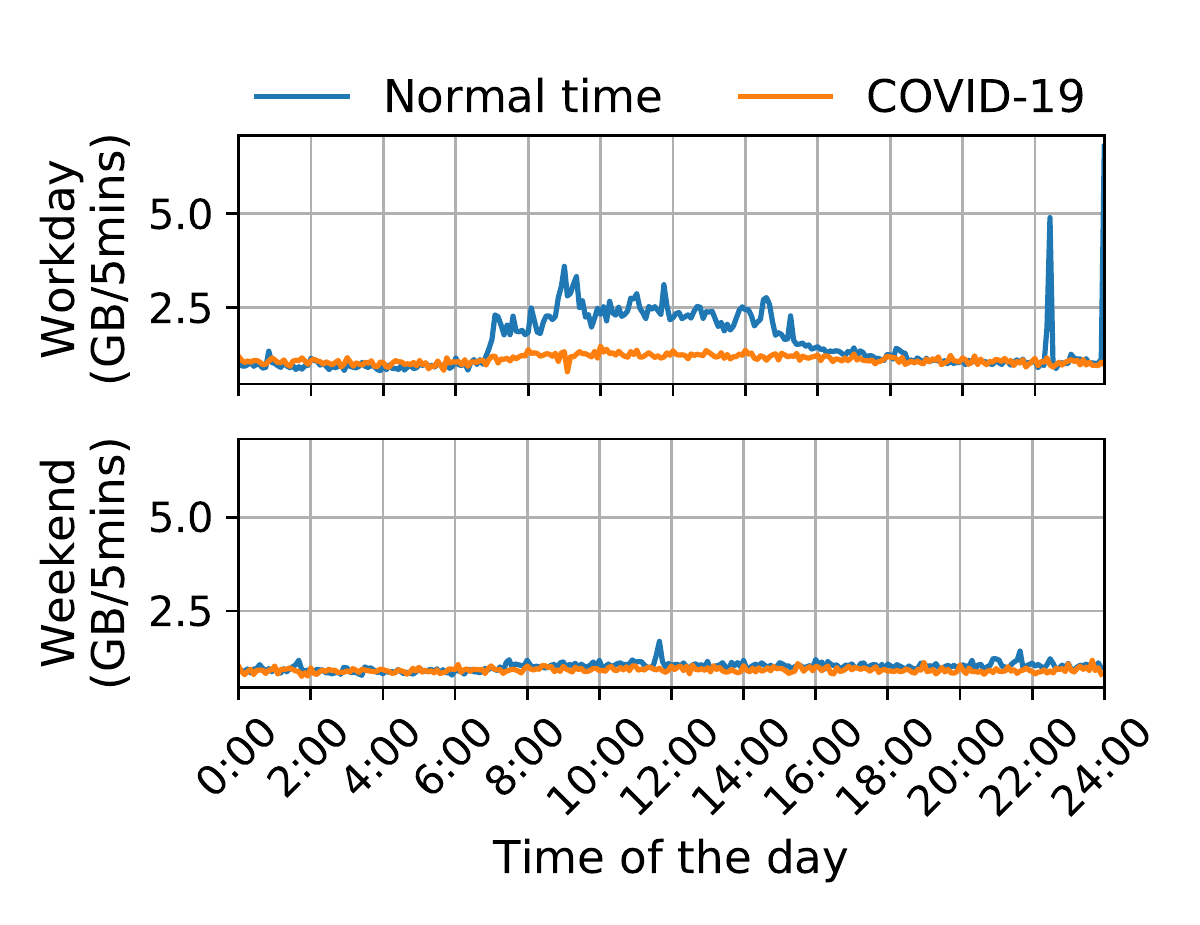}
   \label{fig:daily_dns}
 } 
 \subfigure[Email traffic volume]{
   \includegraphics[width=.23\textwidth]{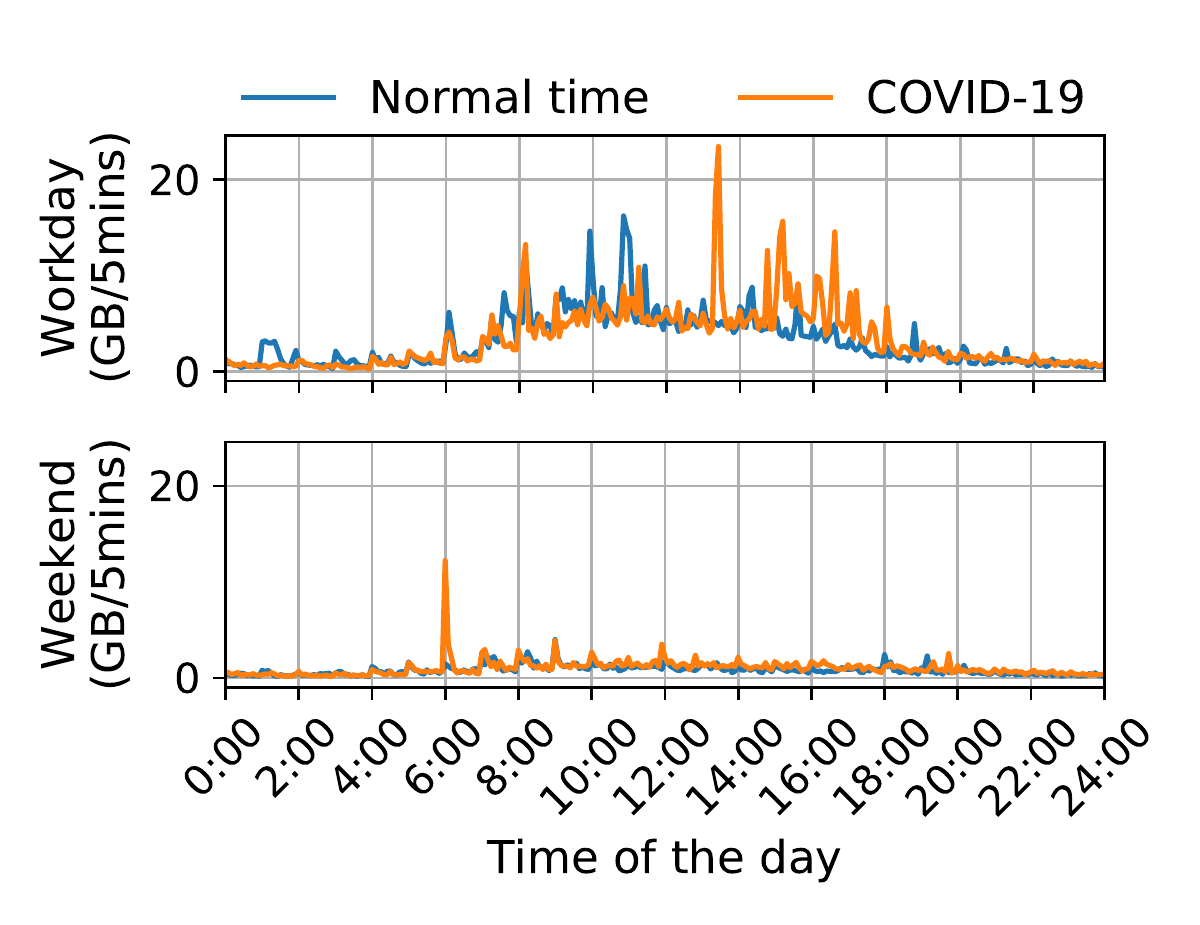}
   \label{fig:daily_smtp}
 } 
 \subfigure[SSH traffic volume]{
   \includegraphics[width=.23\textwidth]{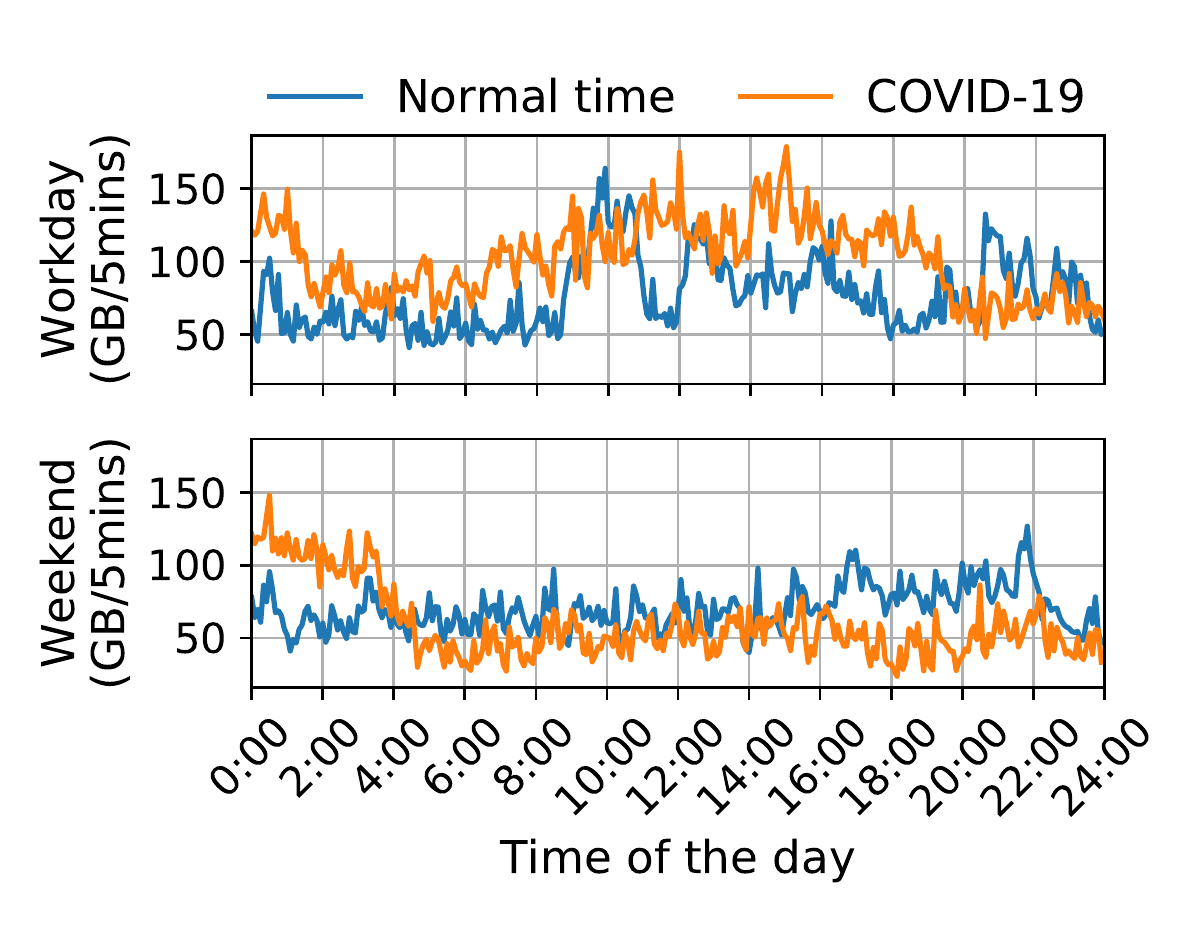}
   \label{fig:daily_ssh}
 }
 
 \subfigure[Google Meet traffic volume]{
   \includegraphics[width=.23\textwidth]{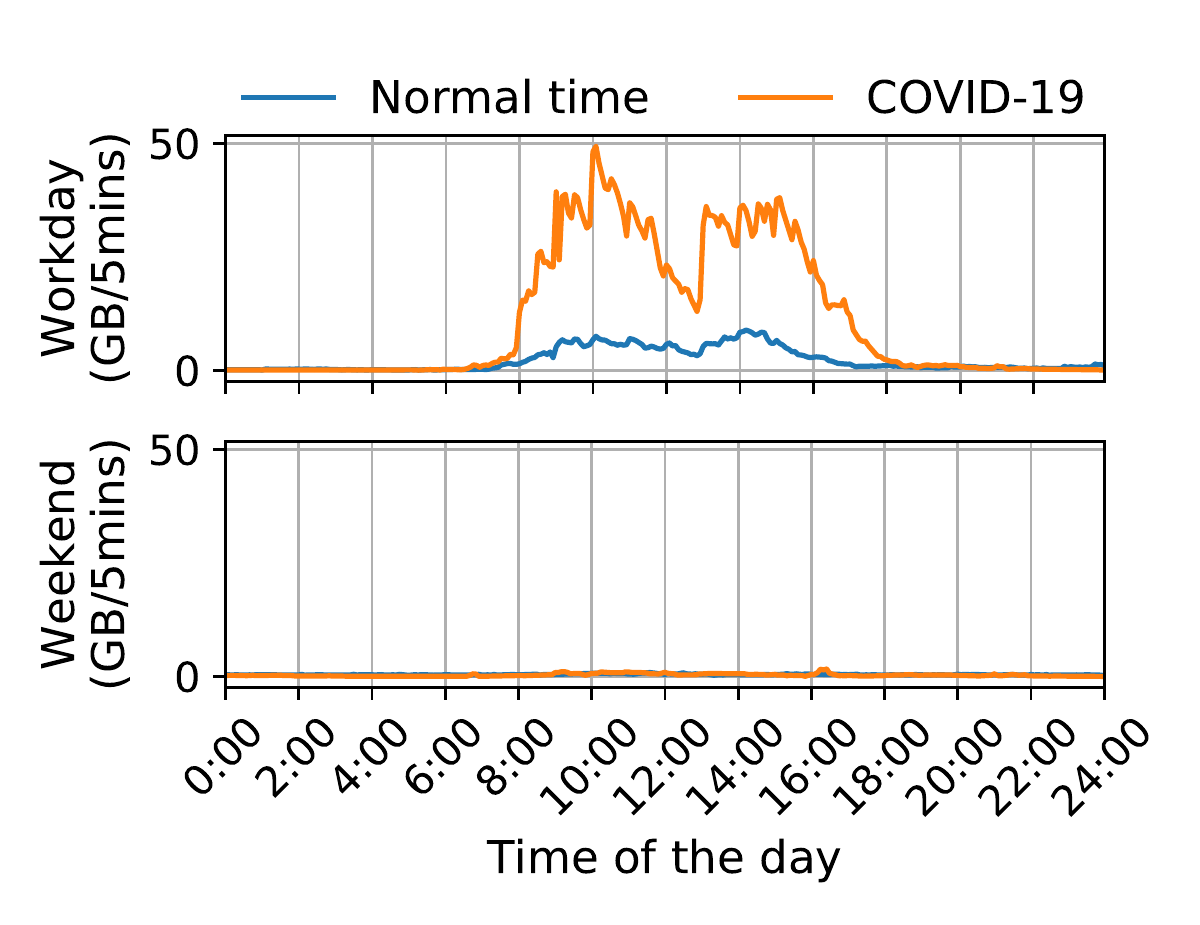}
   \label{fig:daily_gmeet}
 }
 \subfigure[GoTo Meeting traffic volume]{
   \includegraphics[width=.23\textwidth]{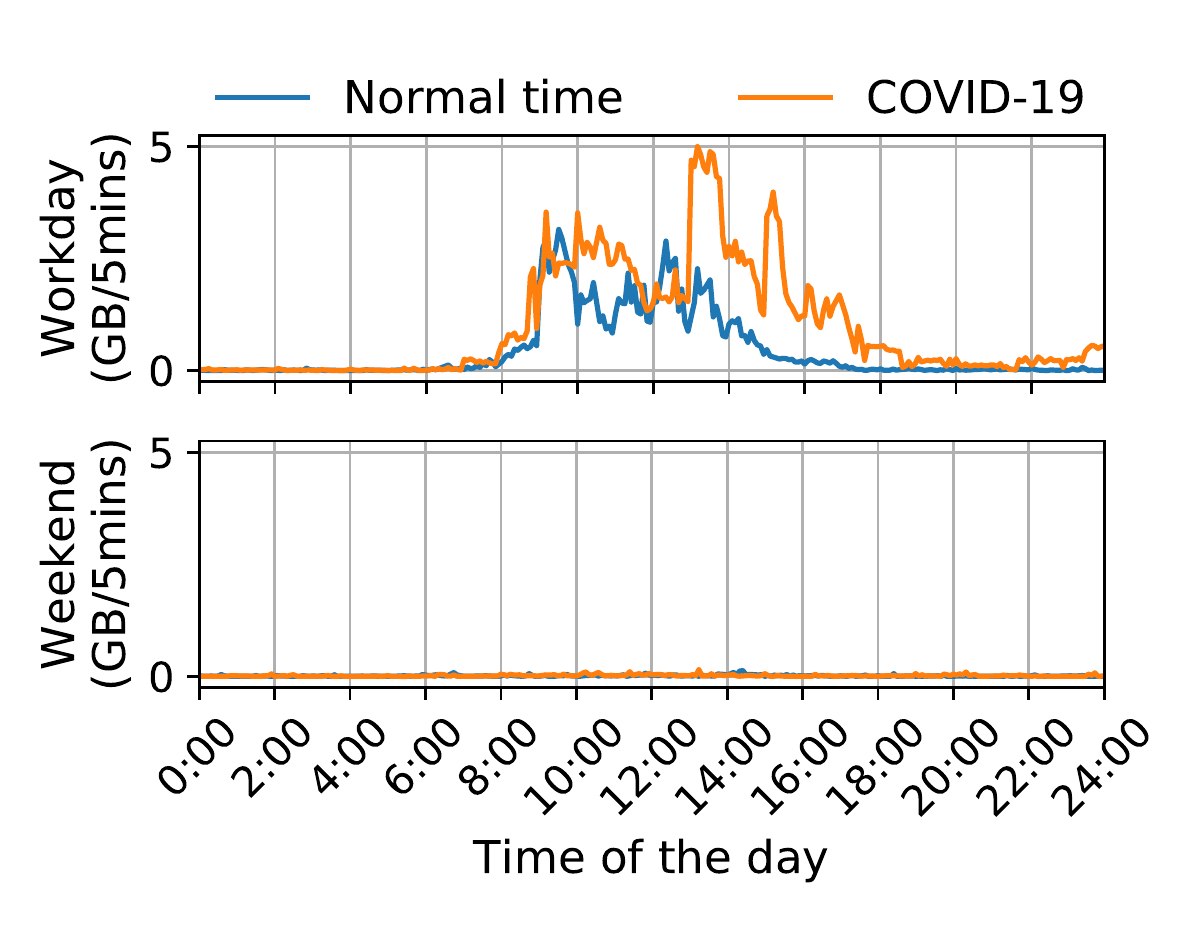}
   \label{fig:goto}
 }
 \subfigure[Webex traffic volume]{
   \includegraphics[width=.23\textwidth]{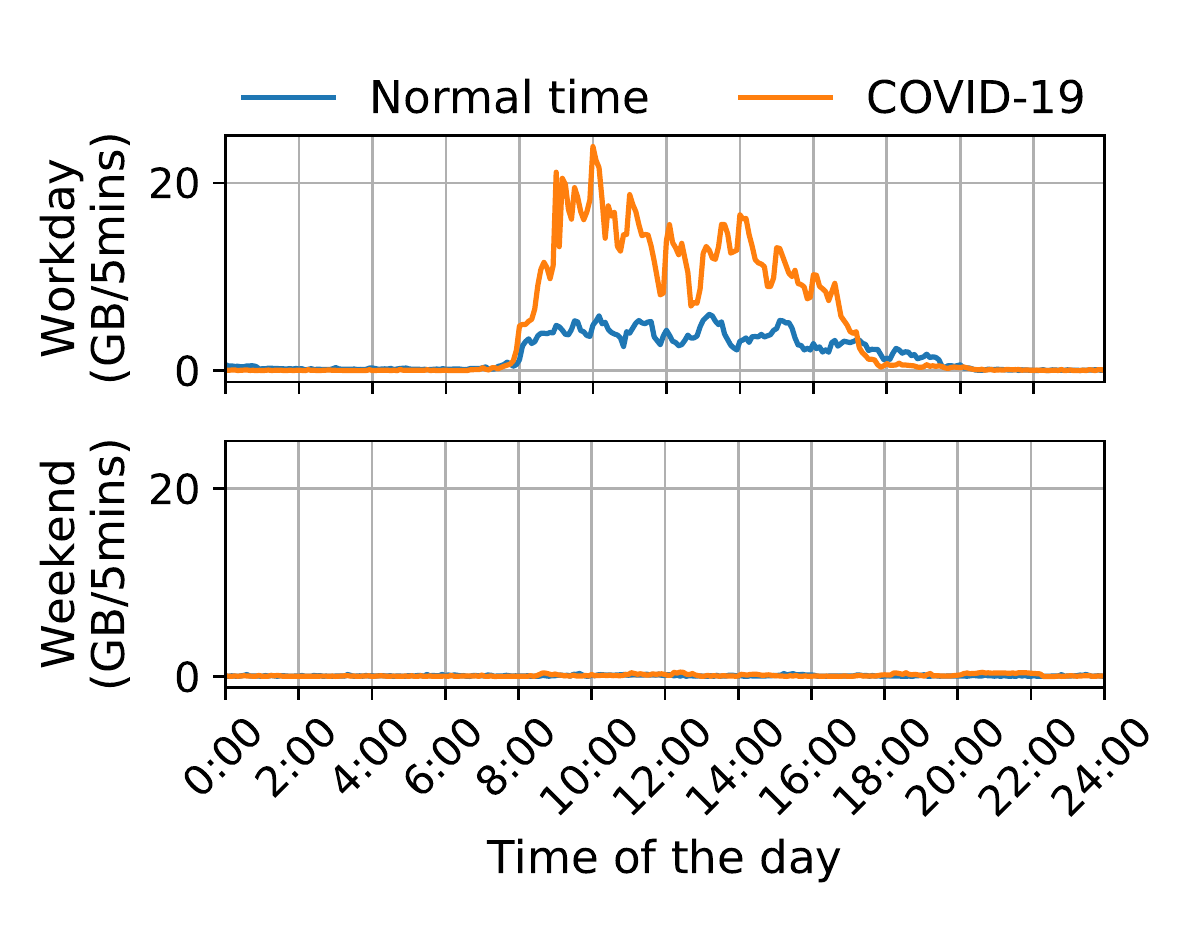}
   \label{fig:daily_webex}
 }
 \subfigure[Skype traffic volume]{
   \includegraphics[width=.23\textwidth]{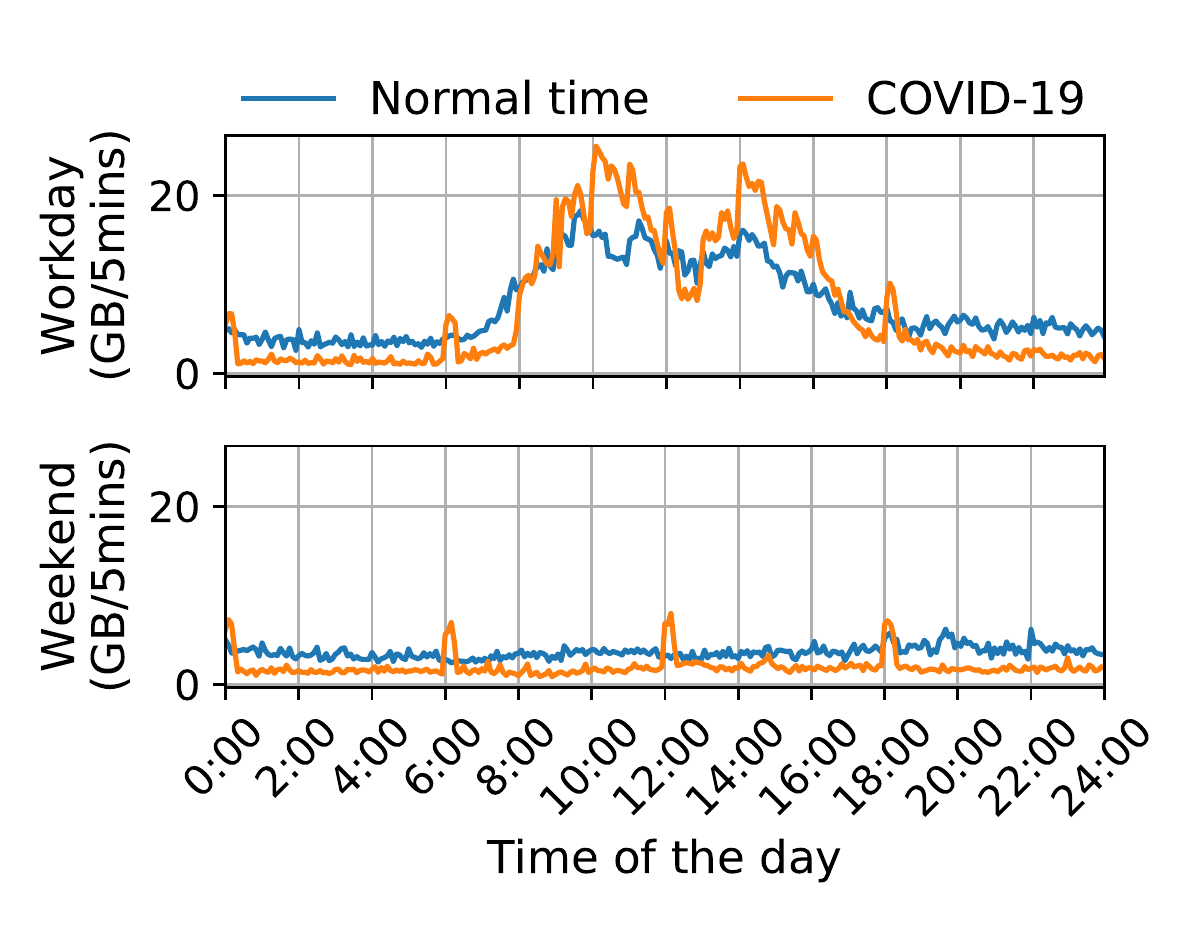}
   \label{fig:daily_skype}
 }
 \subfigure[ICMP traffic volume]{
   \includegraphics[width=.23\textwidth]{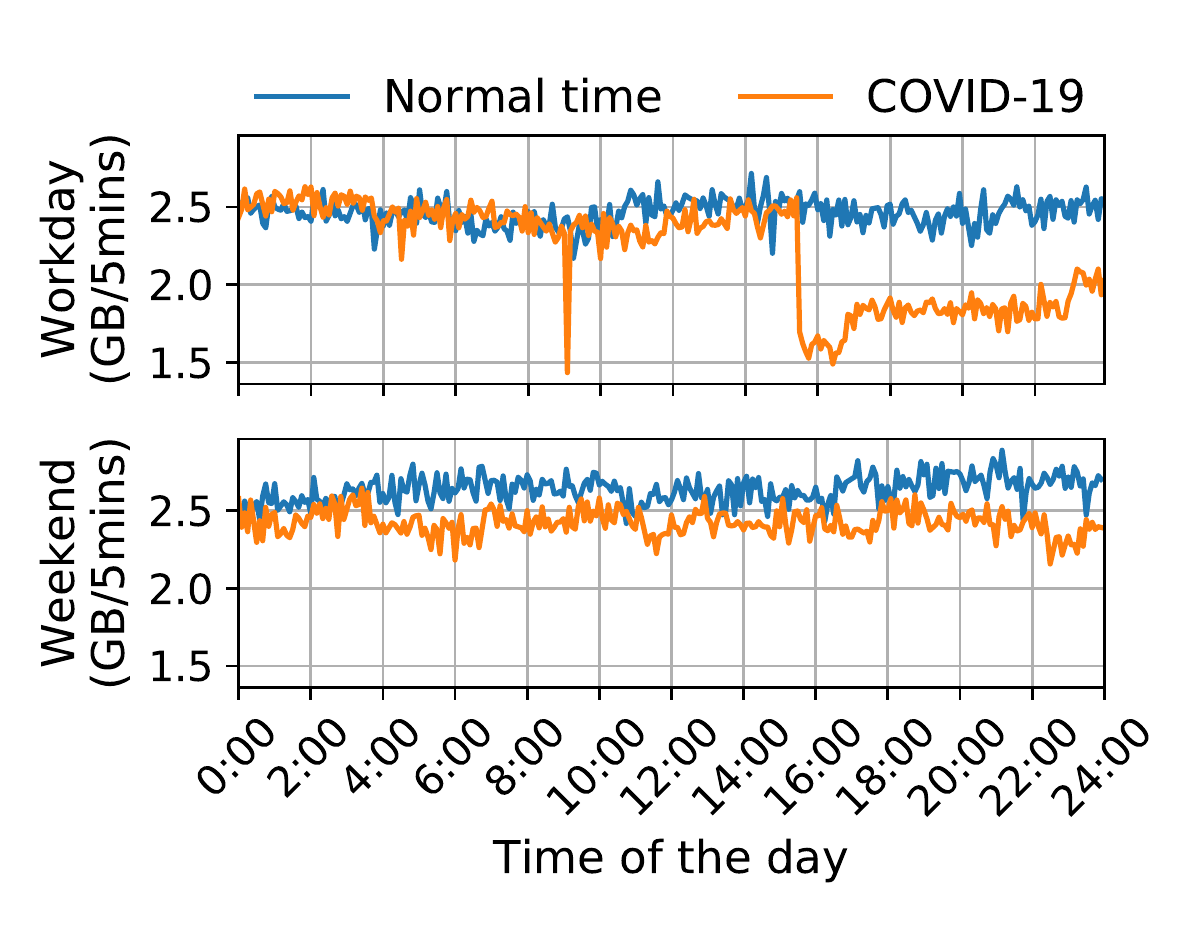}
   \label{fig:daily_icmp}
 }
 \subfigure[SYN traffic volume]{
   \includegraphics[width=.23\textwidth]{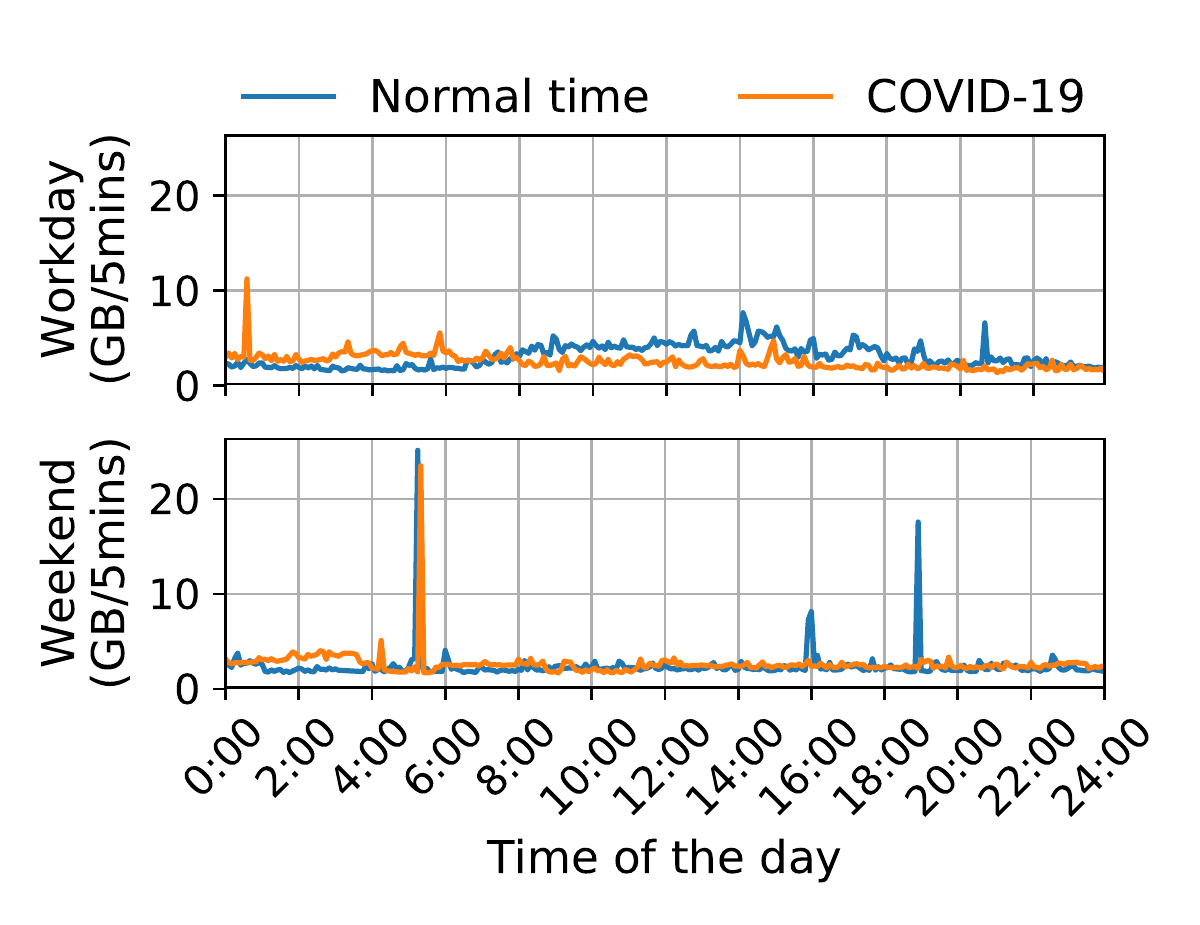}
   \label{fig:daily_syn}
 }
 \subfigure[NTP traffic volume]{
   \includegraphics[width=.23\textwidth]{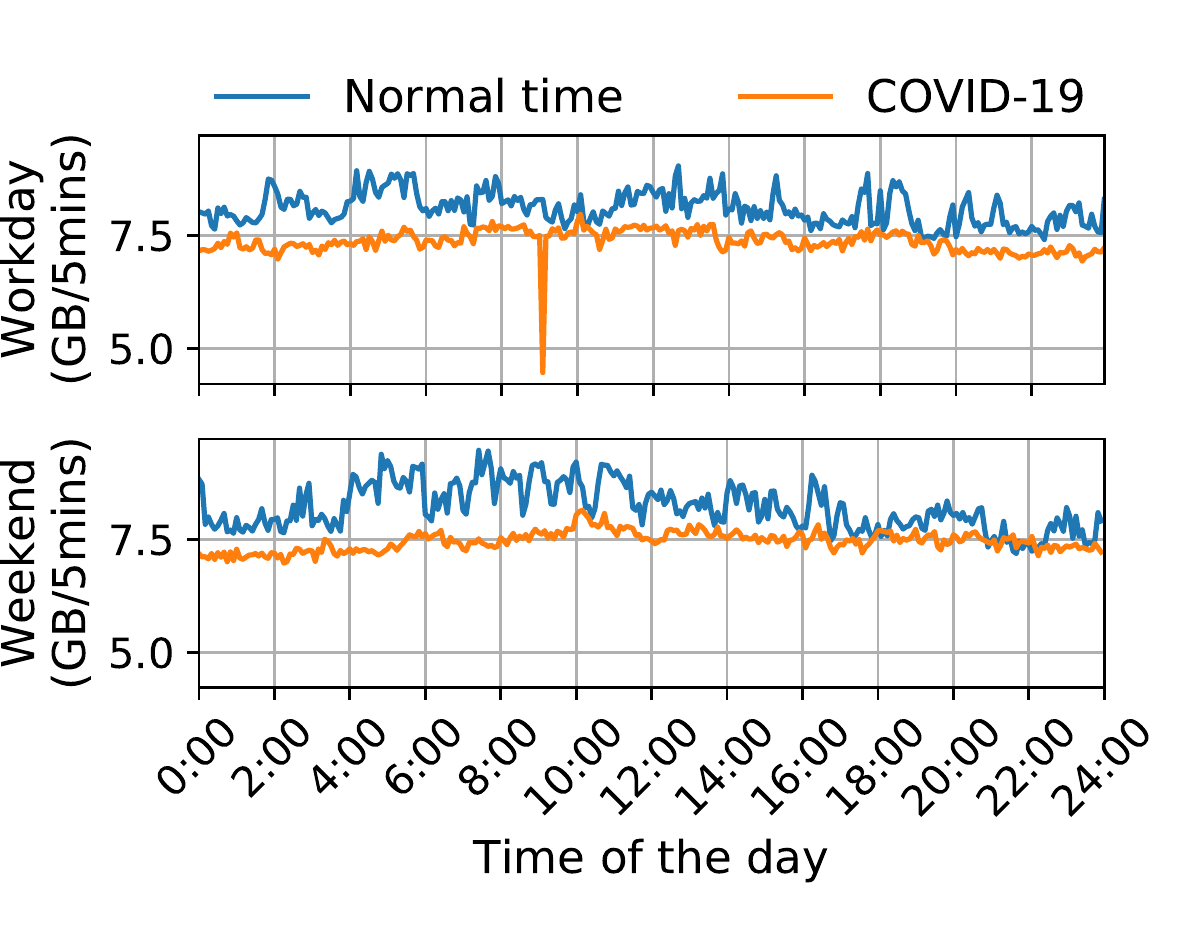}
   \label{fig:daily_ntp}
 } 
\subfigure[HTTP traffic volume]{
   \includegraphics[width=.23\textwidth]{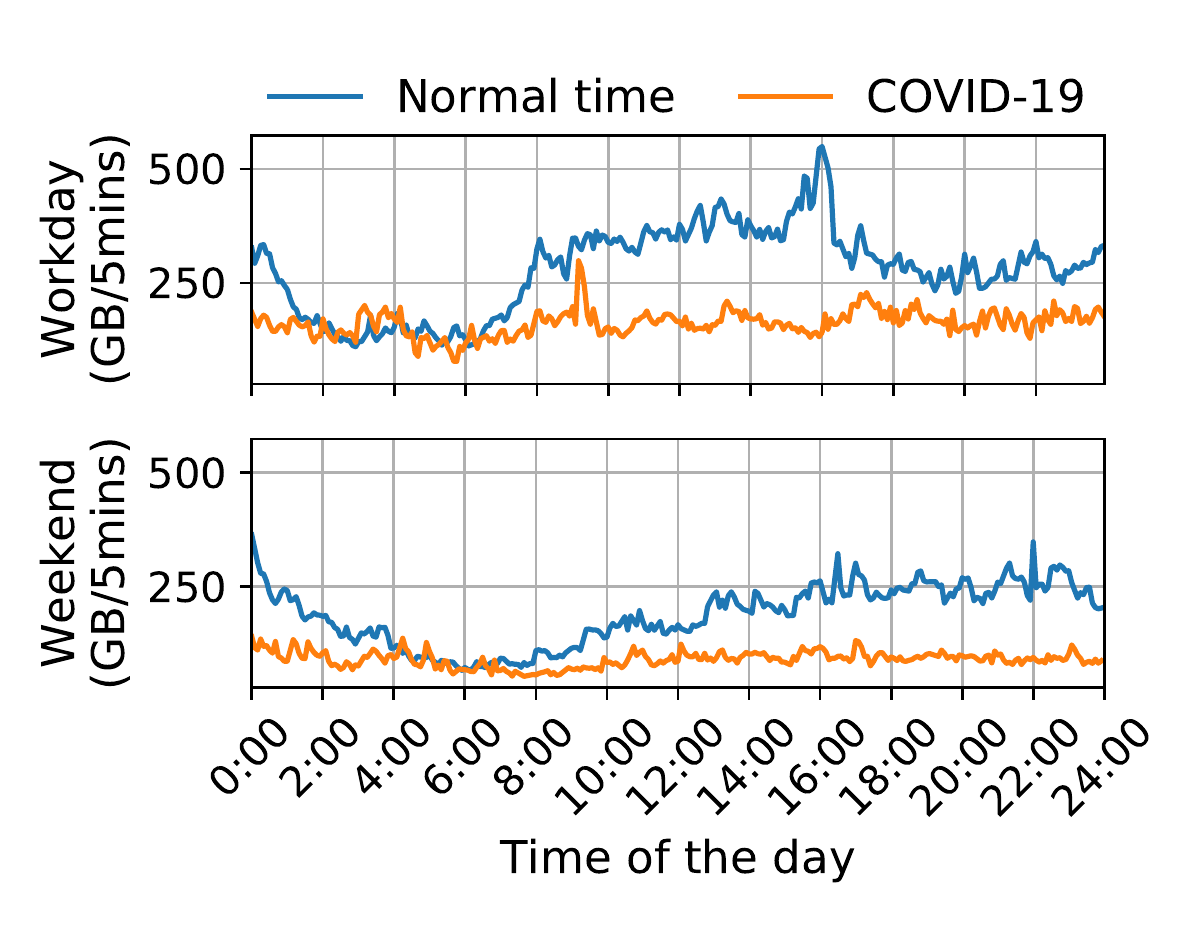}
   \label{fig:daily_http}
 } 
 \subfigure[POP3 traffic volume]{
   \includegraphics[width=.23\textwidth]{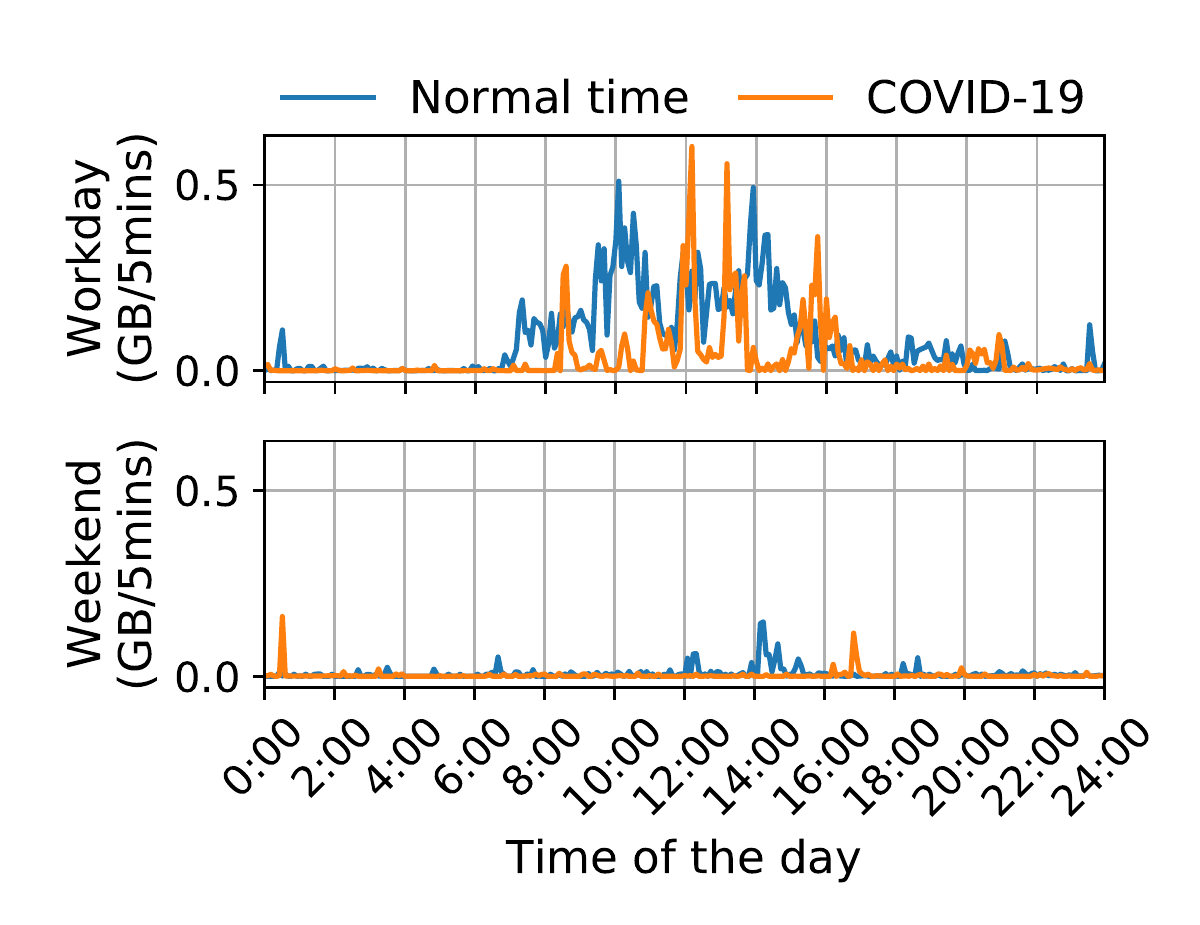}
   \label{fig:daily_pop3}
 }
 \subfigure[Telnet traffic volume]{
   \includegraphics[width=.23\textwidth]{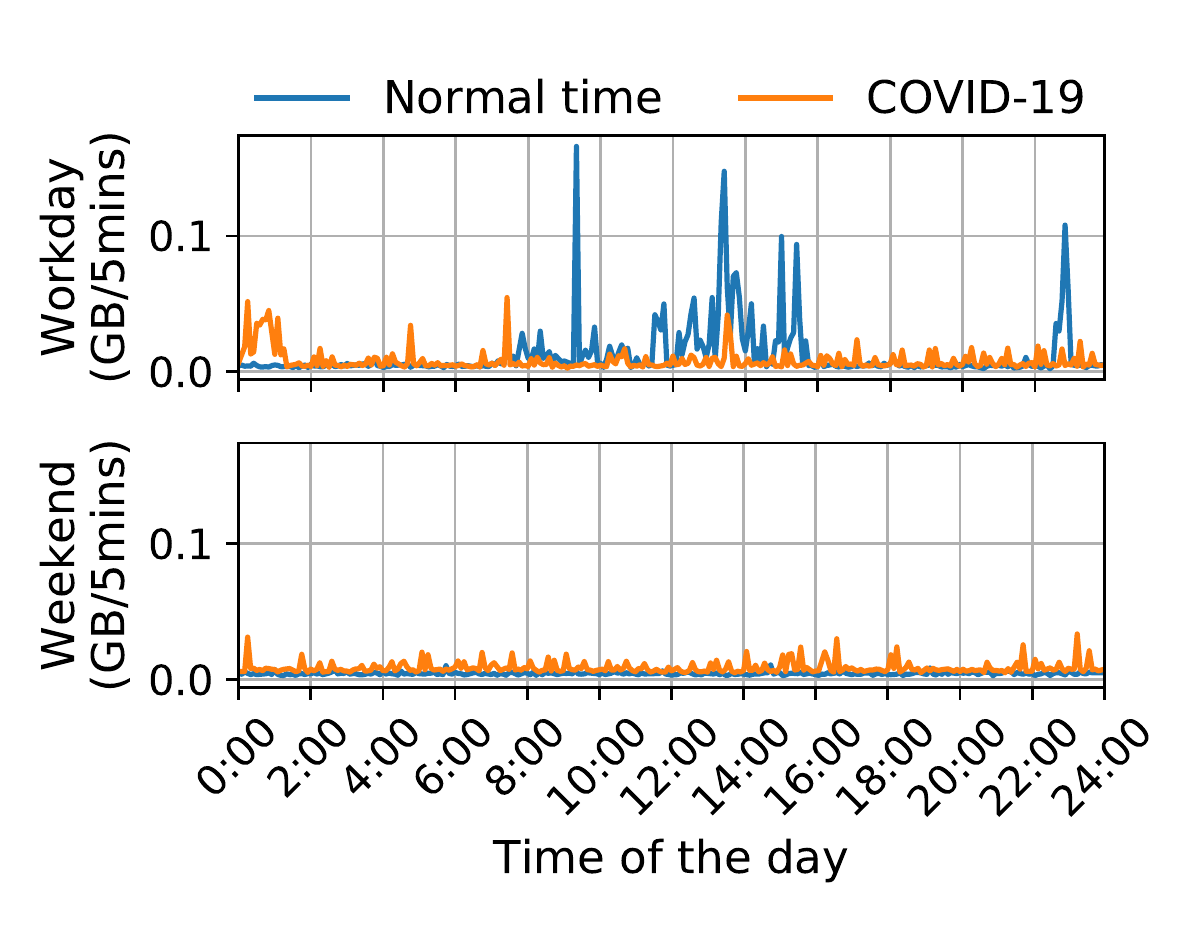}
   \label{fig:daily_telnet}
 }
 \subfigure[FTP traffic volume]{
   \includegraphics[width=.23\textwidth]{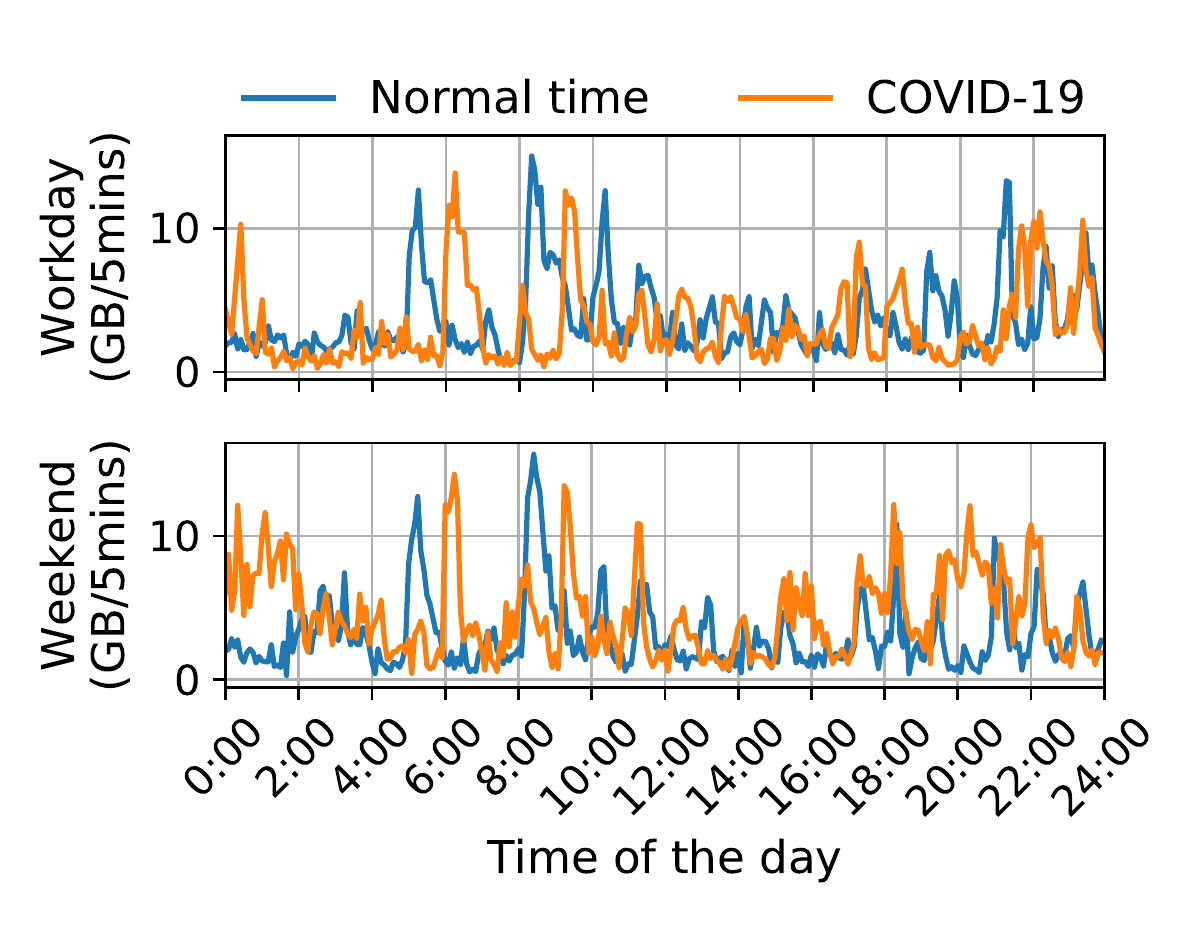}
   \label{fig:daily_ftp}
 }
 \subfigure[BlueJeans traffic volume]{
   \includegraphics[width=.23\textwidth]{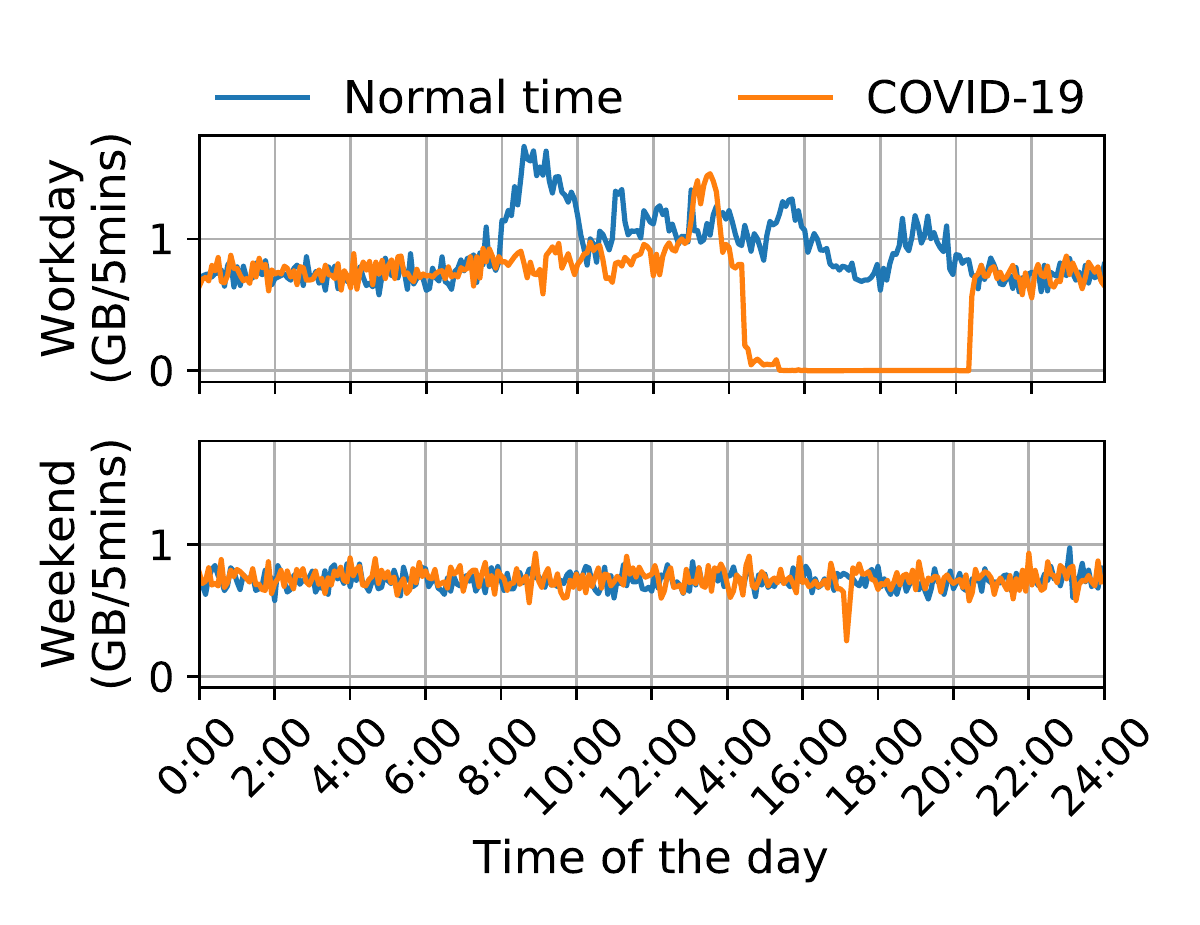}
   \label{fig:daily_BlueJeans}
 }
 \subfigure[Two-service traffic volume]{
   \includegraphics[width=.23\textwidth]{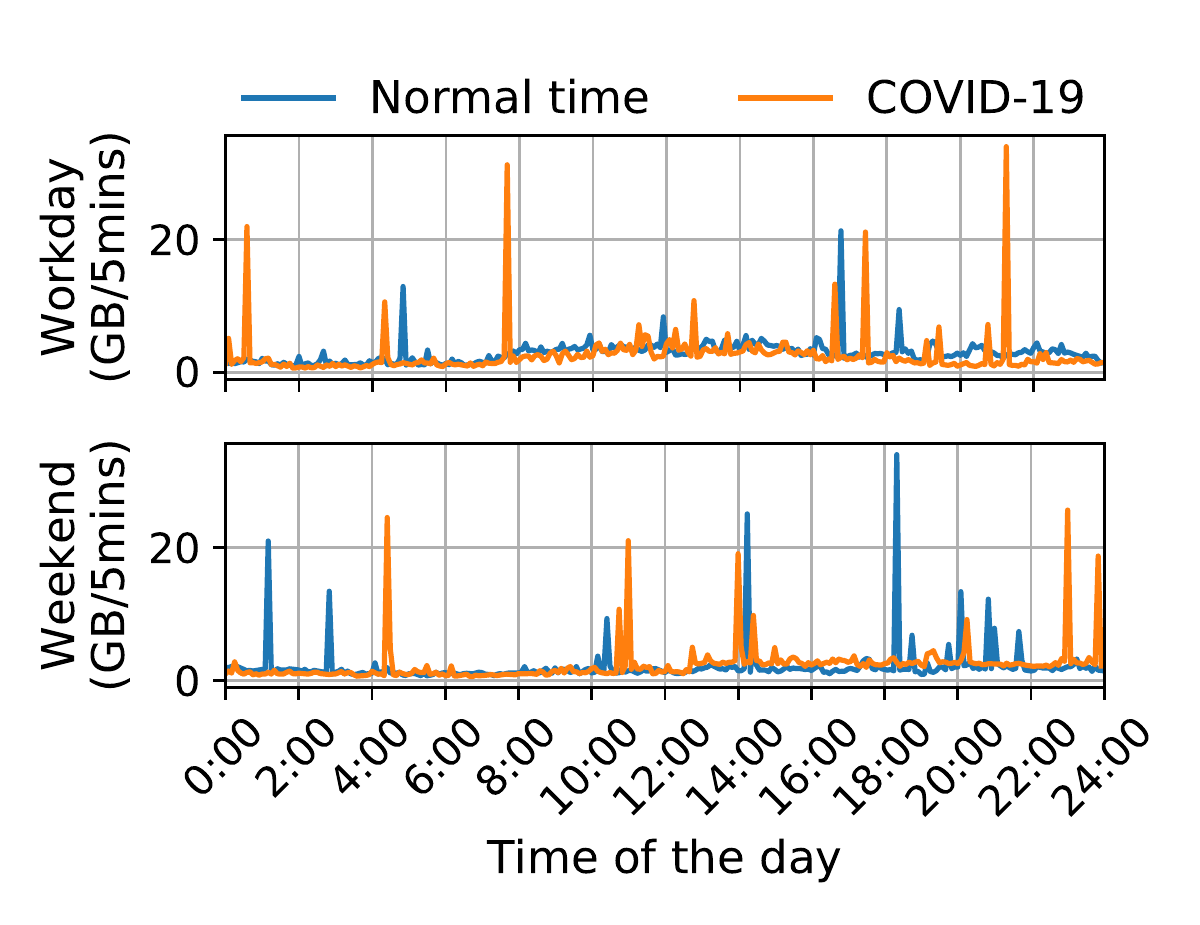}
   \label{fig:daily_twoservice}
 }
 \subfigure[High-high traffic volume]{
   \includegraphics[width=.23\textwidth]{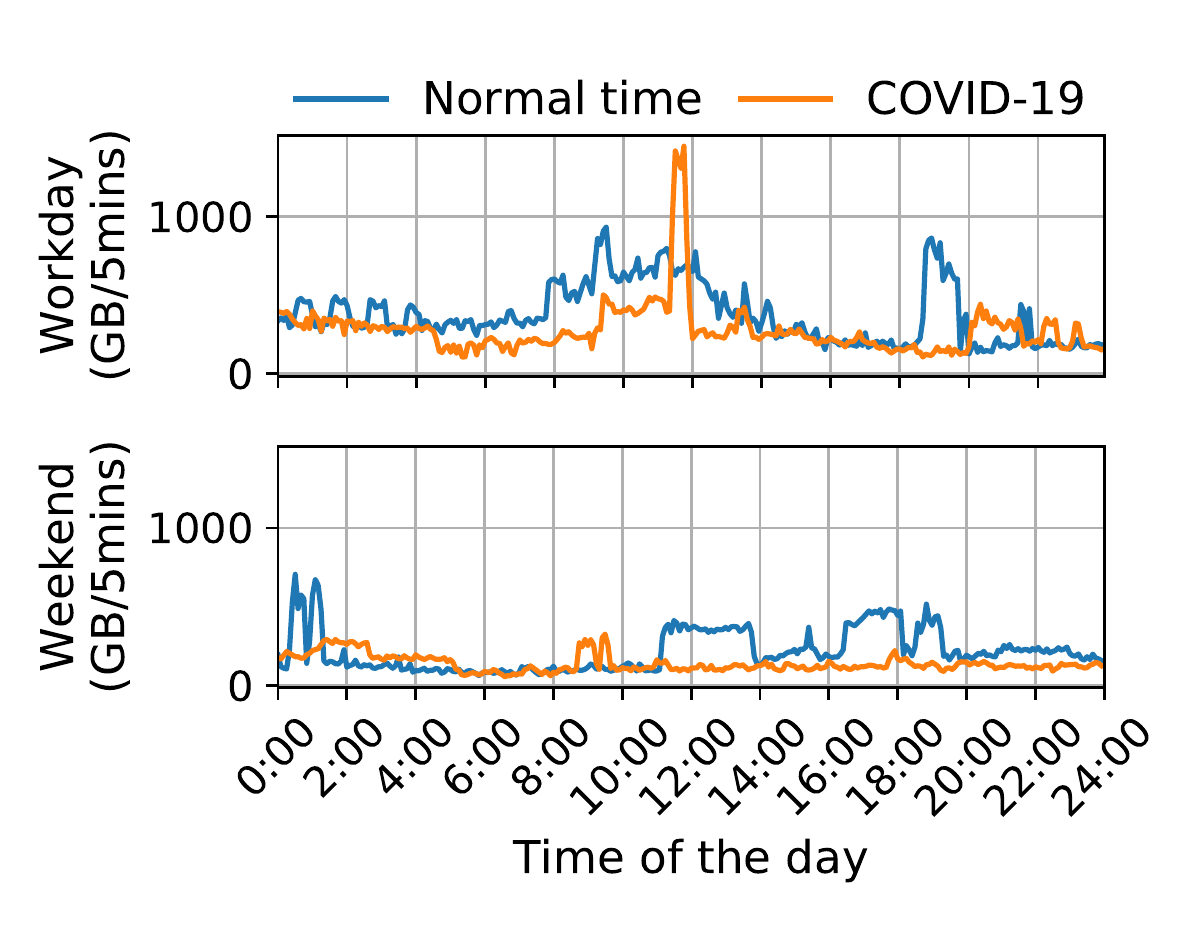}
   \label{fig:daily_highhigh}
 }
 \subfigure[No-service traffic volume]{
   \includegraphics[width=.23\textwidth]{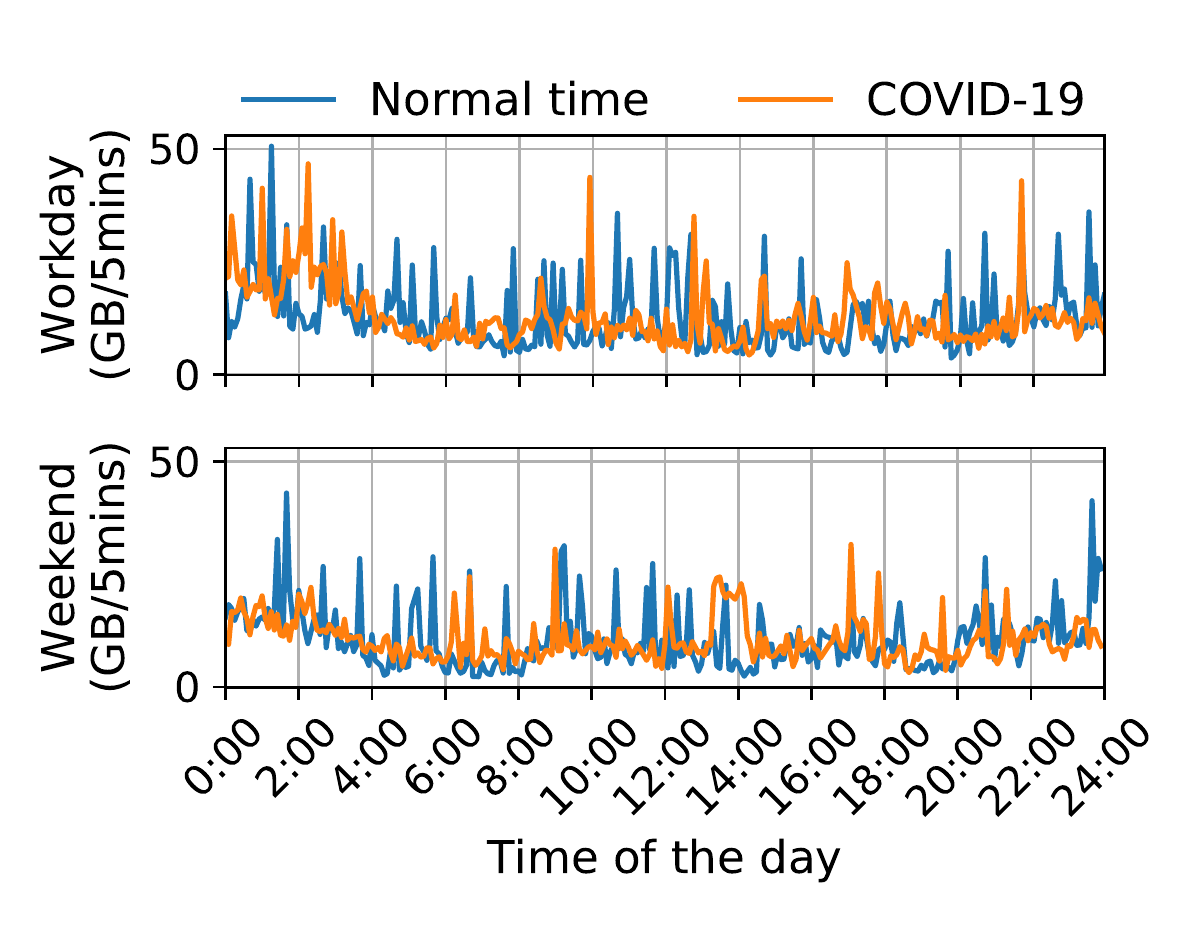}
   \label{fig:daily_noservice}
 }
 \subfigure[Odd-port traffic volume]{
   \includegraphics[width=.23\textwidth]{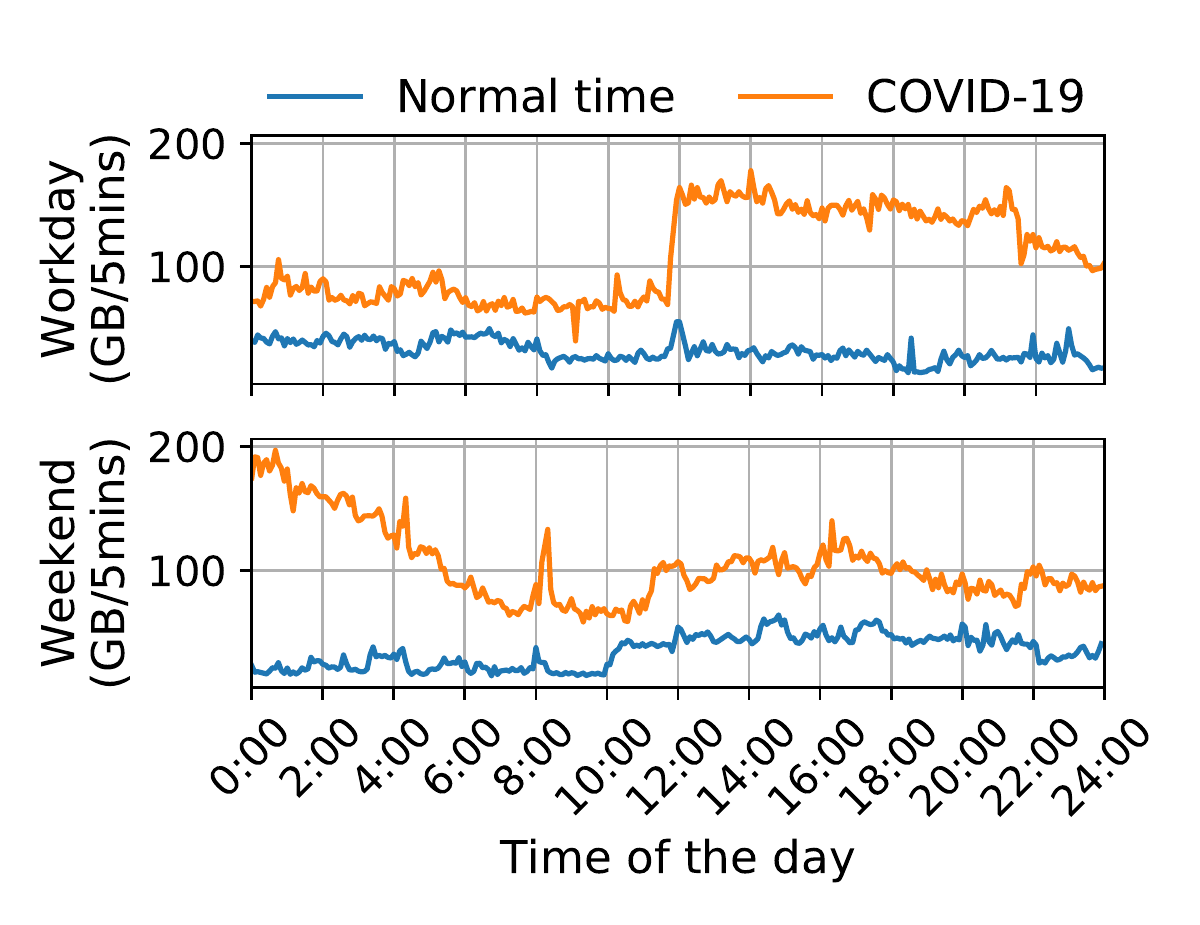}
   \label{fig:daily_oddprot}
 }
\caption{Traffic changes of workday and weekend}
\label{fig:daily_more}
\end{figure*}

Figures~\ref{fig:daily_more} illustrate more examples of the traffic changes for different protocols and applications during a particular day.
The upper subplots show the traffic distributions during a workday and the bottom subplots show the traffic distributions during a weekend.
The same as what we presented in Section~\ref{sec:daily}, for the workday traffic comparisons, we pick a Thursday (March 5) before the transition period and a Thursday (April 9) after the transition period; 
for the weekend traffic comparisons, we pick a Saturday (March 7) before the transition period and a Saturday (April 11) after the transition period.

